\documentclass{acm_proc_article-sp}

\usepackage{epsf}
\usepackage{color}

\usepackage{epsfig}
\usepackage{graphicx}
\usepackage{subfigure}
\usepackage{amsmath,amssymb}
\usepackage{algorithm}

\usepackage[noend]{algorithmic}
\usepackage{multirow}
\usepackage{xspace}
\usepackage{url}
\usepackage[normalem]{ulem}

\newcommand{\red}[1]{\textcolor{black}{#1}}
\newcommand{\blue}[1]{\textcolor{black}{#1}}

\newcommand{\erase}[1]{}

\newtheorem{problem}{Problem}

\newtheorem{definition}{Definition}

\newtheorem{lemma}{Lemma}
\newtheorem{theorem}{Theorem}

\newtheorem{example}{Example}

\def\0{{\mathbf 0}}
\def\1{{\mathbf 1}}

\newcommand{\topk}{top-$k$\xspace}
\newcommand{\RS}{recommender system\xspace}
\newcommand{\hw}{heavy-weight\xspace}
\newcommand{\len}{\mbox {$\ell$}\xspace}
\newcommand{\DP}{Dynamic Programing\xspace}
\newcommand{\Rank}{Rank Join\xspace}
\newcommand{\RJ}{{\sc NextHeavyPath}\xspace}
\newcommand{\Main}{{\sc HeavyPath}\xspace}
\newcommand{\hph}{{\sc HeavyPathHeuristic}\xspace}
\newcommand{\hpp}{HPP\xspace}
\newcommand{\MainPlus}{{\sc HeavyPath+}\xspace}
\newcommand{\RJPlus}{{\sc NextHeavyPath+}\xspace}
\newcommand{\Buf}{{\sc LastBufferIndex}\xspace}

\newcommand{\nop}[1]{}
\newcommand{\from}[2]{{\sc from #1:}{\bf #2}}

\newcommand{\eat}[1]{}
\newcommand{\para}[1]{\medskip\noindent {\bf #1.}}

\newcommand{\wmax}{\mbox {$w_{max}$}}
\newcommand{\wmin}{\mbox {$w_{min}$}}

\eat{
\makeatletter
\def\thm@space@setup{%
  \thm@preskip=4pt \thm@postskip=2pt
}
\makeatother
}

\newtheorem{myprop}{\textbf{PROPOSITION}}
\newtheorem{mytheo}{\textbf{THEOREM}}
\newtheorem{mylemma}{\textbf{LEMMA}}
\newtheorem{mylem}{\textbf{LEMMA}}
\newtheorem{myfact}{\textbf{FACT}}
\newtheorem{myobs}{\textbf{OBSERVATION}}
\newtheorem{myopt}{\textbf{OPTIMIZATION}}

\newlength{\figwidth}
\newlength{\figthree}
\newlength{\figfour}
\begin{document}
\title{
Finding Heavy Paths in Graphs: A Rank Join Approach
}

\author{
\begin{tabular}{ccc}
Mohammad Khabbaz & Smriti Bhagat  & Laks V.S. Lakshmanan
  \end{tabular}
\\
\smallskip
\\
\begin{tabular}{cc}
  University of British Columbia \\
  Vancouver, BC, Canada \\
 \{mkhabbaz, smritib, laks\}@cs.ubc.ca \\
\end{tabular}
}

\date{}

\maketitle

\begin{abstract}
Graphs have been commonly used to model many applications. 
A natural problem which abstracts applications such as 
itinerary planning, playlist recommendation, and flow analysis in information networks 
is that of finding the heaviest path(s) in a graph.
More precisely, we can model these applications as
a graph with non-negative edge 
weights, along with
a monotone function such as sum, which aggregates edge weights into a path weight, 
capturing some notion of quality. 
We are then interested in 
finding the \topk\ heaviest simple paths,
i.e., the $k$ simple (cycle-free) paths with the greatest weight, 
whose length equals a given
parameter $\len$. We call this the \emph{Heavy Path Problem} (HPP).
It is easy to show that the problem is NP-Hard.

In this work,
we develop a practical approach to
solve the Heavy Path problem
by leveraging a strong 
connection with
the well-known Rank Join paradigm. We first present an algorithm by
adapting the Rank Join algorithm.  
We identify its 
limitations and develop a new exact algorithm called \Main\ and  
a scalable heuristic algorithm.
We conduct a comprehensive set of experiments on three real data sets
and show that \Main outperforms the baseline algorithms significantly, with respect 
to both \len and $k$. Further, our heuristic algorithm scales to longer lengths, 
finding paths that are empirically within 50\% of the optimum
solution or better under various settings, and takes only 
a fraction of the running time compared to the exact algorithm.

\eat{
In this work,
we develop a practical approach to \red{\erase{solving this problem}
solve the Heavy Path problem}
by establishing a strong connection \red{\erase{between the \hpp and} with}
the well-known Rank Join paradigm. We present an adaptation of the Rank Join
algorithm and tailor it to self-joins over the list of edges. We present
novel thresholding strategies and develop exact and heuristic algorithms for \hpp. We
conduct a comprehensive set of experiments on three real data sets
and show our algorithms can efficiently find the \topk heaviest paths on these data sets \red{and
significantly outperform baseline algorithms}.
We show our exact algorithm\red{\erase{s}} scale\red{s} well with respect to both \len and $k$.
In addition, our heuristic approach finds paths that are empirically  \red{within} 50\%
of the optimum
solution \red{or better} under various settings\red{, while taking a fraction of the running time
compared to the exact algorithm}.
}
\eat{
up to path lengths of $35$ and
our heuristics go beyond path lengths of $100$.
}
\eat{
In addition, our heuristic approaches find paths that are within 50\% of the optimum
solution in most settings.
}


\end{abstract}

\section{Introduction}
\label{sec:intro}

With increasing availability of data on real social and information
networks, the past decade has seen a surge in the interest in mining and
analyzing graphs to enable a variety of applications.
For instance, the problem of finding dense subgraphs
is studied in the context of finding communities in a social network~\cite{Gibson:05},
and the Minimum Spanning Tree (MST) problem is used to find
 teams of experts in a network~\cite{Lappas:09}. 
\eat{
A problem that has received relatively less attention in the database
community is that of finding heavy paths
of a given length in a graph. Precisely,
given a weighted graph as input, the Heavy Path Problem (\hpp)
is to find \topk heaviest simple (i.e., cycle-free) paths of length \len,
where the weight of a path is defined as the sum 
of weights of edges that compose the path.
} 
\red{Several real applications can be naturally modeled using a 
problem that we call the \emph{Heavy Path Problem} (\hpp): given a 
weighted graph as input, find \topk heaviest simple (i.e., cycle-free) paths of length \len,
where the weight of a path is defined as a monotone aggregation (e.g., sum) 
of weights of the edges that compose the path. Surprisingly, this problem 
has received relatively less attention in the database community.} 

We present a few concrete \red{applications of \hpp } below.

\noindent {\bf (1)}  \red{In \cite{Hansen:09}, Hansen and Golbeck motivate a class of 
novel applications that 
recommend collections of objects, 
\red{for instance, an application that recommends music (or video)
playlists of a given length.} 
They emphasize three key properties for desirable lists: 
value of individual items \red{(say songs)}, 
co-occurrence interaction effects, 
and order effects including placement and arrangement of items. } 
\erase{\red{\erase{Consider an}} In such application\red{s,} 
\red{\erase{that}} \red{one would like to} recommend\red{\erase{s}} music (or video) 
playlists of a given length.}%
\red{\erase{A graph can be abstracted}}%
We can abstract a graph from user listening history, 
where a node represents a song \red{of 
high quality as determined by user ratings} 
and an edge between a pair of songs exists if they were listened to together in one session.
The weight on an edge represents how frequently the songs were listened together by users.
Heavy paths in such a graph correspond to playlists 
\red{of high quality songs} 
\red{that are} frequently enjoyed together. 

\noindent {\bf (2)} Consider an application that recommends an itinerary for visiting a given number
of \red{popular} tourist spots or points of interest (POIs) in a city~\cite{Choudhury:10}.
The POIs  can be modeled as nodes in a graph, and an edge \erase{exists between}%
represents that a pair of POIs \erase{if they are}%
is directly connected by road/subway link. 
The weight on an edge
is a decreasing function of the travel time (or money it costs) to go from one POI to
another.
Heavy paths in such a graph correspond to 
 itineraries with small overall travel time (or cost) for visiting a given number of popular POIs. 
 
\noindent {\bf (3)} Say we want to analyze how a research field has evolved.
Given a citation network of research papers and additional information on
the topics associated with papers (possibly extracted from the session title of the
conference, or from keywords etc.), we can abstract a ``topic graph''.
Nodes in a topic graph represent topics,
and an edge between two topics exists if a paper belonging to one topic cites a paper
belonging to another topic. The weight on an edge may be the normalized
frequency of such citations.
Heavy paths in such a graph capture strong flows of ideas across topics. 

\eat{ 
Apart from these specific applications, as we will show in Section~\ref{sec:probdef}, 
\hpp is closely 
related to the Traveling Salesperson Problem (TSP), more precisely, to its 
variant $\len$-TSP\red{\footnote{In the literature it is called $k$-TSP, where
$k$ is the given length of the path. We refer to it as \len-TSP to avoid confusion
with the parameter $k$ used for number of paths in our \topk 
setting.}}~\cite{Arora:06}. Given a graph with edge weights representing distances, 
the TSP problem is to find a tour of minimum distance that visits every node exactly once; 
the $\len$-TSP problem is to find a tour of minimum distance that visits any $\len$ 
distinct nodes exactly once. 
It is well known that TSP and $\len$-TSP have widespread 
applications across a variety of domains. 

In all these applications, we may be interested in not just the top-$1$ path, but \topk heavy paths.
Finding the \topk heavy paths of a given length \len is not trivial.
Bansal et al.~\cite{Bansal:07} studied a special case of this problem where the input graph
was \len-partite (each partition corresponds to a timestamp) and acyclic.}

\red{In addition, Bansal et al.~\cite{Bansal:07} modeled the problem of 
identifying temporal keyword clusters in blog posts (called Stable Clusters Problem in 
\cite{Bansal:07})   
as that of finding heavy paths in a graph. 
In all these applications, we may be interested in not just the top-$1$ path, but \topk heavy paths, 
for instance, $k$ itineraries or playlists to choose from. 
Finding the \topk heavy paths of a given length \len is not trivial.
The setting studied in~\cite{Bansal:07} is 
a special case of the \hpp } where the input graph
was \len-partite (each partition corresponds to a timestamp) and acyclic. 
They proposed three algorithms based on
BFS (breadth first search), DFS (depth first search) and the TA (threshold algorithm).
Due to the special structure of their graph,
i.e., the absence of cycles and only a subset of  nodes acting as starting points for graph
traversal, they were able to efficiently adapt the BFS and DFS algorithms and show that
these adaptations outperform a TA-based algorithm.
\erase{The problem we consider }%
\red{Our problem} is more general in that the graphs are \emph{not} \len-partite,
and typically contain many cycles. \erase{Thus,
the algorithms proposed by Bansal et al. are not directly applicable in our setting.}%
Adaptations of the algorithms \red{proposed in~\cite{Bansal:07}} 
lead to expensive solutions. For example,
an adaptation of DFS would require performing \len-deep recursion
starting at each node, which is prohibitively expensive for large general graphs.

We present an efficient algorithm for \hpp called \Main,
based on the \Rank paradigm, \red{as detailed in Section~\ref{sec:RJ}}.
\eat{Notice that building a path of length \len\ in a graph  
can be viewed as $\len-1$ self-joins on a table containing the edges stored in a weight
sorted manner, where the join conditions ensure edges match on their end points
and avoid cycles. Thus, this naturally falls within the \Rank paradigm.}
In the last decade, there has been substantial amount of work 
on \Rank~\cite{Ilyas:03,Ilyas:04,Ilyas:05,Schnaitter:8,Schnaitter:10}.
\eat{While there has been discussion of multi-way \Rank, the majority of the work
has focused on binary \Rank. 
Moreover,}
\red{However,} the experimental results reported have
confined themselves to \Rank over a small number of relations.
In the driving applications mentioned above, there is often the need to 
discover relatively long heavy paths.
For example, in playlist recommendation, a user may be interested in getting a list
containing several tens of songs, and in itinerary planning, recommendations consisting of 5-15
POIs for the tourist to visit within a given time interval (e.g., a day or a week) are
useful. 
In general, {\sl it is computationally more demanding to
find \topk heaviest paths as the path length increases.
There is a need for efficient algorithms for meeting this challenge.} 

The approach we develop in this paper
is able to scale to longer lengths compared with classical \Rank.
By exploiting the fact that the relations being joined are identical,
we are able to provide smarter strategies for
estimating the requred thresholds, and hence terminate the algorithm
sooner than classical \Rank.
In addition, we carefully make use of random accesses in the context of \Rank 
as an additional means of ensuring that heavy paths are discovered
sooner, and the thresholds are aggressively lowered. 
Finally, it turns out that all exact algorithms considered (including
 \Rank \red{and}  our proposed \Main algorithm\erase{and its enhancement
\MainPlus}) run out of memory
when the length of the desred path 
\red{exceeds 10 on some data sets.}
\eat{As mentioned earlier, indeed 
\hpp is closely related to the \len-TSP, 
\erase{problem studied in the theory community,}%
which is not only NP-hard but does not have any
bounded approximations \cite{Arora:06}.}
\red{As we will see in Section~\ref{sec:probdef}, \hpp is NP-hard.}
In order to deal with this, we develop a heuristic approach
that works with the allocated memory and allows us to estimate
the distance to the optimum solution for a given problem instance.
We empirically  show
that this heuristic extension scales well even for paths of length 100
on real data sets.

We make the following contributions in the paper: 
\begin{itemize}
\item We formalize the problem of finding \topk heavy paths of a given length for
general  graphs, and establish the connection between \hpp
and the \Rank framework for constructing heavy paths (Section~\ref{sec:probdef}).
\item We present a variety of exact algorithms, including 
\red{two baselines obtained by adapting known algorithms, 
a simple adaptation of \Rank for computing heavy paths, and} 
an efficient adaptation of \Rank called \Main. \erase{We also present improvements
that lead to another efficient exact algorithm called \MainPlus.
With simple modifications we can turn the exact algorithms into heuristics
with the nice property that we can derive an empirical approximation ratio
in a principled manner (Sections~\ref{sec:algo}, \ref{sec:opt}, and \ref{sec:hp}).}%
\red{With simple modifications, we can turn \Main into a heuristic 
algorithm with the nice property that we can derive an empirical approximation ratio
in a principled manner (Sections~\ref{sec:algo}, \ref{sec:opt}, and \ref{sec:hp}).} 
\item We present a comprehensive set of experiments to evaluate and compare the
efficiency and effectiveness of the different algorithms on three real datasets:
last.fm, Cora, and Bay that respectively model the three motivating application\red{s} 
described earlier\erase{in this section}. \erase{We find that our algorithms are orders of magnitude
faster, and can find the exact
solution for paths that are several hops longer as compared with the
\DP and classical \Rank approaches (Section~\ref{sec:expn}).}%
\red{ 
Our results show that \Main is orders of magnitude faster than the baselines and 
\Rank, and can find exact solutions for paths that are several hops longer as compared 
with all other algorithms. In addition, our heuristic algorithm finds paths that are empirically  
within 50\% 
of the optimum
solution or better under various settings, while taking a fraction of the running time 
compared to the exact algorithm (Section~\ref{sec:expn})}.
\end{itemize}
We review the related work in Section~\ref{sec:related} and conclude in Section~\ref{sec:concl}.

\section{Problem Studied}
\label{sec:probdef}


Given a weighted graph $G(V,E,W)$, where
weight $w_{(u,v)}$ \red{$:= W(u,v)$} represents the non-negative weight\footnote{Non-negativity
is not a requirement, but is intuitive in most applications.} 
on edge $(u,v)\in E$,
and parameters $k$ and \len, the {\em Heavy Path Problem (\hpp)} is to find
the \topk heaviest \red{simple} paths of length $\len$,
i.e., $k$ \red{simple} paths of length \len with the highest weight.
\erase{Here, a}%
\red{A simple path} of length \len is \erase{a simple path,} a sequence of 
nodes $P = (v_0, \ldots,
v_{\ell})$ such that $(v_i, v_{i+1})\in E$, $0\leq i < \len$, and there are no
cycles, i.e., the nodes $v_i$ are distinct.
Unless otherwise specified, in the rest of the paper, we use the term
path to mean simple path.
We note that our framework allows path weights defined using any
monotone aggregate function of edge weights.
For simplicity, we define the weight of a path $P = (v_0, ..., v_\ell)$, as 
$P.weight = \sum_{j=0}^{\ell-1} w_{(j,j+1)}$.


We note that the heavy path problem for a given parameter \len is equivalent
to the well-known \len-TSP\footnote{In the literature it is called $k$-TSP, where
$k$ is the given length of the path. We refer to it as \len-TSP to avoid confusion
with the parameter $k$ used for number of paths in our \topk setting.} 
\red{($\ell$-Traveling Salesperson) problem~\cite{Arora:06}}, defined as follows: Given a graph with
non-negative edge weights, find a path of minimum weight that passes through
any $\len+1$ nodes. It is easy to see that for a given length \len,
a path $P$ is a solution to \len-TSP
iff it is a solution to \hpp (with $k=1$) on the same graph but with
edge weights \erase{appropriately} modified \red{as follows}: let
$w_{(u,v)}$ be the weight of an edge $(u,v)$ in the \len-TSP instance, and
\wmax\ be the maximum weight of any edge in that instance;  
then the edge weight in the \hpp instance is
$w'_{(u,v)}=1-w_{(u,v)}/\wmax$. 
\eat{ 
{\sl It is well known that not only is TSP NP-complete but it does not admit
any bounded approximation. }
These properties are inherited by
\len-TSP~\cite{Arora:06}. Thus, HPP is a hard problem and is hard to approximate.
\red{FIX THIS!}
} 
\blue{It is well-known that TSP and $\ell$-TSP are NP-hard and the reduction above 
shows HPP is NP-hard as well. } 
In general, HPP can be defined for both directed and undirected graphs.
Our algorithms and results \red{focus on the undirected case, 
but can be easily extended to directed graphs.
\erase{are applicable to both. In the rest of this
paper, without loss of generality we assume $G$ is undirected. }}

\begin{figure}[t]
  \centering
  \subfigure[]{ 
  \includegraphics[scale = 0.75]{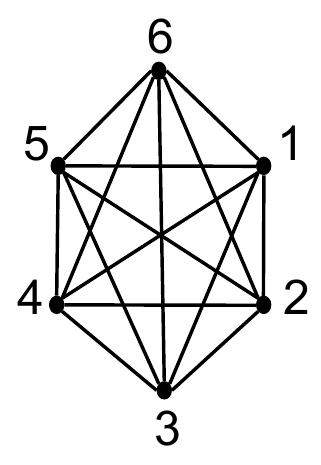}
  \label{fig:cliqueGraph}
}   
\subfigure[]{  
\begin{minipage}[b]{0.4\linewidth}
   \tiny
   \begin{tabular}{|r|c|} \hline 
    Edge & Edge Weight \\ \hline 
    (1,2)& 0.93 \\
    (2,3)& 0.93 \\
    (1,3)& 0.87 \\
    (2,4)& 0.77 \\
    (1,6)& 0.76 \\
    (2,5)& 0.73 \\
    (3,4)& 0.73 \\
    (1,4)& 0.73 \\
    (5,6)& 0.72 \\
    (3,5)& 0.70 \\
    (1,5)& 0.70 \\
    (1,6)& 0.70 \\
    (4,5)& 0.69 \\
    (3,6)& 0.66 \\
    (4,6)& 0.58 \\  \hline 
    \end{tabular}
 \vspace{-3pt}
\end{minipage}
\label{tab:cliqueWts}
}
\vspace{-6pt}
\caption{Example graph for playlist recommendation, and corresponding edge weights table}
\label{fig:Clique}
\vspace{-10pt}
\end{figure}


\red{
We next briefy discuss the utility of heavy paths for applications, compared with cliques. 
Some data sets like 
one that represents the road network (see Bay dataset in 
Section~\ref{sec:expn}) 
naturally do not exhibit large cliques. 
On such data sets, 
heavy paths order the items (i.e., nodes) in a sequence determined by the graph structure. On the 
other hand, some data sets may contain large cliques. 
Figure \ref{fig:Clique} shows an example of the co-listening graph 
for the playlist recommendation application (this is a subgraph induced over 
6 songs from the last.fm dataset used in our experiments -- see Section~\ref{sec:expn}). 
Say the task is to find heavy paths of length $\len=4$. 
Even when the graph is a clique and any permutation of 5 nodes will result in a path of length 4, 
the order in which the nodes are visited can make a significant difference to the overall 
weight of the path. 
Here, the heaviest simple path of $\len=4$ 
is obtained by visiting nodes in the order 6--1--2--3--4 has a weight of 3.35. 
In contrast, a different permutation of the nodes 4--3--6--1--2 has a weight of 3.08. 
}


\section{Finding Heavy Paths}
\label{sec:algo}

\subsection{Baselines}

An obvious algorithm for finding the heaviest paths of length \len
is performing a depth-first search (DFS) from each node, with the search
limited to a depth of \len, while maintaining the
\topk heaviest paths. This is an exhaustive algorithm and is not expected to scale.
A somewhat better approach is dynamic programming. Held and Karp~\cite{Held:62} proposed a
dynamic programming algorithm for TSP, that works with a notion of ``allowed nodes''
on paths between pairs of nodes\red{, which we adapt to HPP as follows.} 
\erase{For the heavy path problem, since \len can be
smaller than the number of nodes, it is more convenient to work with {\em avoidance
\red{\erase{sets} paths} }, 
defined as follows.} 
For a set of nodes $S$, we say
there is an $S$-avoiding path from node $x$ to $y$ provided none of the nodes on this
path other than $x, y$ are in the set $S$. E.g., the
path $(1,2,3)$ is $\{1,4,5\}$-avoiding but not
$\{2,4\}$-avoiding. The idea is to find the heaviest simple path of
length $\len$ ending at $j$ for every node $j$, and \red{\erase{finding} then find}
the heaviest among them. To find the heaviest simple path ending at $j$,
we find the heaviest simple $\{j\}$-avoiding path of length $\len$ that
ends at $j$.

The heaviest path of length $1$ (starting anywhere)
and ending at a given node $j$ is simply the heaviest edge ending
at $j$. In general, the heaviest (simple) $S$-avoiding path of length
$2\le l\le \len$ ending at $j$ is found by picking the heaviest among the following set of
combinations: concatenate the heaviest $S\cup\{y\}$-avoiding
path (from anywhere) of length $l-1$ to any
neighbor $y$ of $j$ where $j\in S$ and $y\not\in S$, with the edge $(y,j)$.

\erase{
Essentially, dynamic programming constructs all simple paths of length $\len$
in order to find the heaviest among them. But unlike DFS, it aggregates
path segments early, thus achieving some pruning. 
} 
\eat{For example, suppose there
are $p$ paths $P_1, ..., P_p$ of length $l-1$ that end at $y$, and all of them avoid the node $j$,
which is a neighbor of $y$.
Then for the purpose of finding the heaviest path of length $l$ ending at $j$, we just need
to remember the heaviest among $P_1, ..., P_p$. Another key observation exploited by the
dynamic program is that the start nodes of the paths we consider are immaterial and can be
dropped from further consideration.}
\erase{ 
More precisely, here are the
equations of the dynamic program. 
Letters $i,j,y$
denote variables which
will be instantiated to actual graph nodes when the program is run. $S$ denotes the
\red{\erase{avoidance set} set of nodes to be avoided}. 
The variable $l$ will be instantiated in the range $[2, \len]$.} 
\eat{ 
\vspace{-3pt}
\begin{align*}
P^l_{i,j,S} &= \textsc{MAX}\{P^{l-1}_{i,y,S\cup\{y\}}\circ e(y,j) \mid (y,j) \in E, \\
  & y\not\in S, j\in S\} \\
P^1_{i,j,S} &= \begin{cases}
		\textsc{create}(P, e(i,j)) & \mbox{if} (i,j) \in E \\
		\textsc{null} & \mbox{otherwise}
		\end{cases}
\end{align*}
} 

\erase{ 
In the above equations, we can think of $P^l_{i,j,S}$ as a ``path object''
with properties ${path}$ and ${weight}$. Here, $P^l_{i,j,S}.{path}$ denotes the
heaviest $S$-avoiding path from $i$ to $j$ of length $l$, and $P^l_{i,j,S}.{weight}$ denotes
its weight. The operator
\textsc{MAX} takes a collection of path objects and  finds the object
with maximum weight among them. The ``$\circ$'' operator takes
a $P$ object and an edge \red{\erase{$e(u,v)$} $e(i,j)$}, concatenates the edge with the
$P.{path}$ and updates $P.{weight}$ by adding the weight of the
edge. Finally, $create(P, e(i,j))$ creates a
path object $P$ whose ${path}$ property is initialized to $(i,j)$ and whose
${weight}$ property is initialized to the edge weight of $(i, j)$. 
}

\erase{ 
We invoke the dynamic program above using $P^{\len}_{\$i, j, \{j\}}$
every node $j \in V$ , where we have left the start node as a variable $\$i$.
The heaviest path of length $\len$ in
the graph is the heaviest among the paths found above for various $j$. 
} 
%
\eat{
\begin{figure}
\vspace{-3mm}
  \centering
  \includegraphics[width=3in, height=1in]{example.eps}\\
\vspace{-2mm}
  \caption{Example Co-occurrence Graph. }\label{fig:example}
\vspace{-3mm}
\end{figure}

\begin{example}
\label{ex:ex1}
Figure~\ref{fig:example} shows an example that will be used as a running example
to illustrate various algorithms.
Consider $P^3_{I,5,\{5\}}$, i.e., asking for the heaviest simple path ending
at $5$. Note the start node is left as a variable and the avoidance set includes
only the end node $5$. A ``run'' of the dynamic program on this ``query'' is illustrated
in Figure~\ref{fig:dprun}. Notice the applications of \textsc{MAX} operator
(indicated via arcs under tree nodes) which allow multiple paths to be pruned
at multiple stages. Observe also the fact that start nodes are completely
``ignored'' in that among path segments starting at different start nodes, the
\textsc{MAX} operator picks only one, the heaviest one.
\end{example}
}
%

\erase{ 
\red{\erase{For finding the \topk heavy paths of a
given length for $k > 1$, all we need to do is} 
To find the \topk heavy path, for $k>1$, we can re}define
the \textsc{MAX} operator such that it works in an iterative mode and finds the next
heaviest path segment (ending at a certain node, of a given length, and avoiding
certain set of nodes).
}  
We can apply this idea recursively and easily extend the
dynamic program for finding \topk heaviest paths of a given length. We skip  the
details for brevity. The equations and details of the dynamic 
programming algorithm can be found in Appendix~\ref{app:dp}. 
Clearly, both dynamic programming and DFS  have an
exponential time complexity, \erase{and we only use them as baselines in this paper.} 
\red{although unlike DFS, dynamic programming aggregates
path segments early, thus achieving some pruning. Both algorithms are used 
as baselines in this paper.} 

\eat{
\begin{figure}
\begin{center}
\scalebox{0.5}{\input{dprun.pstex_t}}
\end{center}
\caption{\label{fig:dprun} A run of the dynamic program. Arcs under tree nodes
indicate application of \textsc{MAX} operation.}
\end{figure}
}

\subsection{Rank Join Algorithm for \hpp}
\label{sec:RJ}
\eat{
The methods discussed in the previous section potentially explore all
paths of length \len. In particular, they lack the ability to ignore 
paths that have no hope of making it to the \topk\ and perform early
termination. Indeed, \Rank~\cite{Ilyas:03,Schnaitter:10} is an efficient algorithmic
framework specifically designed for determining \topk\ results of a
join query. We briefly review this framework next and discuss how it
may be adapted to the \hpp\ problem.
}

\begin{figure}[t]
\subfigure[No random access]{
\includegraphics[scale=0.5]{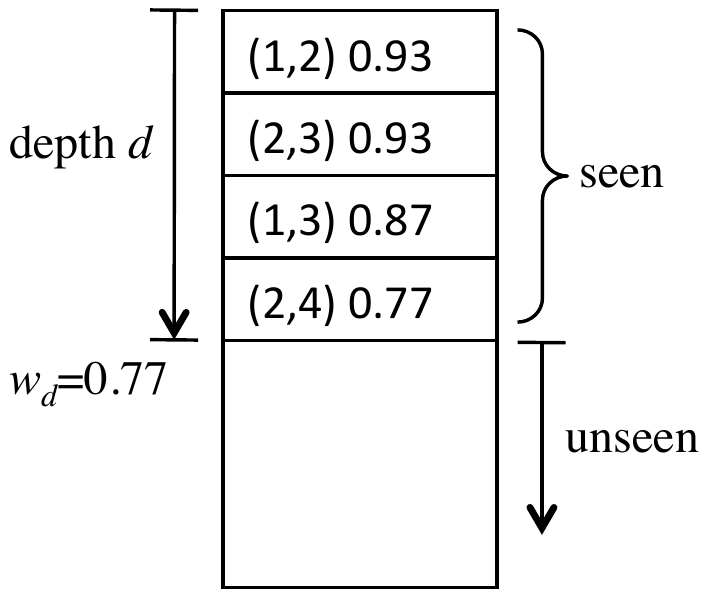}
\label{fig:depth1}
}
\subfigure[With random access]{
\includegraphics[scale=0.5]{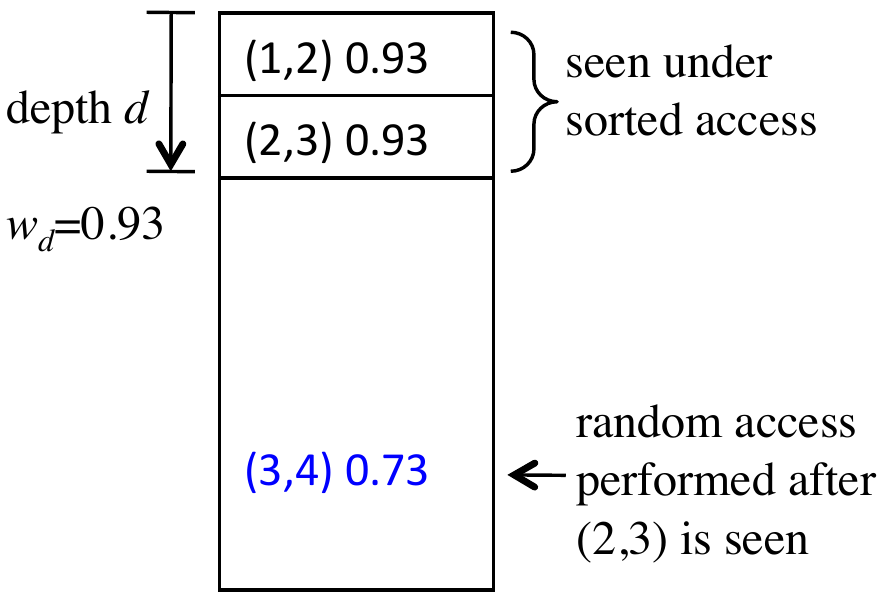}
\label{fig:depth2}
}\label{fig:access}
\vspace{-10pt}
\caption{Adapting Rank Join for HPP}
\vspace{-10pt}
\end{figure}

The methods discussed above are unable to prune paths
that have no hope of making it to the \topk. Rank Join~\cite{Ilyas:03,Schnaitter:10} 
 is an
efficient algorithmic framework specifically designed for finding
\topk results of a join query. We briefly review it and discuss how
it may be adapted to the HPP problem.

\para{Background}
We are given $m$ relations $R_1, ..., R_m$. Each relation contains a
\emph{score} attribute and is sorted on non-ascending score order.
The score of each tuple in the join
$R_1 \Join \cdots \Join R_m$ is defined as $f(R_1.score,
\ldots, R_m.score)$ where $R_i.score$ denotes the score of the
current tuple in $R_i$ and $f$ is a monotone aggregation function.
The problem is to find the \topk\ join results for a given
parameter $k$. The \Rank algorithm proposed by Ilyas et al.~\cite{Ilyas:03}
works as follows. (1) Read tuples in sorted order from each relation
in turn. \red{This is called \emph{sorted access}.}  
(2) For each tuple read from $R_i$, join it with all tuples from
other relations read so far and compute the score of each result. Retain the $k$
result tuples with the highest score. (3) Let $d_i$ be the number of tuples
read from $R_i$ so far and let $t_i^j$ be the tuple at position $j$ in $R_i$.
Define a threshold $\theta := \max\{
f(t_1^{d_1}.score, t_2^1.score, \ldots, t_m^1.score), \ldots,$
$f(t_1^1.score, \ldots,$\\  $t_i^{d_i}.score, \ldots, t_m^1.score),$ $\ldots,
f(t_1^1.score, \ldots,$  $t_i^1.score, \ldots,$\\   $t_m^{d_m}.score)\}$.
This threshold is the maximum possible score of any future join result.
(4) Stop when there are $k$ join results with score
at least $\theta$. \red{It is clear no future results can have a score 
higher than that of these $k$ results.}  

In the rest of the paper, 
we will assume $f$ is the sum function.
Our results and algorithms carry over to any monotone
aggregation function.

\para{\blue{Adapting \Rank for \hpp}}
Given a weight sorted table $E$ of edges, \hpp can be solved using
the \Rank framework. Indeed, paths of length \len\ can be
found via an \len-way self-join of the table $E$, where the
join condition can ensure the end node of an edge matches the
start node of the next edge and cycles are disallowed. In
particular, whenever a new edge is seen from $E$ (under \red{\emph{sorted
access}}), it is joined with every possible combination of $\len-1$
edges that have already been seen to produce paths of length $\len$. Only the
$k$ paths with the highest weight are retained in the buffer.
Let $d$ be the depth of the edge (tuple) last seen under sorted access and
let $w_d$ denote the weight of the edge seen at depth $d$ and \wmax\ be the
maximum weight of any edge.
The threshold is updated as $\theta := w_d + (\len-1)\wmax$. We stop when
$k$ paths of length \len\ are found with weight at least $\theta$.
For simplicity, we refer to this adapted \Rank algorithm 
as just \Rank.

\red{Consider the example graph in Figure~\ref{fig:Clique}. 
A \Rank performed on 4 copies of the edge weight table, with an appropriately defined join 
condition can be used to compute \topk heavy paths of length $\len=4$. 
The \Rank algorithm proceeds by scanning the edge table in sorted order of the edge weights. 
Figure~\ref{fig:depth1} illustrates a snapshot during the execution, where the algorithm 
has ``seen'' or scanned 4 tuples from the edge table, and therefore the depth $d=4$. Also, 
the weight of edge seen at depth 4 $w_{d}=0.77$ and $\wmax=0.93$. 
At this depth, the threshold is updated as $\theta = 0.77 + (4-1)*0.93 = 3.56$.
For this particular example, \Rank scans all edges in the table before being able to output
the top-1 path of $\len=4$ as the weight of the top path (i.e., 3.35) 
is not over the threshold $\theta=0.58+(4-1)*0.93=3.37$ even at depth  $d = |E|$.}

\section{\red{Limitations and Optimizations}}
\label{sec:opt} 

\red{ 
In this section, we make some observations about the limitations of
using \Rank for finding \topk\ heavy paths in general graphs and discuss possible
optimizations. We establish some key properties of (the above adaptation of)
\Rank, which will pave the way for a more efficient algorithm in 
Section~\ref{sec:hp}.  
} 

\begin{myobs}
Let $P$ be a path of length \len\ and suppose $e$ is the lightest
edge on $P$, i.e., its weight is the least among all edges in $P$.
Then until $e$ is seen under sorted access, the path $P$ will not
be constructed by \Rank.
\end{myobs}

\begin{figure}
  \centering
  \includegraphics[scale=0.8]{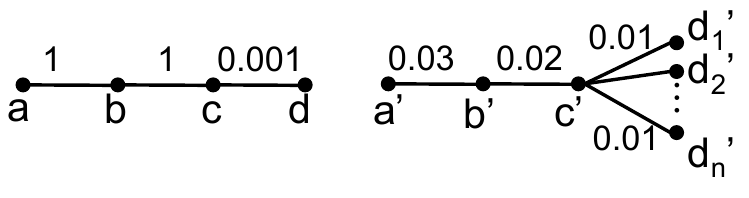}
    \vspace{-10pt}
  \caption{Example instance of HPP. Graph has one heavy path and $n$ lighter paths.}
  \label{fig:ex1}
  \vspace{-10pt}
\end{figure}

\eat{
Consider the graph in Figure~\ref{fig:ex1}.
It consists of one path of weight $2.001$
(the top path) and $n$ paths of weight $0.03+0.02+0.01 = 0.06$.
Clearly, the top path (call it $P$) is the unique heaviest path of length $3$.
The observation above suggests that until the edge $(c,d)$ 
is seen, \Rank
cannot determine the top path.
However, by the time $(c,d)$ 
is seen, \Rank will
produce every path of length \len\ made up of edges that are no lighter
than $0.001$.
}
\red{ We illustrate this observation with an example graph in Figure~\ref{fig:ex1}
for the task of finding the heaviest path of length 3. 
\erase{For example, consider finding the heaviest path of length $3$ in
the graph of Figure~\ref{fig:ex1}.}}%
The graph consists of one path of weight $2.001$
(the top path) and $n$ other paths of weight $0.03+0.02+0.01 = 0.06$, all of length 3.
Clearly, the top path (call it $P$) is the unique heaviest path of length $3$. However,
\Rank must wait for the edge with weight $0.001$
(the lightest in the graph) to be seen before it can find and report $P$. Until
then, it would be forced to produce (and discard!) the $n$ paths of length
$3$, each with weight $0.06$. Since $n$ can be arbitrarily large, \Rank
is forced to construct an arbitrarily large number of paths most of which are
useless. Indeed, as an extreme case, we can show the following result on \Rank,
for the special case $k=1$.

\begin{mylem}
\label{theo:rjcost}
Consider an instance of HPP where the weight of the heaviest path of length \len is smaller than
$w_{min} + (\len - 1) \times w_{max}$, where \wmax and \wmin are the weights of the
heaviest and lightest edge in the graph, respectively. For this instance, \Rank produces every path
of length \len.
\end{mylem}

\begin{proof}
The \Rank algorithm stops when there is a path of length
\len in the buffer whose weight is not smaller than the threshold. Assume the heaviest path
$P$ is lighter than  $w_{min} + (\len - 1) \times w_{max}$, and \Rank stops
after seeing an edge $e$ with weight $w \ge \wmin$. In this case, 
the threshold is
$\theta = w + (\len-1)\wmax \ge \wmin + (\len - 1)\wmax > P.weight$, so by definition,
\Rank cannot terminate, a contradiction! 
This shows, it must see all edges with weight \wmin before
halting. By this time, by definition, \Rank will have produced all paths of length
\len. 
\end{proof}

The observation and lemma motivate the following optimization for finding heaviest
paths.

\begin{myopt}
We should try to avoid delaying the production of a
path of a certain length until the lightest edge on the path is seen
under sorted access. One possible way of making this happen is via
random access. However, random accesses have to be done with care in
order to keep the overhead low.
\end{myopt}

\red{Figure~\ref{fig:depth2} shows that an edge (3,4) that joins with the edge (2,3) 
can be accessed sooner by using random access.}
It is worth noting that Ilyas et al.~\cite{Ilyas:03} mention the value of
random access for achieving potentially tighter thresholds. Of course,
the cost of random access is traded for the pruning power of the improved
threshold. Indeed, in recent work Martinenghi and Tagliasacchi~\cite{RA:11} study the role of random
access in \Rank from a cost-based perspective.

Following this optimization, suppose we use random accesses to find ``matching''
edges with which to extend heaviest paths of length $\len - 1$ to length \len.
This is a good heuristic, but heaviest paths of length $\len-1$ may not always lead to
heaviest paths of length \len. A natural question is whether the use of random
accesses in this manner will lead to a performance penalty w.r.t. \Rank.
Our next result shows that \Rank will produce heaviest paths of length $\len-1$
before it produces and reports the heaviest path of length \len.

\begin{mylem}
\label{theo:nested}
Run \Rank on an instance of HPP first with input length $\len - 1$, and
then with input length $\len$. Suppose $d'$ (resp., $d$) is the depth at which
\Rank finds the heaviest path of length $\len-1$ (resp., $\len$) for the first
time in the respective runs, then, $d \geq d'$. Thus, by the time \Rank produces
the heaviest path of length \len, it also produces the heaviest path of length
$\len-1$.
\end{mylem}

\begin{proof}
Suppose $P$ is the heaviest path of length \len and $Q$ is the heaviest
path of length $\len - 1$\footnote{
\erase{If there is more than one of either kind, we break the ties}%
Ties are broken
arbitrarily.}.
Suppose $w_{d'}$ and $w_{d}$ are the edge weights at depths $d'$ and $d$ resp. in the edge 
table $E$.
Assume $d < d'$.
We know that
$w_{d'} + (\len - 2) \times w_{max} \le Q.weight \le w_{d'-1} + (\len - 2) \times w_{max}$
and
$w_d + (\len - 1) \times w_{max} \le P.weight$.
Therefore,
$P.weight - w_{max} \ge w_d+(\len - 2) \times w_{max} \ge w_{d'-1}+
		(\len - 2) \times w_{max} \ge Q.weight$.
Now, $P.weight - w_{max}$ is a lower bound on the heaviest sub-path of $P$ of
length $\len - 1$. This means $P$ has a sub-path of length $\len - 1$ that is at least as
heavy as $Q$, and can be returned by \Rank at depth $d < d'$. This contradicts
the fact that $Q$ is the first heaviest path of length $\len - 1$ discovered by \Rank.
Thus, the heaviest path of length $\len-1$ is found no later than the heaviest path of
length $\len$.
\end{proof}

\begin{figure}[t]
\includegraphics[scale=0.45]{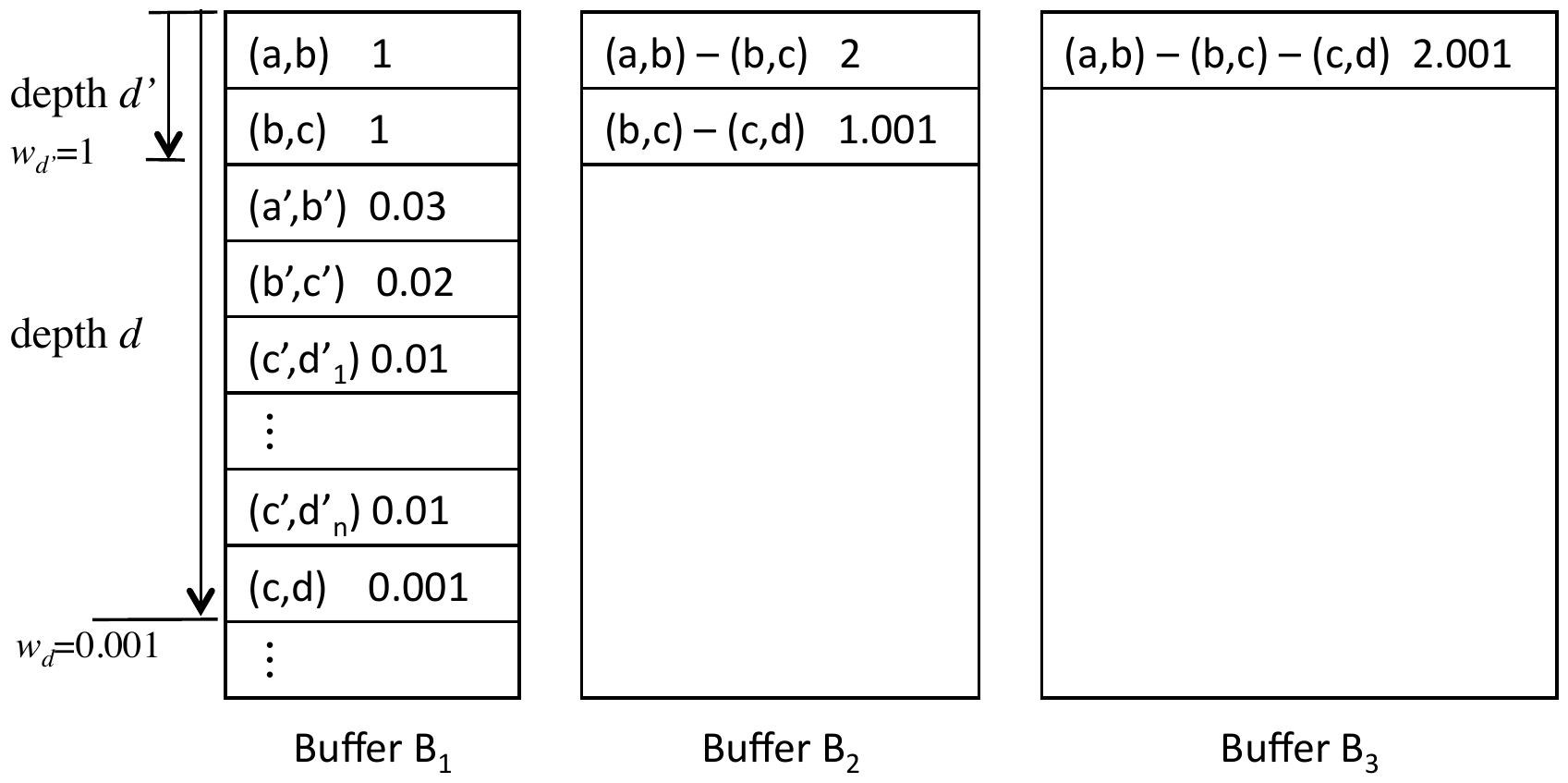}
\vspace{-10pt}
\caption{\Main using buffers to extend paths for Figure~\ref{fig:ex1}}
\label{fig:buffer}
\vspace{-10pt}
\end{figure}

\red{Figures~\ref{fig:ex1} and \ref{fig:buffer} illustrate the notions of 
depth and edge weight at a given depth.}  
\red{ 
The leftmost table in Figure~\ref{fig:buffer} represents the 
sorted edge list for the graph in Figure~\ref{fig:ex1}. 
While computing the heaviest path of length $\len=3$, \Rank 
obtains the heaviest path of $\len-1$ (i.e., 2) at depth $d'$. 
Without random accesses, the heaviest path of length $\len=3$
is obtained after scanning edge (c,d) at depth $d$. }

The real purport of the above lemma is that by the time \Rank reports a heaviest path of
length \len, it has already seen heaviest paths of length $\len-1$. Thus, any heaviest
paths of length $\len-1$ we try to extend to length \len (using random access)
are necessarily produced by \Rank as well\red{, albeit later than if we were to use 
no random access}.

\eat{
Hence, we do not incur any performance
penalty against \Rank by considering such paths.
}
\begin{myobs}
\Rank uses a very conservative threshold $w_d + (\len-1)\wmax$, by supposing
the new edge seen may join with $\len-1$ edges with maximum weight. In many
cases, the new edge just read may not join with such heavy edges at all.
\end{myobs}

One way of making the threshold tighter is by keeping track of shorter paths.
For example, if we know $P$ is a heaviest path of length $\len - 1$,
we can infer that the heaviest path of length \len cannot be heavier than
$P.weight + \wmax$, a bound often much tighter than
$w_d + (\len-1)\wmax$\footnote{Our algorithm makes use of an even tighter threshold as
described shortly.}.
For this, the (heaviest) paths of length $\len-1$ have to be maintained.
Pursuing this idea recursively leads to a framework where we maintain
heaviest paths of each length $i$, $2 \leq i \leq \len$. More precisely, we
can perform the following optimization.

\begin{myopt}
Maintain a buffer $B_i$ for the  heaviest paths of length $i$,
and a threshold $\theta_i$ on the weight of all paths of length $i$
that may be produced in the future, based on what has been seen so far.
When a new heaviest path $P$ of length $i-1$ is found,
i.e., $P.weight \ge \theta_{i-1}$, update the threshold
$\theta_{i}$ for buffer $B_i$ as
$\theta_i = \max(\theta_{i-1}, B_{i-1}.{\it topScore\/}) + \wmax$
where $B_{i-1}.{\it topScore\/}$ is the weight of the heaviest path in $B_{i-1}$
\emph{after} the current heaviest path $P$ with $P.weight \ge \theta_{i-1}$ is
removed from $B_{i-1}$\footnote{$B_{i-1}.{\it topScore\/}$ may be greater or less
than $\theta_{i-1}$.}.
The rationale is that, in the best case, a heaviest future path of length $i$ 
may be obtained by joining the current heaviest path in $B_{i-1}$ with an
edge that has the maximum weight, or joining a future path in $B_{i-1}$ with such an edge.
\end{myopt}

\red{Figure~\ref{fig:buffer} schematically describes the idea of using buffers 
for different path lengths. Buffer $B_{1}$ is the sorted edge list for the graph in 
Figure~\ref{fig:ex1}. Buffers $B_{2}$ and $B_{3}$ store paths of length 2 and 3 respectively.  
When random accesses are performed to extend paths of length $l-1$ to 
those of length $l$, the buffers can be used to store these intermediate paths.
For instance, when the heaviest path of length 1 is seen, say edge $(a,b)$ is seen, 
it can be extended to paths of length 2 by accessing edges connected to its end points, 
as represented in $B_{2}$ in Figure~\ref{fig:buffer}. Similarly, the heaviest 
path of length 2 can be extended by an edge using random access to obtain 
path(s) in buffer $B_{3}$.  }

\begin{myobs}
\Rank, especially when performed as a single \len-way operation, tends to produce
the same sub-path multiple times. For example, when \Rank is required to
produce the heaviest path of length $3$ for the graph in Figure~\ref{fig:ex1},
for paths $(a',b',c',d'_i)$, $i\in[1,n]$, it produces the
length $2$ sub-path $(a',b',c')$ $n$ times as it does not maintain shorter
path segments.
\end{myobs}

This observation motivates the following optimization. 

\begin{myopt}
Shorter path segments can be maintained so they are created once and used
multiple times when extended with edges to create longer paths.
For instance, in Figure~\ref{fig:ex1},
we can construct (and maintain) the length $2$ path segment $(a',b',c')$ once
and use it $n$ times if needed as edges are appended to extend it into $n$ different
length $3$ paths.
\end{myopt}

A concern surrounding a random access based approach to extending
heaviest paths of length $l-1$ into paths of length $l$
is that it may result in extending too many such paths. Our next result
relates these paths to those produced by \Rank as part of its processing.

\begin{mylem}
\label{theo:nestedThresh}
Let $P$ be the heaviest path of length \len.
When \Rank terminates, every path of length $\len - 1$
that has weight no less than
$P.weight - \wmax$, will be created.
\end{mylem}

\begin{proof}
Suppose \Rank finds $P$ at depth $d$. This means
$P.weight \ge w_d + (\len -1) \times \wmax$, and,
$P.weight - \wmax \ge w_d + (\len -2)\times \wmax$. Notice
that $P.weight - \wmax$ is a lower bound on the weight
of the heaviest path of length $\len-1$.
Therefore, if there is a path of length $\len -1$ that is heavier
than $P.weight - \wmax$, it will be produced by \Rank
by depth $d$.
\end{proof}

The results, observations, and optimizations discussed in this
section suggest an improved algorithm for finding heaviest paths,
which we present next.

\section{\red{HeavyPath Algorithm for HPP}}
 \label{sec:hp}

We start this section by providing an outline of our main algorithm
for solving HPP, and subsequently explore some refinements.
Our algorithm maintains a buffer $B_i$ for storing paths
of length $i$ explored so far, where $2\le i\le \len$.
Let threshold $\theta_i$ denote an upper bound on the
weight of any path of length $i$ that may be constructed in the future.
Each buffer $B_i$ is implemented as a sorted set of paths
of length $i$ sorted according to non-increasing weight
and without duplication
(e.g., paths $(a,b, c)$ and $(c,b,a)$ are considered duplicates).
\eat{
Each buffer is also a set and does not allow duplicates.
Sorted set is one of the data structures provided
by Java programming language and cost of insertion
in a sorted set is $log(n)$ where $n$ is the number of paths
existing in the buffer.
The main function is provided by Algorithm~\ref{algo:main}.
Every buffer is initialized as an empty sorted set.
Each threshold is initialized to the maximum possible value at start.
}%
Algorithm~\ref{algo:main} describes the overall approach.
It takes as input a list of edges $E$ sorted in non-increasing order of edge weights,
\red{\erase{a graph $G$,}} and parameters \len and $k$.
It calls the \RJ method (Algorithm~\ref{algo:HP})
repeatedly until the \topk heaviest paths of length \len
are found.

\begin{algorithm}[t]
\begin{algorithmic}[	1]
\caption{\Main$(E, \red{\erase{G,}} \len, k)$}
\label{algo:main}
\REQUIRE Sorted edge list $E$, 
path length $\len$, number of paths $k$
\ENSURE top-$k$ heaviest paths of length \len \red{\erase{in $G$}}
\FOR{$l=2$ to \len}
\STATE $B_{l}\gets\emptyset$  \COMMENT empty sorted set 
\STATE $\theta_l = \wmax \times l$
\ENDFOR
\STATE $topPaths\gets\emptyset$ \COMMENT empty sorted set
\WHILE{$\mid topPaths\mid < k$}
\STATE $topPaths\gets topPaths$ $\cup$ \RJ$(E,\red{\erase{G,}}\len)$
\ENDWHILE
\end{algorithmic}
\end{algorithm}
\begin{algorithm}[t]
\begin{algorithmic}[1]
\caption{\RJ$(E, \red{\erase{G,}} l)$}
\label{algo:HP}
\REQUIRE  Sorted list of edges $E$, \red{\erase{a graph $G$,}}  and
path length $l$
\ENSURE Next heaviest path of length $l$
\IF{$l = 1$}
\STATE $P^{1} \gets$ {\sc ReadEdge}$(E)$
\STATE $\theta_2 = 2 \times P^{1}.weight$
\STATE \textbf{return} $P^{1}$
\ENDIF
\WHILE{$B_{l}.topScore \le \theta_{l}$}
\STATE $P^{l-1} \gets$ \RJ$(E,\red{\erase{G,}} l-1)$	\COMMENT recursion
\STATE $s,t \gets$ {\sc EndNodes}$(P^{l-1})$
\FORALL{$y\in V\mid(y,s) \in E$}
\STATE $B_{l}\gets B_{l}\cup((y,s)+P^{l-1})$  \COMMENT avoiding cycles
\ENDFOR
\FORALL{$z\in V\mid(t,z) \in E$}
\STATE $B_{l}\gets B_{l}\cup(P^{l-1}+(t,z))$  \COMMENT avoiding cycles
\ENDFOR
\ENDWHILE
\STATE $P^{l} \gets$ {\sc RemoveTopPath}$(B_{l})$
\IF{$l < \len$}
\STATE $\theta_{l+1} = \max(B_{l}.topScore, \theta_l) + \wmax$
\ENDIF
\STATE \textbf{return} $P^{l}$
\end{algorithmic}
\end{algorithm}
\eat{
\IF{$B_{l}.topScore > \theta_{l}$}
\STATE $X = B_{l}.removeTopPath()$
\IF{$l < \len$}
\STATE $\theta_{l+1} = MAX(B_{l}.topScore, \theta_l) + W_{max}$
\ENDIF
\STATE \textbf{return} $x$
\ENDIF
\STATE $nextPath_{l-1} = \RJ(E,l-1,G)$
\STATE $s = nextPath_{l-1}.StartNode$
\STATE $t = nextPath_{l-1}.EndNode$
\FORALL{nodes $z$ s.t. $(t,z) \in G$ avoiding cycles}
\STATE $B_{l}.insert(nextPath_{l-1}+t\leadsto z)$
\ENDFOR
\FORALL{nodes $z'$ s.t. $(z',s) \in G$ avoiding cycles}
\STATE  $B_{l}.insert(z'\leadsto s+nextPath_{l-1})$
\ENDFOR
\STATE $ \RJ(E,G,l)$
}%
%
Algorithm~\ref{algo:HP} describes the \RJ method.
It takes as input a list of edges $E$ sorted in non-increasing
order of edge weights,\red{\erase{a graph $G$,}}  and
the desired path length $l$, \red{$l \ge 2$}.
It is a recursive algorithm that produces heaviest paths of shorter lengths
on demand, and extends them with edges to produce paths of length \len.
The base case for this recursion is when $l = 1$ and
the algorithm reads the next edge from the sorted list of edges. 
The {\sc ReadEdge} method 
returns the heaviest unseen edge in $E$ \red{(sorted 
access, line 2)}.
If $l < \len$, the path of length $l$ obtained as a result of the recursion
is extended by one hop 
to produce paths of length $l + 1$.
Specifically, a path of length $l<\len$ is extended using edges 
\red{(random access, lines 8 and 10)} 
that can be appended to either one of its ends (returned by method
{\sc EndNodes}).  The ``+'' operator for appending an edge
to a path is defined in a way that guarantees no cycles are created.
The threshold $\theta_l$ is updated appropriately and when it
becomes smaller than the weight of the heaviest
path in buffer $B_l$, 
the next heaviest path of length $l$ that
is greater than the threshold is returned. 
This is done by calling the method {\sc RemoveTopPath}
for buffer $B_{l}$ and returning the resulting path.
If $l < \len$ and the next heaviest path of length $l$ has been obtained,
$\theta_{l+1}$ is updated.

\para{Updating the thresholds}
To start, Algorithm~\ref{algo:HP} explores
the neighborhood of the heaviest edge for finding the heaviest path of length $2$.
Heavy edges are explored and the threshold $\theta_2$ is updated until
the weight of the heaviest explored path of length $2$ is greater than $\theta_{2}$.
The \RJ algorithm updates $\theta_2$ aggressively when the next heaviest edge is seen.
It uses the fact that any path of length $2$ created later from currently unseen edges
cannot have a weight greater than $2 \times P^{1}.weight$,
where $P^{1}$ is the lightest edge seen so far.
For $l>1$, $\theta_{l+1}$ is updated when the next heaviest path of length $l$ is obtained.
Therefore, we can use $\theta_l$ as an upper bound on the weight of any path
of length $l$ that can be created \emph{in the future} from the buffers $B_{i}$, where $i<l$.
The maximum of $\theta_l$ and the weight of heaviest path in $B_l$ (i.e., $B_{l}.topScore$) 
provides an upper bound on the weight of any path
of length $l$ that can be created in the future 
(after the previous heaviest path of length $l$).
Adding $\wmax$ to the obtained upper bound provides a new (tighter) threshold for
paths of length  ${l+1}$.

\begin{mytheo}
\label{algo:HPcorrectness}
Algorithm \Main correctly finds top-$k$ heaviest paths of length \len.
\end{mytheo}

\begin{proof}
The proof is by induction. The base case is for going from edges
to paths of length $2$. Given that all of the edges above depth $d$
are extended, the heaviest path that can be created from them is already
in $B_2$. The weight of the heaviest path that can be created from lighter edges is
at most $2 \times w_d$.
If the heaviest path in $B_2$ is heavier than $2 \times w_d$, then it must be the
heaviest path of length $2$.

Assuming the heaviest
paths of length $l$ are produced correctly in sorted order, we show the heaviest path of length
$l+1$ is found correctly.
Suppose $P$, the heaviest path of length $l+1$, is created for the first time from by
extending $Q$, which is the $n^{\rm th}$ heaviest path of length $l$.
The next heaviest path of length $l$ is either already in
$B_{l}$ or has not been created yet.
Therefore, $\max(\theta_{l}, B_l.topScore)$ is an upper bound
on the next heaviest path of length $l$ that has not been extended and
any edge that can join this path can have weight at most $\wmax$.
Suppose when the $m^{\rm th}$ heaviest path of length
$l$ is seen, $\max(\theta_{l}, B_l.topScore) + \wmax$ is updated to a value
smaller than $P.weight$. It is guaranteed that $P$ is already
in $B_{l+1}$ and has the highest weight in that buffer.
In other words, when the threshold is smaller than $P.weight$,
the difference between the weight of $P$ and next heaviest path of length $l$
is more than $\wmax$. Now,
paths of length $l+1$ that can be created from heavier paths of
length $l$ are already in the buffer, and no unseen path of length $l$ can
be extended to create a path heavier than $P$.
Therefore, $P$ is guaranteed to be the heaviest path.
The preceding arguments hold for \topk heaviest paths where $k>1$.
\end{proof}

Algorithm~\ref{algo:HP} extends the heaviest paths in sorted order,
to avoid their repeated creation. However, since
paths are extended by random accesses, it is possible to create
a path twice, which is unnecessary. For instance,
a path of length $l$ may be created
while extending its heaviest sub-path and again 
while extending its lightest subpath
of length $l-1$. 
\erase{Next, we provide a
strategy for minimizing duplicates as much as possible.}

\subsection{\red{Duplicate\erase{Elimination} Minimization by Controlling Random Accesses}}
\label{sec:random}
 
Algorithm \RJ extends a path of length $l$ to one of length $l+1$ by
appending all edges that are incident on either end node of the path.
Since these edges are not accessed in any particular order,
in the literature of top-$k$ algorithms, \erase{it is} they are referred to as
random accesses. 
\red{
In this section, we develop a strategy for controlling random 
accesses performed by Algorithm \RJ for minimizing duplicates. }
Duplicate paths of length $l+1$ can be created either due to
extending the same path of length $l$,
or by extending two different subpaths of the same path of
length $l+1$.
Our solution for avoiding duplicates of the first kind
is implementing every buffer as a sorted set. This avoids propagation
of duplicates during execution.
Further, the threshold update logic
guarantees that if there is a copy $P'$ of some path $P$ that is already in the buffer
$B_{l}$, the path will not be returned before its copy $P'$
makes it to $B_{l}$. The algorithm ensures that when $P$ is returned,
$P'$ either does not exist or has been constructed and eliminated.

In addition to eliminating duplicates that have been created,
we take measures to reduce the\blue{ir very}  creation.
Suppose $P$ is a path of length $l+1$ whose right sub-path of length $l$
is the heaviest path of length $l$ and its left sub-path of length $l$ is
the second heaviest path of length $l$.
Since random accesses are performed at both ends of a path, $P$ will be
created twice, using each of the top-$2$ paths of length $l$. 

One possible 
solution is to perform random accesses at one of the ends of a path.
Although this prevents duplicate creation of $P$,\erase{but} it 
does not allow fully exploring the neighborhoods
of heavier edges. 
For example, consider a path with edges $\{(a, b),(b,c),(c, d)\}$
that is the heaviest path of length $3$, \blue{with} 
 $(b, c)$ 
 the heaviest in the graph. If
the addition of edges is restricted to the beginning of the path,
the construction of the heaviest path of length $3$
will be delayed until $(a,b)$ is observed.
There can be graph instances for which this can happen at an arbitrary depth.
Therefore, it is advantageous to extend paths on both ends, and we dismiss 
the idea of one sided extension.

\begin{mylem}
\label{lem:ra}
Suppose $Q$ is the $n^{\rm th}$ heaviest path of length $l$ with $(a, b)$ as
its heaviest edge. 
No new path of length $l+1$ can be created from $Q$ by adding an edge
which is heavier than $(a, b)$ using the \RJ method.
\end{mylem}
\begin{proof}
Let $P$ be a new path of length $l+1$ created from $Q$. If $P$ is
derived for the first time, it can not have a subpath of length $l$ that is
heavier than $Q$. Otherwise, the heavier subpath is one of the $n-1$ paths
created \erase{ and explored }before $Q$.
On the other hand, adding any edge to the end of $Q$ which is
heavier than $(a, b)$ results in a path of length $l+1$
that has a subpath of length $l$ \erase{ which is }heavier than $Q$.
The  lemma  follows.
\end{proof}
\vspace{-6pt}
Therefore, no new path of length $l+1$ can be created by
adding an edge to $Q$, that is heavier than the 
heaviest edge of $Q$.
This leads to the following theorem.

\begin{mytheo}
\label{theo:ra}
Using \RJ, every new path of length $l+1$ is created only by extending its
heavier sub-path of length $l$. No path is created more than twice.
A path of length $l+1$ is created twice iff both its sub-paths of length $l$ have the
same weight. \qed
\end{mytheo}

\eat{
As a result of the above lemma, every path of length $l+1$ is created only by extending its
heavier sub-path of length $l$. This strategy leads to creating duplicate
paths only when both the first and the last edges of a path of length $l$
have equal weight.
}

\red{The strategy for controlling random accesses embodied in Theorem~\ref{theo:ra} 
can be generalized
to a stronger strategy as follows.} 

\begin{myfact}
\label{fact:ra1}
Using \RJ, given a path of length $l$, no new path of length $l+1$ can be created 
by adding an edge to its rightmost node that is heavier that its leftmost edge, 
or adding an edge to its leftmost node that is
heavier than its rightmost edge.
\end{myfact}

\red{In the rest of the paper, we refer to the strategy for controlling random accesses 
described in Fact~\ref{fact:ra1} as \emph{random access strategy}. Notice that 
Algorithm \Main always employs random access in addition to sorted 
access. Additionally, we have the option of adopting (or not) the random access strategy 
above for controlling when and how random accesses are used. We have:} 

\begin{myfact}
\label{fact:ra2}
If all of the edge weights in the graph are distinct, every path is created only once
when the random access strategy mentioned above is followed.
\end{myfact}

In the rest of this paper, we follow the random access strategy of Fact~\ref{fact:ra1}  for performing random accesses unless otherwise specified.
\eat{In the rest of this paper, we follow Fact~\ref{fact:ra1} 
for performing random accesses unless otherwise specified.  }
We refer to this way of performing random accesses as 
\emph{random access strategy}. \red{In our experiments, we measure 
the performance of \erase{Algorithm}\Main both without and with\erase{the 
random access strategy adopted.} this strategy.}%
\eat{
In the rest of this paper, we follow this random access strategy that extends
a path with only those edges that are lighter than the heaviest edge in that path,
unless otherwise specified.

\begin{proof}
Proof is done by induction. The base case is going from edges to
paths of length $2$ and this is obviously true for the base case.
Suppose all created paths of length $l$ are distinct. We will
show no duplicate are possible for paths of length $l+1$. Suppose
$X$ is a path of length $l+1$ that is created twice. Therefore, $X$
can be created from two distinct paths of length $l$ which we call
$Y$ and $Y'$. This means $Y$ and $Y'$ have a sub-path of length $l-1$
in common. $Y$ is the beginning sub-path of length $l$ of $X$ and $Y'$
is the ending. If $X$ is constructed by adding its first edge to $Y'$
it means $X.firstEdge.weight < X.lastEdge.weight$.
If it is created from $Y$ the opposite of the above and this
is a contradiction. Which means if all paths of length $l$ are distinct,
all paths of length $l+1$ will also be distinct.
\end{proof}
If there are many equal edge weights our algorithm generates some
paths more than once. Such cases are not the most likely cases
in most practical situations when edge weights represent probabilities
and are decimal numbers. Despite the fact that these cases are not
very likely, one could avoid the propagation of such duplicates using
the initial idea of keeping a history corresponding to each buffer
in order to minimize duplicate paths.
}
\subsection{\red{HeavyPath Example}}
\label{sec:example} 
In this section, we present a detailed example 
that illustrates the creation of paths by \Main and \Rank
for finding the heaviest path (i.e., $k=1$) of length $3$ for the graph in
Figure~\ref{fig:ex1}. The idea of using buffers was already illustrated in 
Figure~\ref{fig:buffer}, to which we refer below. 
Later in Section~\ref{sec:expn}, we present several examples of heavy paths found by our algorithms for the applications of finding playlists and topic paths.

\eat{
\begin{figure}
  \centering
  \includegraphics[scale = 0.25]{ex1.eps}\\
  \caption{Example Instance of HPP. One heavy path and $3$ lighter paths.}\label{fig:ex1}
\end{figure}
}

\eat{ 
\begin{figure}[t]
  \centering
  \includegraphics[scale = 0.7]{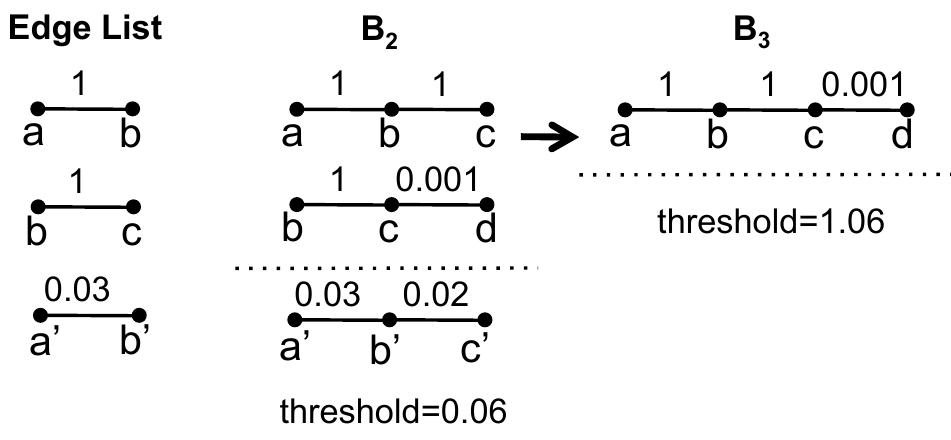}\\
      \vspace{-6pt}
  \caption{Running \Main on\erase{ the graph of } Figure~\ref{fig:ex1}}\label{fig:HP}
    \vspace{-6pt}
\end{figure}
} 

\begin{figure}[t]
  \centering
  \includegraphics[scale = 0.36]{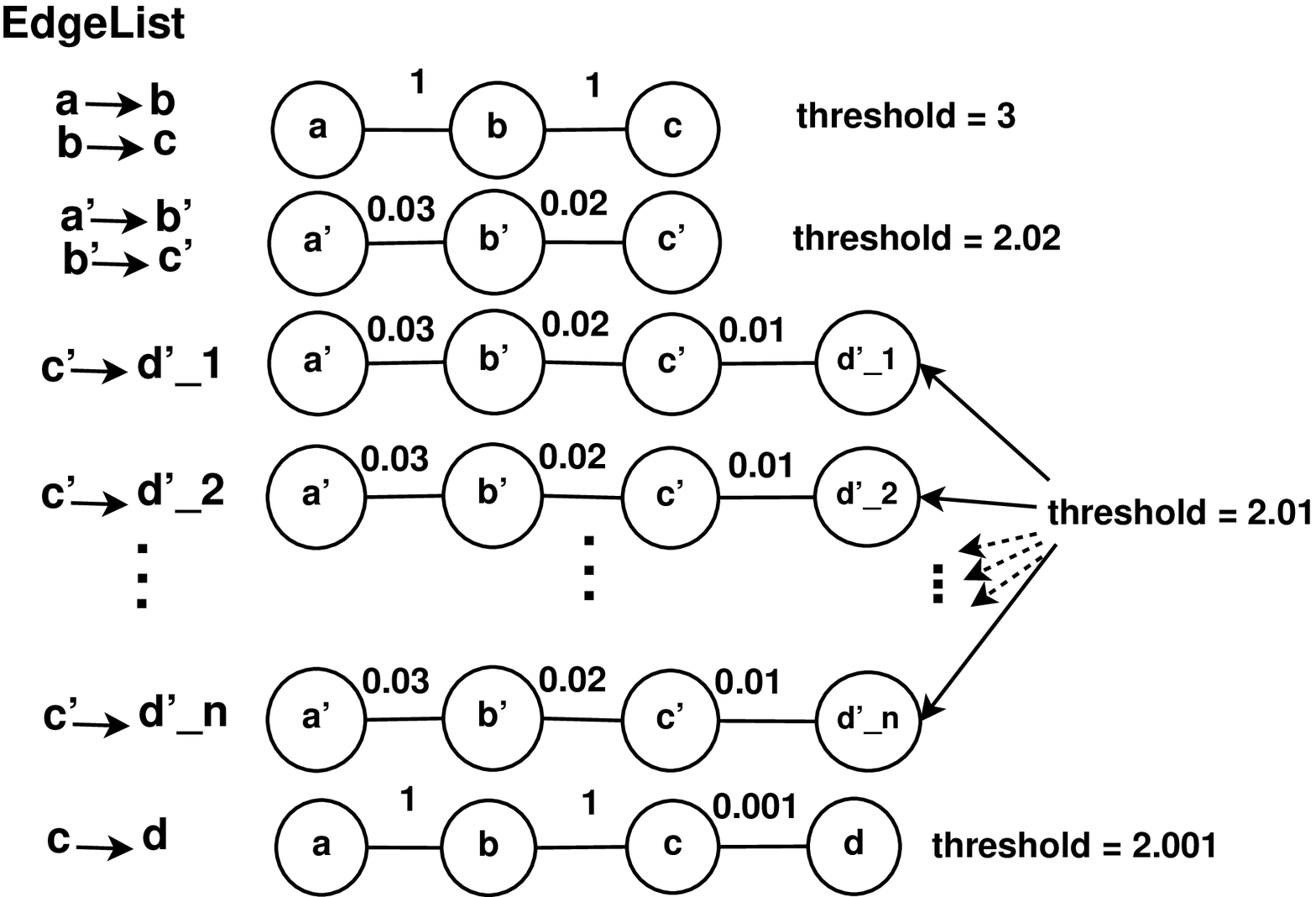}\\
    \vspace{-6pt}
  \caption{Paths created by \Rank on\erase{ the graph of} Figure~\ref{fig:ex1}}\label{fig:RJ}
  \vspace{-10pt}
\end{figure}

\Main first reads the first heaviest edge $(a,b)$ and then extends it using a
random access to edge $(b,c)$ into the path $(a,b,c)$ of length $2$. It then reads edge
$(b,c)$ again under sorted access and tries to extend via a random access to $(a,b)$.
The duplicate derivation of the path $(a,b,c)$ is caught and discarded. Edge $(b,c)$ is extended
with another random access into the path $(b,c,d)$. At this point, paths $(a,b,c)$ of weight $2$
and $(b,c,d)$ of weight $1.001$ are added to buffer $B_2$. %
\red{The threshold $\theta_2$ at this point is $2$ and is updated to $0.06$ 
when the next edge $(a', b')$ is visited under sorted access.} 
At this time, the two heaviest paths of length $2$ are both above the threshold
and are returned.
\blue{Of these two paths,} $(a,b,c)$ is extended with a
random access to edge $(c,d)$ to form a length $3$ path.
\eat{
The extension of $(b,c,d)$ using random access results in
the creation of another copy of the same path of length $3$
which is ignored.
}%
\red{If we do not adopt the random access strategy (see Fact~\ref{fact:ra1}, 
Section~\ref{sec:random}), then
 $(b,c,d)$ will be similarly extended and again the duplicate derivation 
would be discarded. If we adopt the random access strategy, random 
access is restricted to edges whose weight is no more than that of the edges
at either end of the path, so $(b, c, d)$ will \emph{not} be similarly extended.} 
\eat{The heaviest path of length $2$ is extended with random access
that results in the heaviest path of length $3$.
However, the second heaviest path of
length $2$ (i.e., $b \leadsto d$) does not satisfy the
random access condition described in Section~\ref{sec:random} and is thus ignored. }%
\erase{On returning these two length $2$ paths, }%
\blue{Now, }$\theta_2$ is updated to
 $2 \times 0.03$ and $\theta_3$ is updated to $1.06$. 
The heaviest path of length $3$ found so far, which has weight $2.001$, 
is reported.

\Main performs $5$ joins in total before reporting the heaviest
path of length $3$. 
That includes the joins for paths of length $2$ and $3$ that are created, and the
additional 
duplicate path that is created and removed during the execution.

Figure~\ref{fig:RJ} illustrates \Rank for the same graph and parameter settings.
\Rank is not able to produce the heaviest length $3$ path until it scans
every edge under sorted access for this graph instance.
Each edge is joined twice with the partial list of edges that are scanned before it,  to
construct paths of length $3$. \Rank produces $n+1$ paths of length
$3$ and two paths of length $2$ in the order shown in Figure~\ref{fig:RJ}.
It performs a total of $2n+4$ join operations before finding the heaviest length $3$ path.
This example demonstrates that the performance of \Rank can be 
significantly worse than \Main in terms of the number of join operations
it performs. 

\erase{
\subsection{HeavyPath+ Algorithm for \hpp}
\label{sec:Stats}
} 

\erase{ \red{The \Main algorithm described in 
Section~\ref{sec:hp} 
defines $\theta_{l+1}=\max(B_l.topScore, \theta_l) + \wmax$, 
where $\wmax$ is the weight of
the heaviest edge, and $\max(B_l.topScore, \theta_l)$ 
is an upper bound on the
heaviest path of length $l$ that can be created in the future. 
Thus, $\theta_{l+1}$ is a 
threshold on heaviest path of
length $l+1$ that is yet to be created.  
We can do better than this by exploiting the random access 
strategy based on Fact~\ref{fact:ra1}. 
Consider a scenario where every path of length $i < l+1$
that can be created from the heaviest edge that has weight $\wmax$, 
has already been created. 
Then, we can use $\max(B_i.topScore, \theta_i) + w'_{max}$
for all $i \le l$ to update $\theta_{i+1}$ where $w'_{max}$ is the 
\emph{current} 
heaviest edge weight that has not been explored up to length \len yet. 
Note that $w'_{max} \le \wmax$. 
To do this efficiently, instead of scanning all paths in each buffer for 
finding $w'_{max}$, we keep track of such statistics 
as paths are added and removed from buffers. 
This is the central idea of our next algorithm, 
which gives a tighter bound for $\theta_{l+1}$. 
We call this algorithm \MainPlus, which calls the recursive method \RJPlus to get the 
next heaviest path (similar to Algorithms~\ref{algo:main} and~\ref{algo:HP} resp.). }}

\erase{ \red{In addition to initializing buffers and thresholds,
\MainPlus also initializes a ``$Stats$'' table. This table has $\len-1$ columns
corresponding to different path lengths in the range between $[1, \len-1]$ and rows
corresponding to different edge weights. 
In the worst case, it can have as many rows as the number of possible edge weights, 
however, these rows are only populated on demand.
The rows are ordered in decreasing edge weights. 
\eat{The first row corresponds to
the heaviest edge weight and the last row corresponds to the lightest edge weight.}
Column 1 of this table 
corresponds to paths of length $1$ (i.e., edges).  
\MainPlus scans the list of edges once to initialize each row of column 1 to the number of
edges having the corresponding edge weight. Intuitively, column 1 can be thought of 
as a frequency table that has the number of times a particular edge weight appears. 
In practice, 
$Stats$ can be implemented as a hash table 
that maps edge weights to vectors of length $\len-1$.
Let $Stats(w,l)$ denote the number of paths in $B_l$ whose heaviest edge weight is $w$.
During an execution, when an edge with weight $w$ is returned,
\RJPlus decrements $Stats(w,1)$ and adds the number of simple paths of length $2$ that
are created from the heaviest edge in the graph to $Stats(w,2)$. 
This process continues when longer
paths are returned  from $B_l$ and are expanded and added to $B_{l+1}$. 
In an update for $\theta_l$, $\wmax$ can be replaced with
the heaviest edge weight whose corresponding row includes at least
one non-zero entry among its columns. 
This results in a tighter threshold.
As an example, consider a scenario where all $Stats(\wmax,i)$ entries are $0$ for all $i < 5$.
Then following the random access strategy, 
no path of length $4$ or shorter can be extended with $\wmax$ 
(since the paths are required to be simple). 
In particular, this thresholding scheme is more effective when the neighborhood of the heaviest
edge is not dense, or when the graph is disconnected and the heaviest edge can only contribute
to the score of a relatively smaller number of paths. 
We omit the modified pseudocode for brevity. }}
\eat{Figure (CITE HERE) shows an example of such cases.}

\subsection{Memory-bounded Heuristic Algorithm}
\label{sec:heuristic}
As described in the earlier sections, the problem of finding the heaviest path(s) is NP-hard. 
Even though \Rank uses a fixed buffer space by storing only the \topk heaviest
paths at any time, it needs to construct many paths 
and may run out of allocated memory. 
On the other hand, \Main  explicitly stores intermediate
paths in its buffers, and in doing so it may run out of memory
(see Section~\ref{sec:expn} for performance of various algorithms and 
their memory usage). 
From a practical viewpoint, having a memory bounded algorithm for
\hpp would be useful. 
\eat{
In our experimental results
section we show using fixed memory, \Main significantly
outperforms rank join. It is useful to
provide memory-bounded algorithms in practice from a systems point of view.
We propose a memory-bounded
version of our \Main to achieve this. }%
Thereto, we propose a heuristic algorithm for \hpp called \hph 
 that takes an input parameter $C$ which
is the 
\emph{total} number of
paths the buffers are allowed to hold, in all. If \Main
fails to output the exact answer using the maximum allowed collective buffer size $C$,
its normal execution stops and it switches to a greedy post-processing heuristic.
Algorithm~\ref{algo:post} provides the pseudocode for this heuristic. 
The post-processing
requires only constant additional memory $O(d_{max} \times \len)$, where 
$d_{max}$ is the maximum node degree in the graph.  
This is negligible in most practical scenarios, but  
should be factored in while allocating memory.

\begin{algorithm}[t]
\begin{algorithmic}[1]
\caption{{\sc HeavyPathHeuristic}$(\len, B, \red{\erase{G}})$}
\label{algo:post}
\REQUIRE Path length $\len$, \red{\erase{graph $G$,}} buffers $B$ from a run of \Main 
\ENSURE a heavy path of length $\len$ and ratio $\rho$ 
\STATE $i =$ \Buf$()$     \COMMENT last non-empty buffer
\STATE $j =$ \Buf$()$    \COMMENT last non-empty buffer
\WHILE{$i<\len$}
\STATE $P^{i} \gets$ {\sc RemoveTopPath}$(B_{i})$
\STATE  $s,t \gets$ {\sc EndNodes}$(P^{i})$
\FORALL{$y\in V\mid(y,s) \in E$}
\STATE $B_{i+1}\gets B_{i}\cup((y,s)+P^{i})$  \COMMENT avoiding cycles
\ENDFOR
\FORALL{$z\in V\mid(t,z) \in E$}
\STATE $B_{i+1}\gets B_{i}\cup(P^{i}+(t,z))$  \COMMENT avoiding cycles
\ENDFOR
\STATE $i = $ \Buf$()$     \COMMENT last non-empty buffer
\ENDWHILE
\STATE \red{$q \gets \lfloor\len/(j-1)\rfloor$}
\STATE \red{$r \gets \len-q \times (j-1)$} 
\IF{$r > 0$}
\STATE $U_{\len} \gets U_{j-1}\times q + U_{r}$ 
\ELSE
\STATE $U_{\len} \gets U_{j-1}\times q$ 
\ENDIF
\STATE $\rho = B_{\len}.topScore / U_{\len}$
\STATE \textbf{return}({\sc RemoveTopPath}$(B_{\len}), \rho)$
\end{algorithmic}
\end{algorithm}

Suppose $i$ is the index of the last non-empty buffer 
(returned by the routine \Buf)
 when \Main  
runs out of \erase{memory} \red{the total allocated buffer size $C$}. 
At this time, the heaviest path
of length $i-1$ has already been created and extended. 
However, $\theta_i$ is still larger than the weight of the
heaviest path in $B_i$. 
\eat{  \RJ needs more memory in order to
produce at least one and possibly more heaviest paths of length $i-1$.
We disallow this because memory usage has reached its limit ($C$).  }%
\hph takes the current heaviest path of length $i$
and performs random accesses to create longer simple paths.
Note that these 
random accesses do not follow the 
rule described in Section~\ref{sec:random} in order to have a 
better chance of finding heavier paths. 
Having done this, it calls \Buf in a loop.  
\Buf returns $i+1$ if 
the previous path on the top of $B_i$ has been successfully extended 
to create at least one path of length $i+1$. The while loop continues
until a path of length \len is found. 
In case 
none of
the existing paths in $B_i$ can be extended to create paths of length \len,
it \emph{backtracks} to a buffer of shorter path length. 
The whole process is guaranteed to run in 
$O(C+d_{max} \times \len)$ buffer size.

\noindent{\bf Empirical Approximation Ratio.} 
In addition to returning
a path of length \len, this heuristic can estimate the worst-case
ratio between the weight of the path found and the maximum possible weight of the
heaviest path of length \len. 
We call this the \emph{empirical approximation ratio}. 
\eat{
In addition
to a path of length $\len$, this algorithm calculates and outputs
an upper bound on the distance to the optimal answer that is obtained
by stitching the heaviest paths of shorter lengths. }%
Let $U_{l}$   
denote the\erase{weight of the heaviest path
of length $l$} \red{maximum possible weight of a path of length $l$.} 
The empirical approximation ratio, denoted $\rho$, is 
defined as $\rho = B_{\len}.topScore / U_{\len}$. 
If \Main makes it to the $j^{th}$ buffer, 
the heaviest path of length $l\le j-1$\erase{have} \red{has} already been found and can
be used to provide an approximation ratio along with the output.
If $j$ is the last non-empty buffer
when \Main terminates, $U_l$ is not known for $l \ge j$.
Since the approximation ratio mentioned above is the worst case, 
we refer to $U_l$ as a pessimistic upper bound on the weight of the heaviest path of
length $l$, for $l \ge j$. The main idea
behind this calculation is the following (see lines 11-17): 
if \len is divisible by $j-1$, 
then no path of length \len can be heavier than $U_{j-1}\times(\len/(j-1))$, 
since $U_{j-1}$ is an upper bound on the weight of any path of length $j-1$; 
if \len is not divisible by $j-1$,\erase{the remainder of the division ($r$) is 
guaranteed to be smaller than $j-1$} \red{let $r$ be the remainder of the 
division.}
\erase{Therefore, $U_r$ is also tight} By the same reasoning as above, 
$U_{j-1}\times\lfloor\len/(j-1)\rfloor + U_r$ is an upper bound on the weight of 
any path of length $\len$. \erase{The upper bound is calculated as shown in line $14$.}

\section{Experimental Analysis}
\label{sec:expn}

\subsection{Experimental Setup}
\label{sec:data}
\eat{
\begin{figure*}[t]
\centering
} 

\eat{\subfigure[Summary of datasets]{
\begin{minipage}[b]{\figwidth}
\centering
\small
\begin{tabular}{|l@{ }|l@{ }|l@{ }|l@{ }|}
\hline Measures & Cora & last.fm & Bay  \\ \hline
Nodes  &  70 & 40K & 321K  \\
Edges  &  1580 & 183K & 400K  \\ \hline
Average Degree & 22.6 & 4.5 & 1.2  \\ \hline
Number of components & 1 & 6534 & 1 \\ \hline
\end{tabular}
\vspace*{2mm}
\end{minipage}
\label{fig:data}
}}

\begin{figure*}[t]
\centering
\subfigure[Cora]{
\includegraphics[width=\figthree]{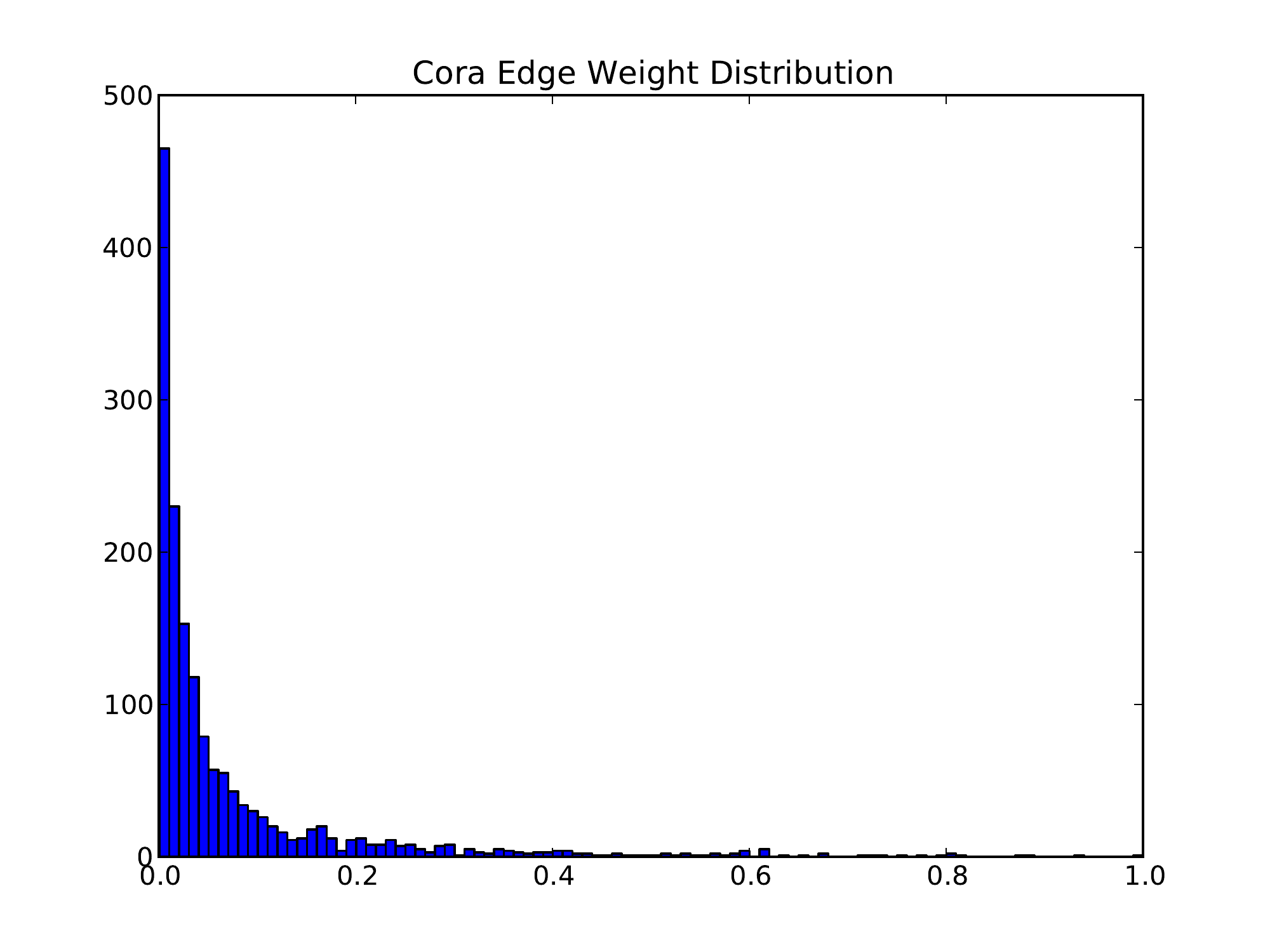}
\label{fig:distC}
}
\subfigure[last.fm]{
\includegraphics[width=\figthree]{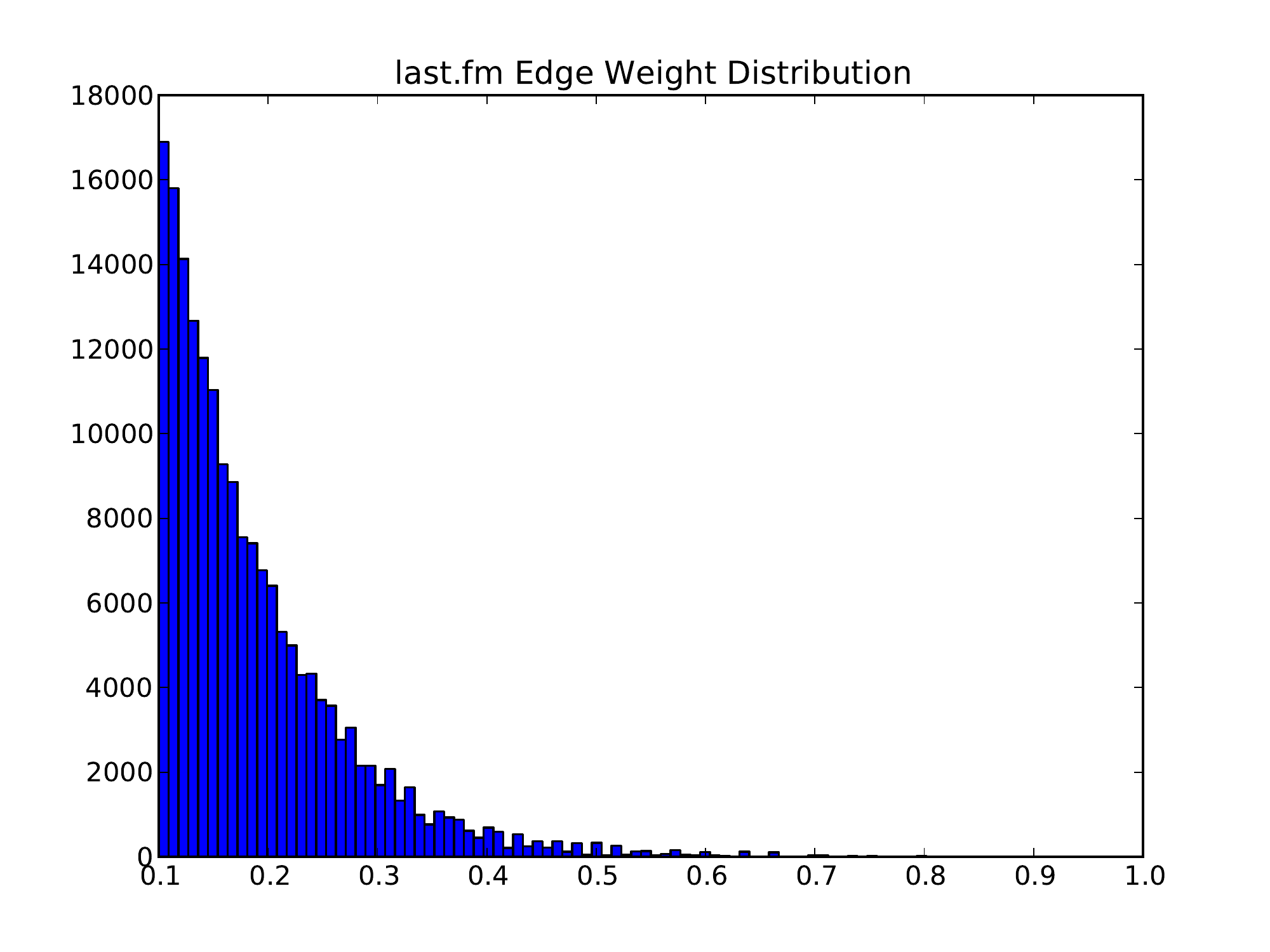}
\label{fig:distL}
}
\subfigure[Bay]{
\includegraphics[width=\figthree]{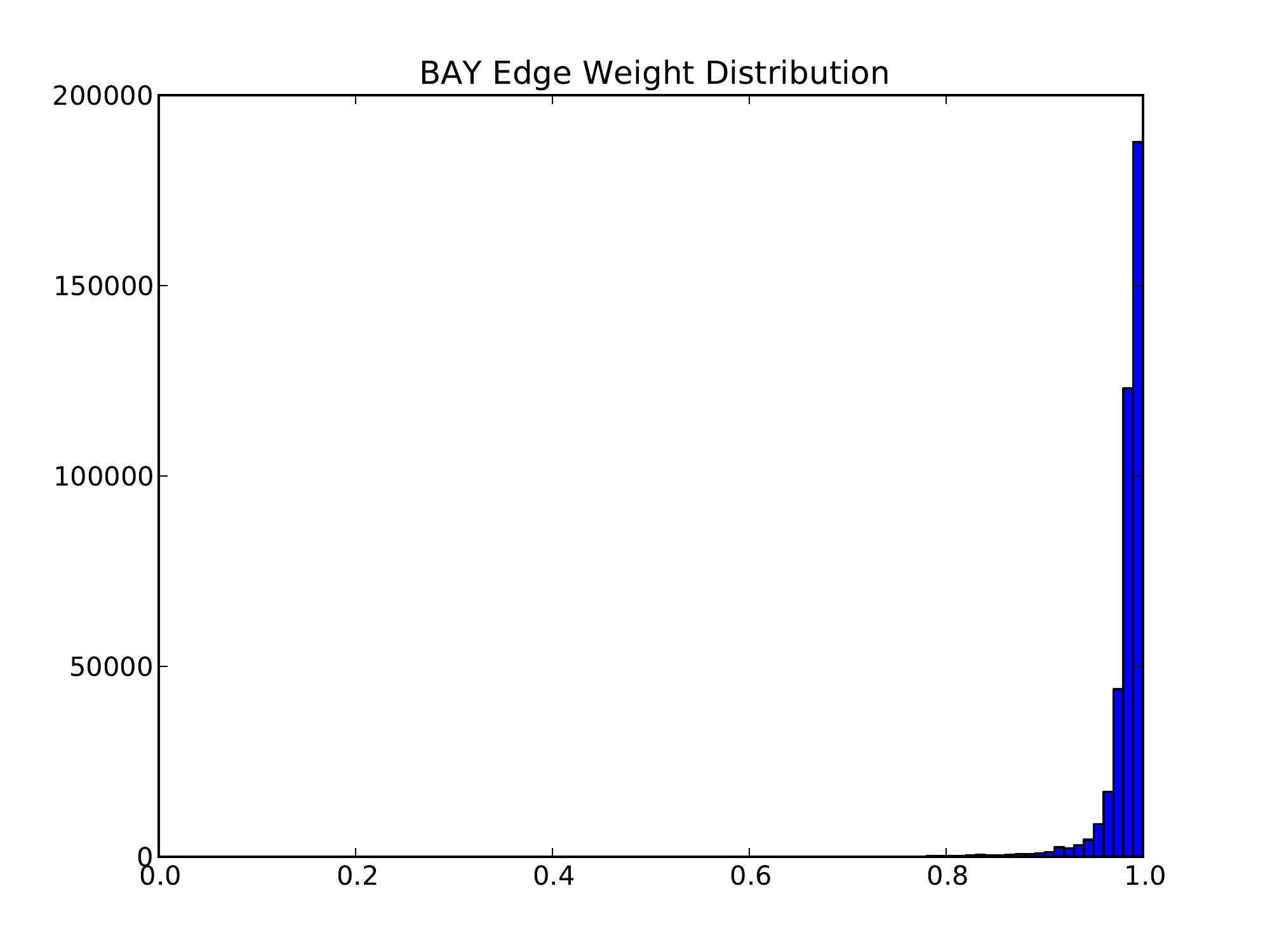}
\label{fig:distB}
}\vspace{-2pt}
\caption{Edge weight distributions for the Cora, last.fm and Bay datasets}
\vspace{-2pt}
\label{fig:data}
\end{figure*}

\begin{table}[t]
\small
\centering
\begin{tabular}{|l@{ }|l@{ }|l@{ }|l@{ }|}
\hline Measures & Cora & last.fm & Bay  \\ \hline
Nodes  &  70 & 40K & 321K  \\
Edges  &  1580 & 183K & 400K  \\ \hline
Average Degree & 22.6 & 4.5 & 1.2  \\ \hline
Number of components & 1 & 6534 & 1 \\ \hline
\end{tabular}
\vspace{-6pt}
\caption{Summary of datasets}
\label{tab:data}
\end{table}

\para{\red{Algorithms Compared}} 
\red{We implemented the algorithms \DP, \Rank  and \Main (without and 
with the random access strategy). We also implemented the heuristic algorithm 
\hph and a simple Greedy algorithm to 
serve as a quality baseline for \hph. 
We evaluate our algorithms over three real datasets: Cora, last.fm and Bay,
summarized in Table~\ref{tab:data}. The distributions of edge weights for the
three datasets can be found in Figure~\ref{fig:data}.
} 

\para{Cora}
We abstract a topic graph from the Cora Research Paper Classification
dataset\footnote{\url{http://www.cs.umass.edu/~mccallum/data}}. 
Nodes in the topic graph represent research topics into which research papers are classified,
and an edge between two topics $a,b$ represents that a paper belonging to topic $a$
cited a paper belonging to topic $b$ or vice versa or both. The weight on an edge is computed as the
average of the fraction of citations from papers in topic $a$ to papers
in topic $b$ and vice versa.
A heavy path in the topic graph captures the flow of ideas across topics.

\para{last.fm} The last.fm data was crawled using their API service\footnote{\url{http://www.last.fm/api}}.
Starting with the seed user ``RJ'' (Richard Jones was one of the co-founders
of last.fm), we performed a breadth first traversal and crawled 400K users.
Of these, 163K users had created at least one playlist, for a total of
173K playlists with 1.1M songs in all.
We use these playlists as a proxy for listening sessions 
of users to build a co-listening graph.
A node in the co-listening graph is \red{a} song and an edge represents that a pair of
songs were listened to together in several playlists.
The weight of an edge is defined as the Dice
coefficient\footnote{$\texttt{dice}(i,j) = \frac{2|i\cap j|}{|i|+|j|}$}. 
We filtered out 
edges that had a dice coefficient smaller than 0.1.
\red{The graph obtained has 6534 connected components, which implies that
there are many pairs of songs that are not heard together frequently.} A heavy 
path in the co-listening graph captures a popular playlist of songs
with a strong ``cohesion'' between successive songs.

\para{Bay}
Our third dataset is a road network graph of the San Francisco Bay area.
It is one of the graphs used for the $9^{\textrm{th}}$ DIMACS Implementation
Challenge for finding Shortest Paths\footnote{\url{http://www.dis.uniroma1.it/$\sim$challenge9/}}.
The nodes in the graph represent locations and edges represent existing roads connecting
pairs of locations. The weight on an edge represents the physical distance between
the pair of locations it connects. We normalize the weights as described in
Section~\ref{sec:probdef} to solve the \emph{lightest path problem} on this graph.
\eat{
A heavy path represents the
longest (in terms of physical distance) path connecting a given number of locations.
A more natural abstraction would be the following: consider the edge weights as
representing the traffic volume between a pair of locations, then a heavy path represents the
busiest (in terms of traffic volume) route that connects a given number of locations.
}


\para{Implementation Details}
All experiments were performed on a Linux machine with
64GB of main memory and 2.93GHz-8Mb Cache CPU.
To be consistent, we allocated 12GB of memory for each run,
unless otherwise specified.
The algorithms were implemented in Java using built-in
data structures (that are similar to priority queues) for implementing buffers as sorted
sets.
Our implementation of \Rank was efficient in that it maintained a
hash table of scanned edges to avoid scanning the edge list
every time a new edge was read under sorted access.
Similarly, for efficient random access, we maintained the
graph as a sparse matrix.
\eat{
Similarly we use java library to create hash tables when needed.
We try  to be as efficient as possible in implementing
multi-way join operator for \Rank. We avoid scanning
the edge list multiple times and maintain two in memory
hash tables to index the scanned edges using both
their start and end nodes. Once a new edge is
read under sorted access it is joined with what
is read previously using both of these hash tables.
This joining continues for the create path segments as well until
every path of length \len containing the newly scanned
edge and the edges scanned previously is created.
This approach works efficiently and correctly
for self-joins.
As for our algorithms that need the adjacency matrix
of the graph, we use a sparse matrix in order to
avoid unnecessary overhead. We implement every row
of the sparse matrix as a hash table that maps column
indices to edge values.
}

\subsection{Experimental Evaluation}
\begin{figure*}[t]
\centering
\subfigure[Vary $\len, k=1$, Cora ]{
\includegraphics[width=\figthree]{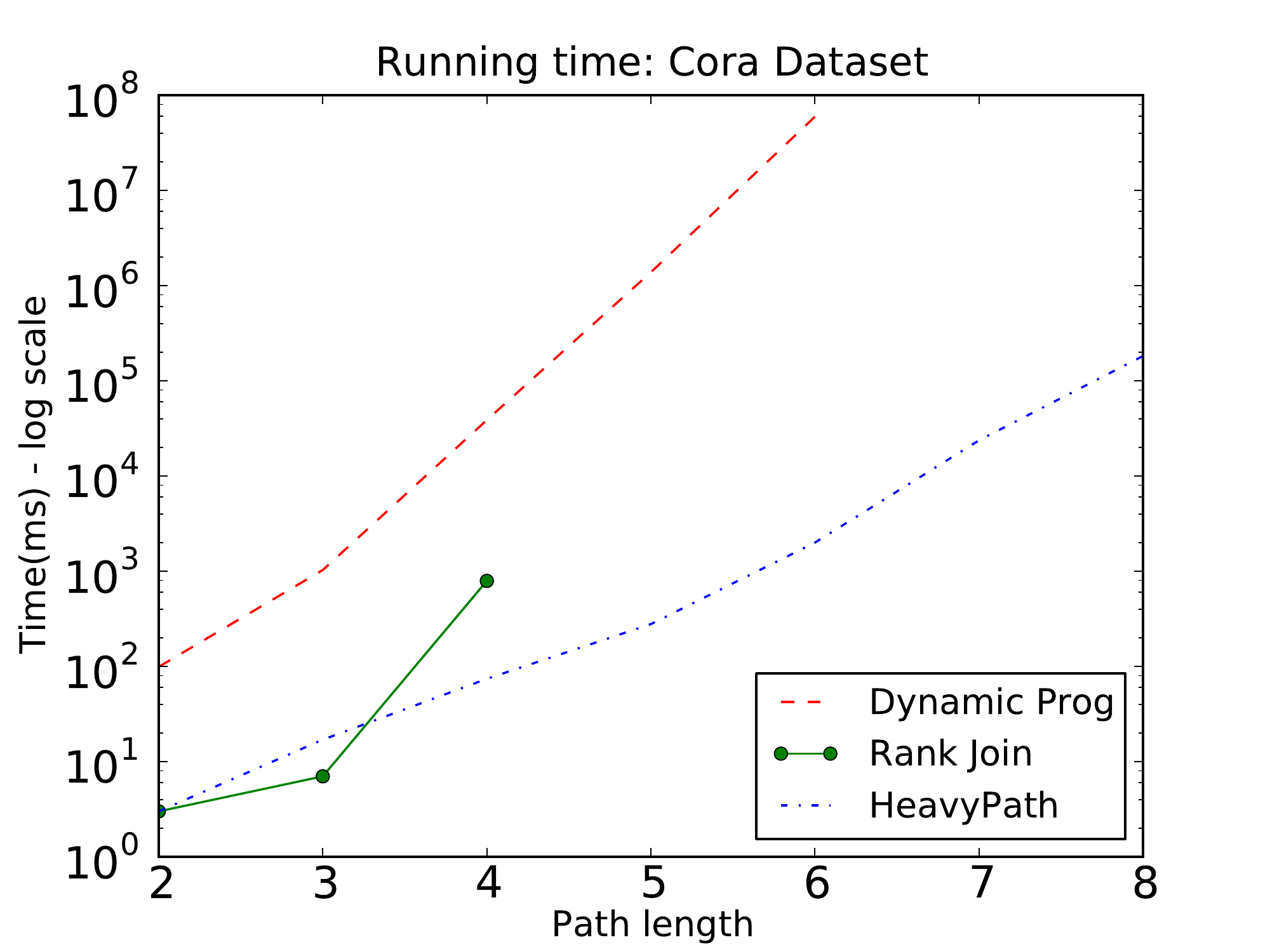}
\label{fig:Lcora}
}
\subfigure[Vary $\len, k=1$, last.fm ]{
\includegraphics[width=\figthree]{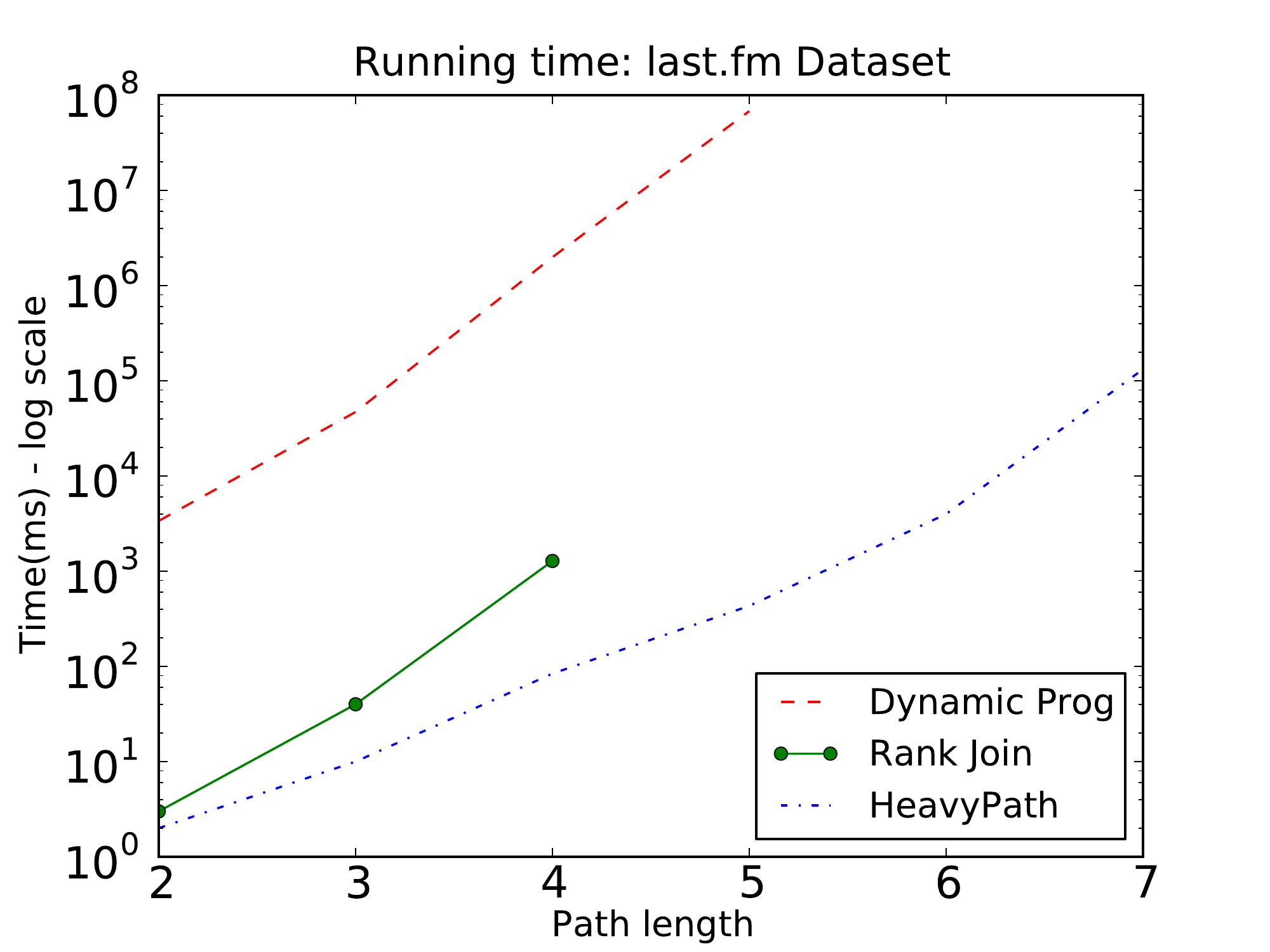}
\label{fig:Llastfm}
}
\subfigure[Vary $\len, k=1$, Bay ]{
\includegraphics[width=\figthree]{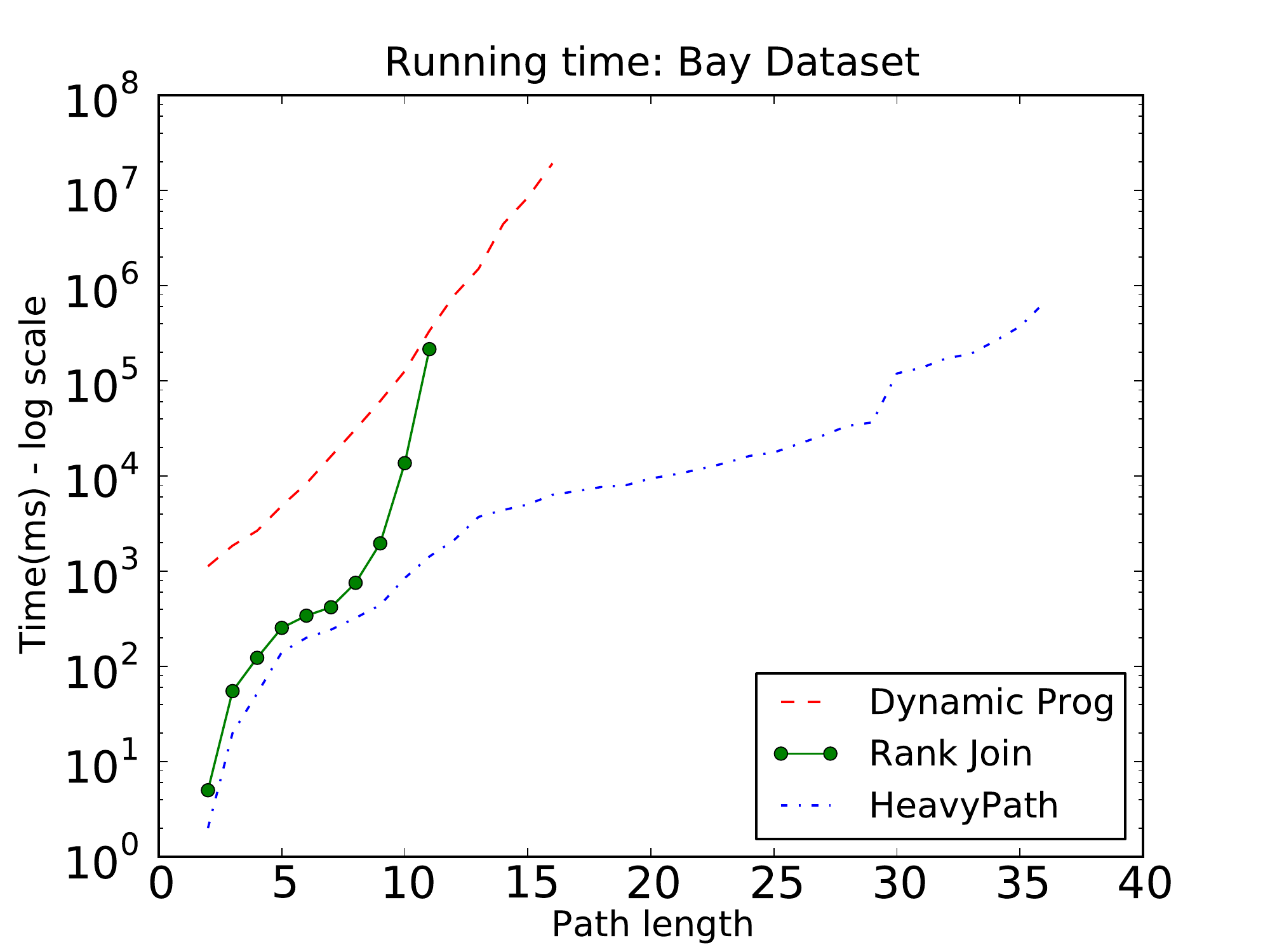}
\label{fig:Lbay}
}
\subfigure[$\len=4$, vary $k$, Cora ]{
\includegraphics[width=\figthree]{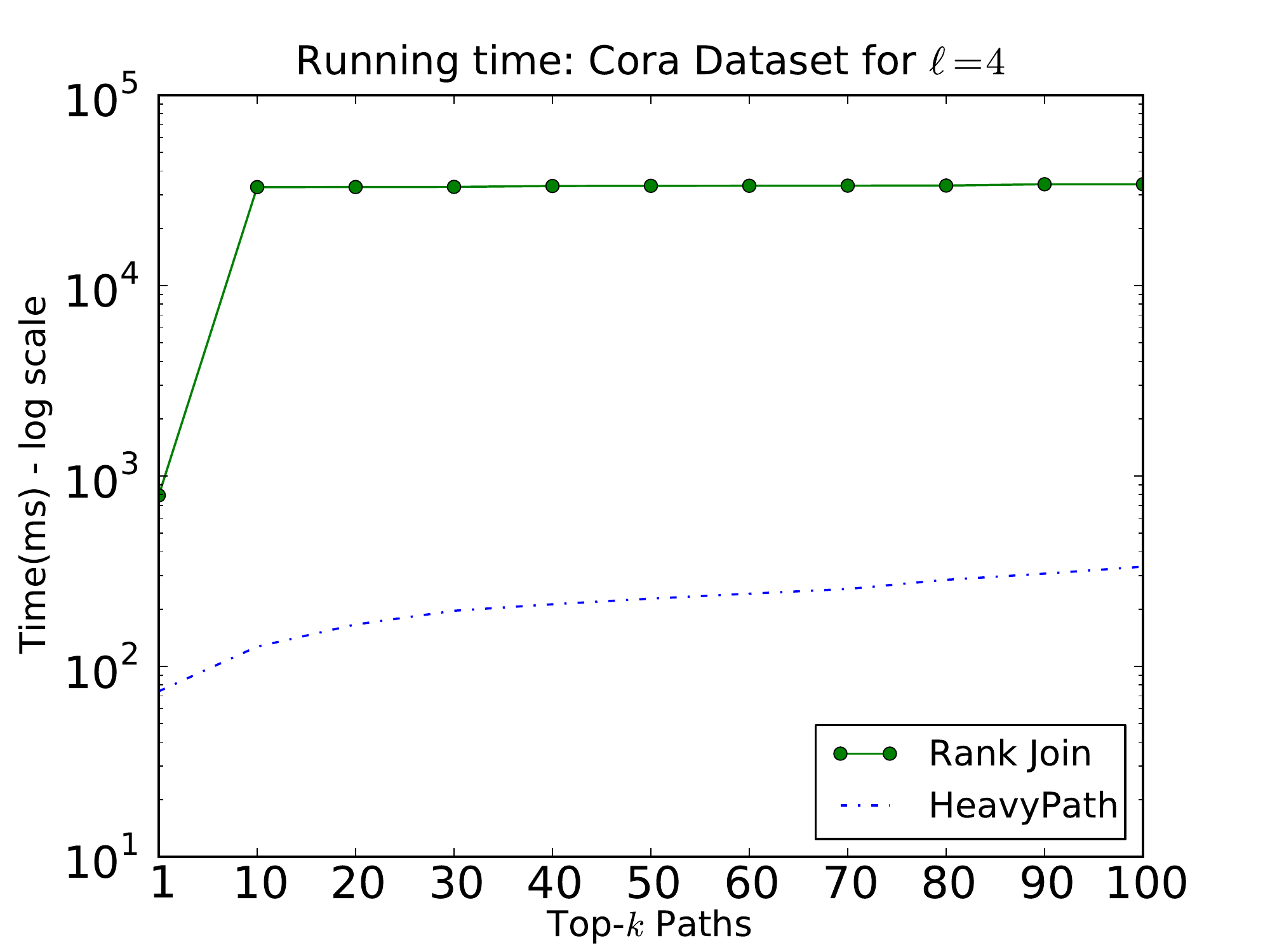}
\label{fig:kcora}
}
\subfigure[$\len=4$, vary $k$, last.fm ]{
\includegraphics[width=\figthree]{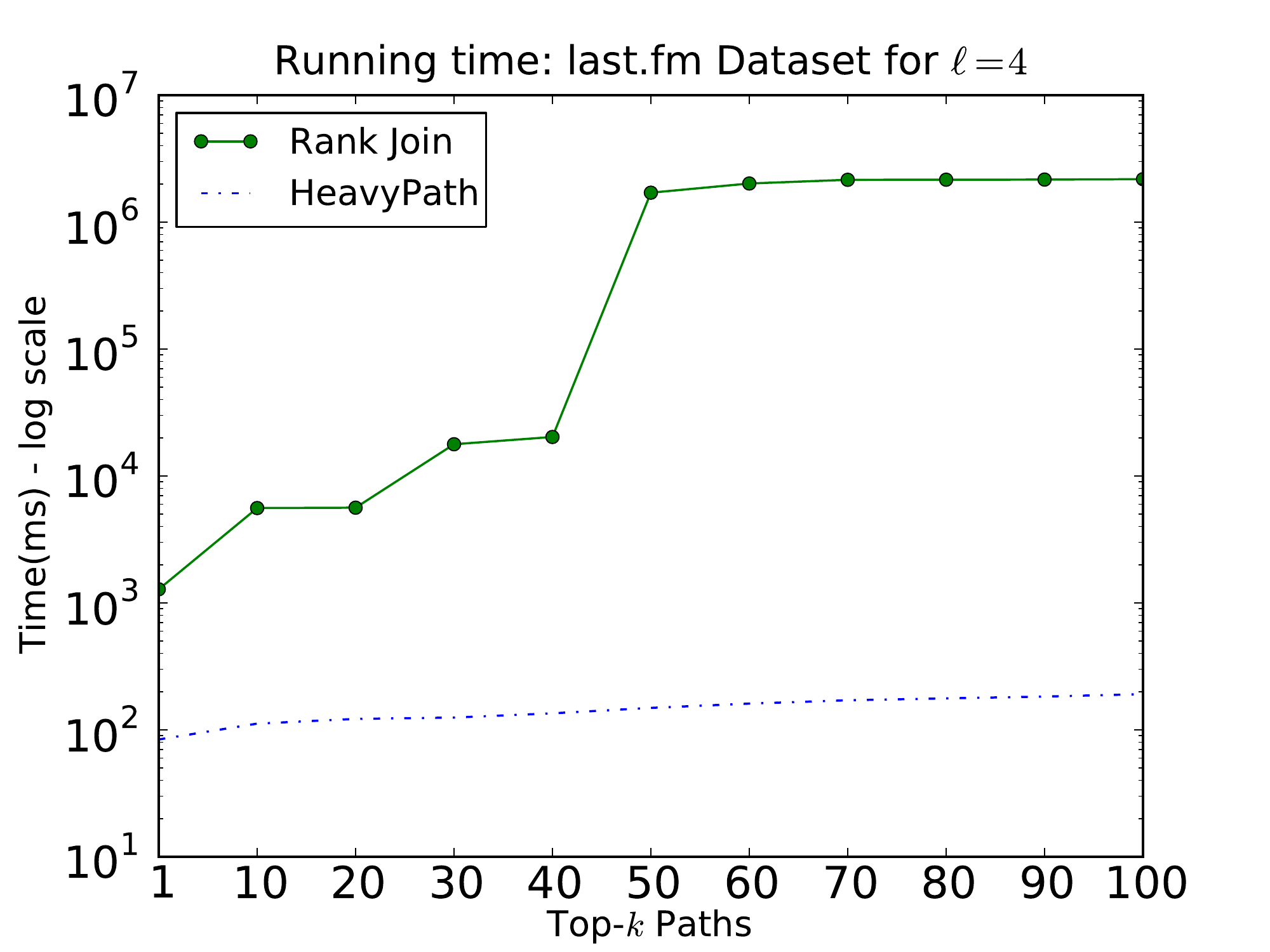}
\label{fig:klastfm}
}
\subfigure[$\len=10$, vary $k$, Bay ]{
\includegraphics[width=\figthree]{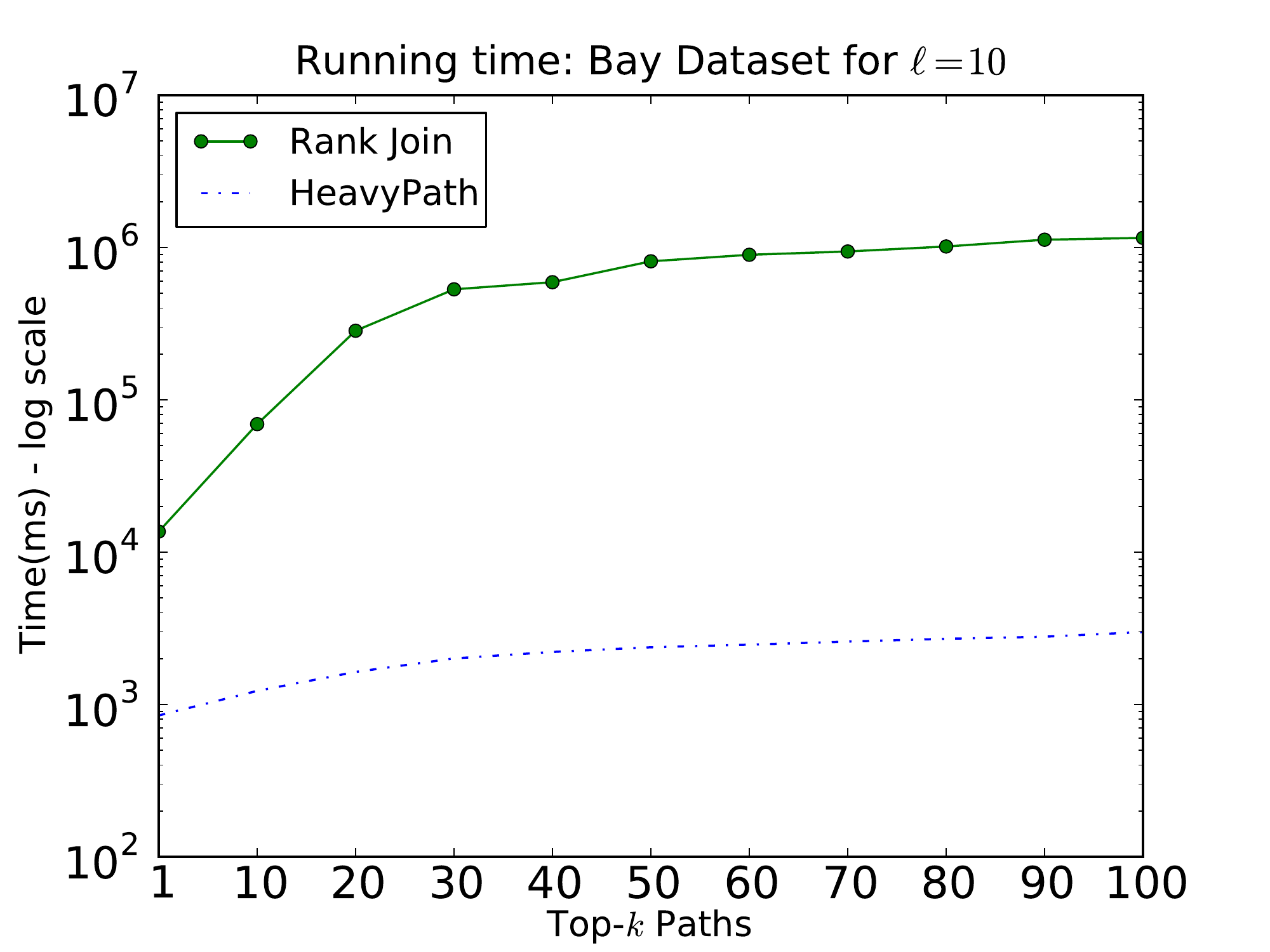}
\label{fig:kbay}
} \vspace{-2pt}
\caption{Running time comparisons for exact algorithms with different parameter settings}
\vspace{-2pt}
\label{fig:exact}
\end{figure*}

\begin{figure*}[t]
\centering
\subfigure[Vary $\len, k=1$, Cora ]{
\includegraphics[width=\figthree]{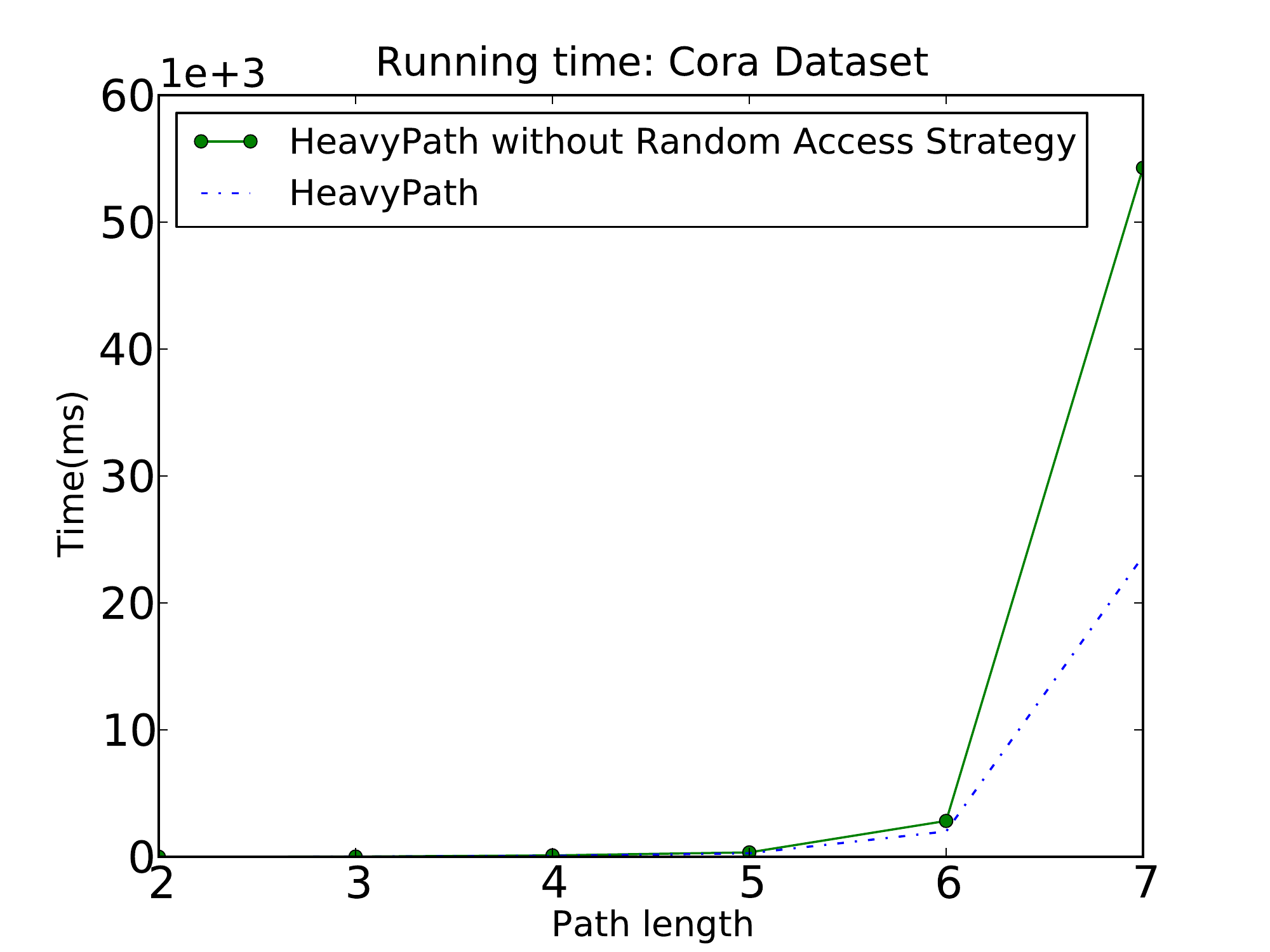}
\label{fig:LcoraNRA}
}
\subfigure[Vary $\len, k=1$, last.fm ]{
\includegraphics[width=\figthree]{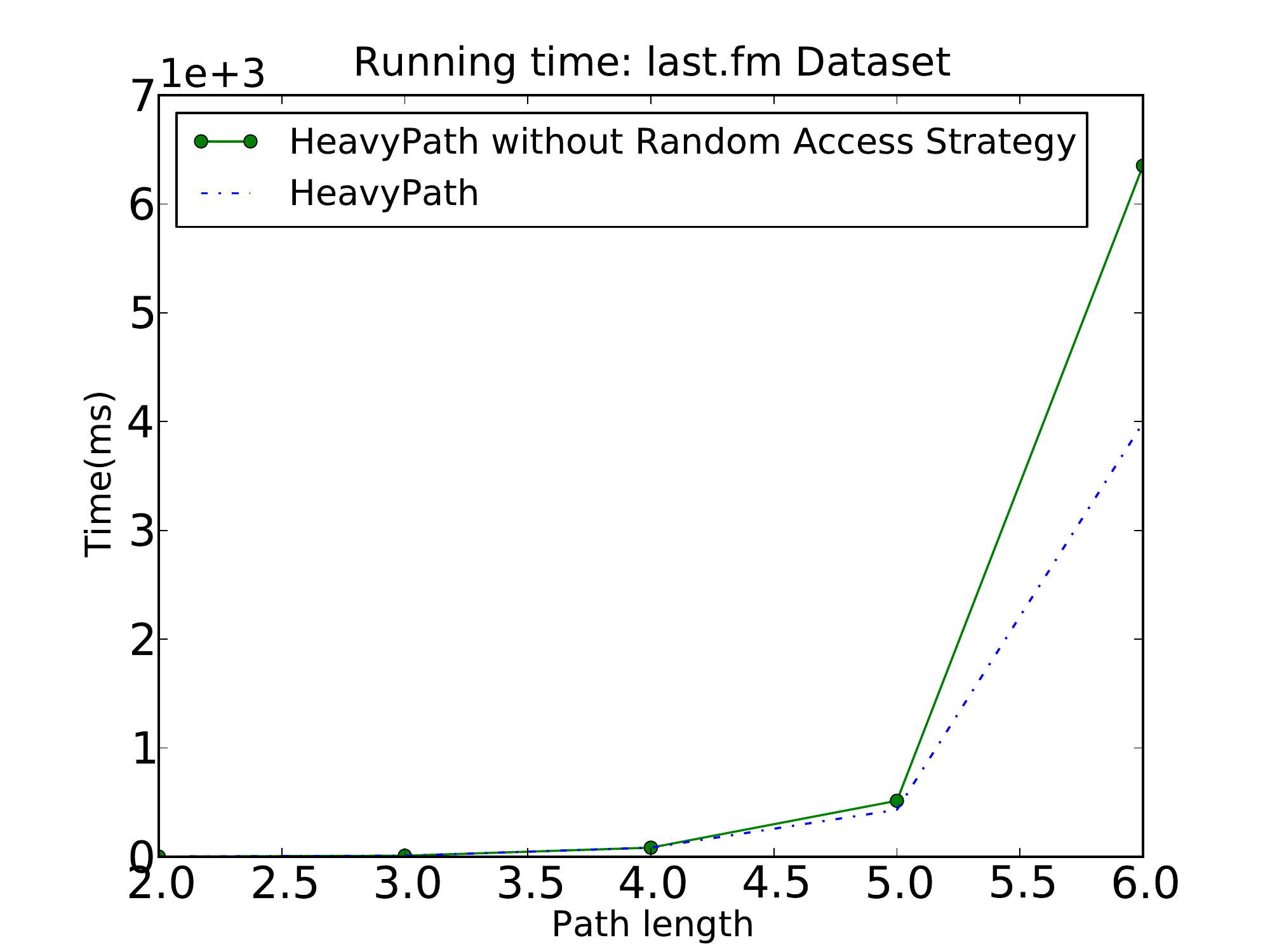}
\label{fig:LlastfmNRA}
}
\subfigure[Vary $\len, k=1$, Bay ]{
\includegraphics[width=\figthree]{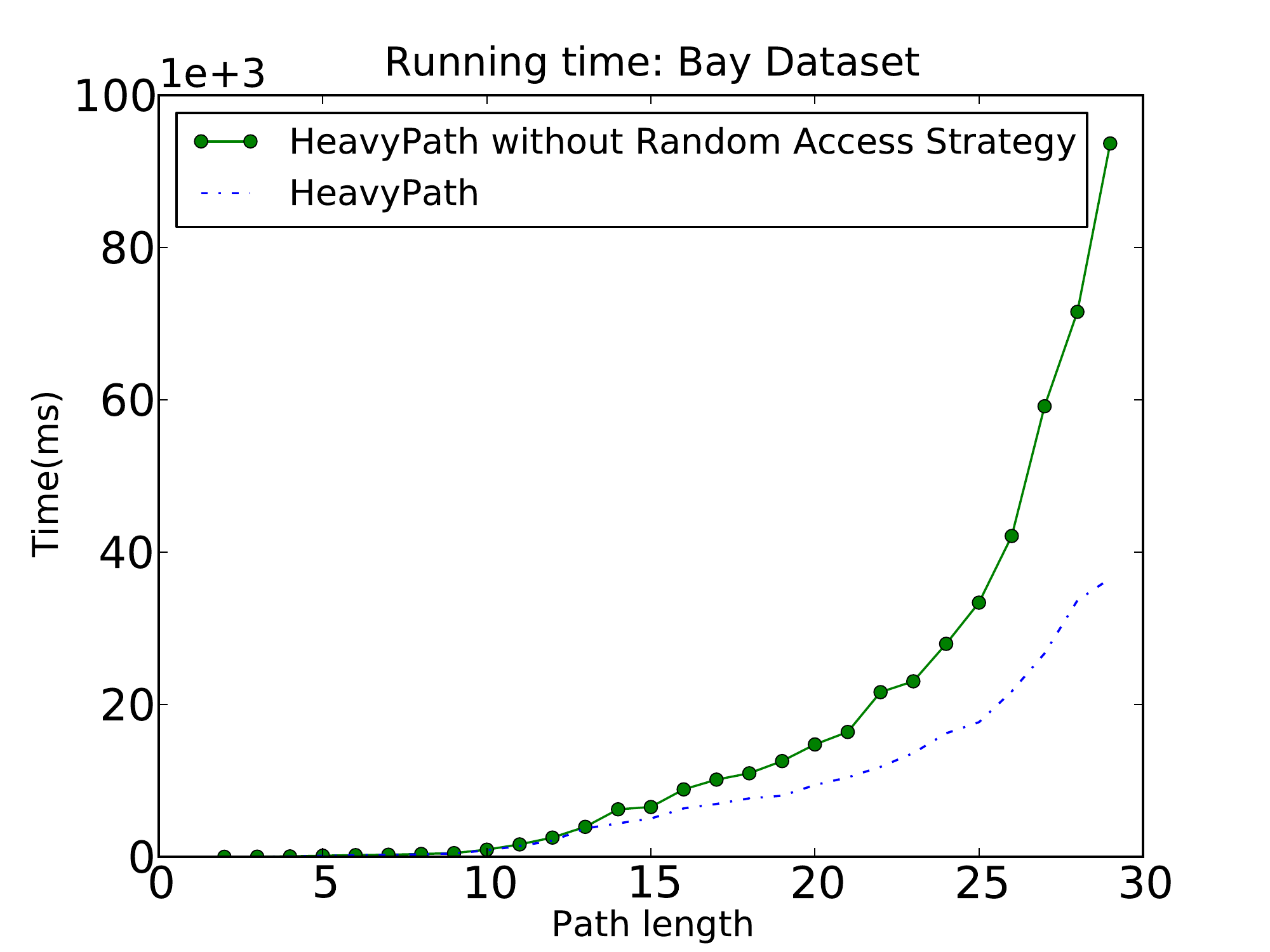}
\label{fig:LbayNRA}
}
\vspace{-2pt}
\caption{Improvement obtained with Random Access Strategy}
\label{fig:LNRA}
\vspace{-2pt}
\end{figure*}

\begin{figure*}[t]
\centering
\subfigure[Path quality using heuristics on Cora]{
\includegraphics[width=\figthree]{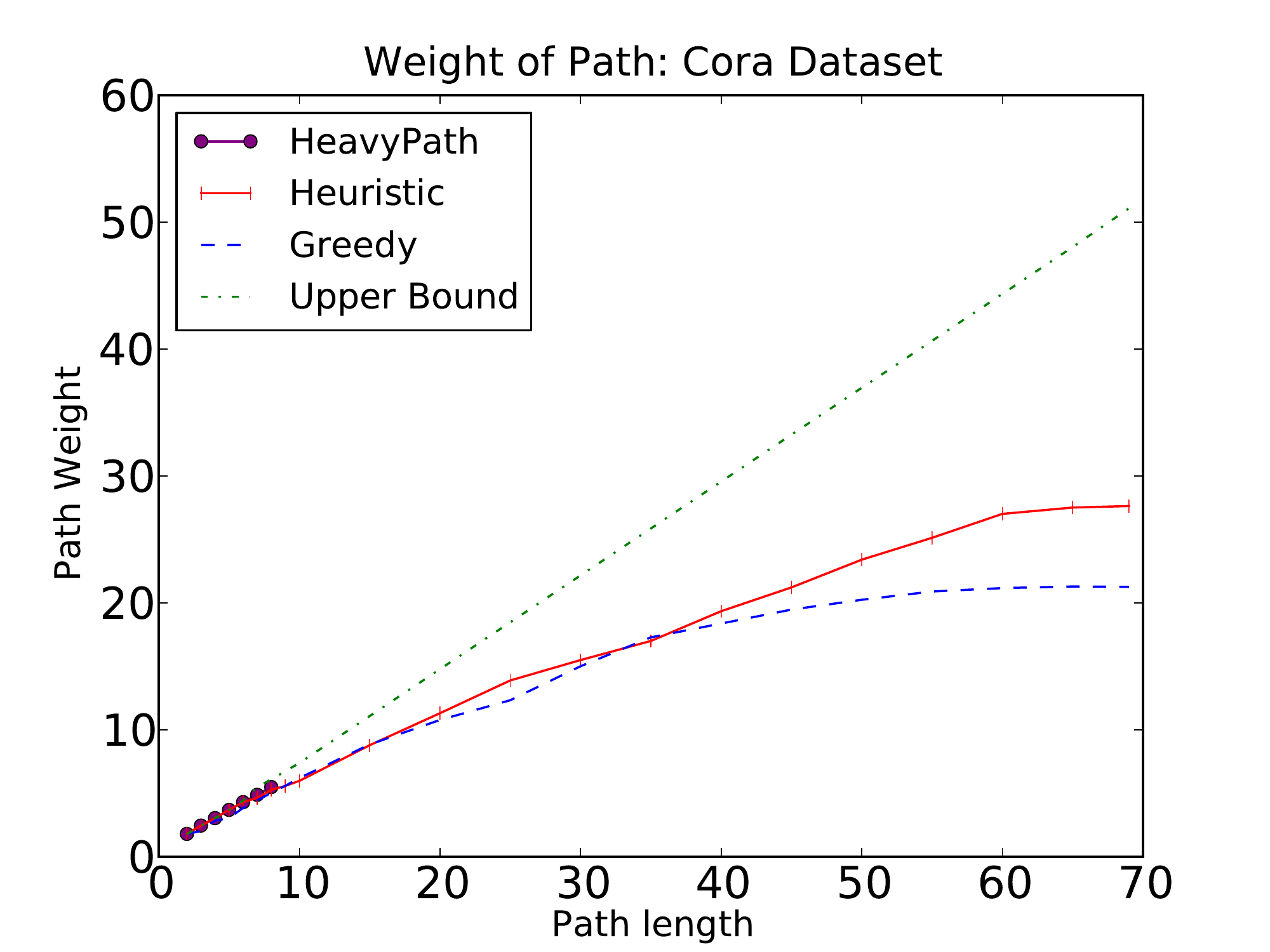}
\label{fig:Hcora}
}
\subfigure[Path quality using heuristics on last.fm]{
\includegraphics[width=\figthree]{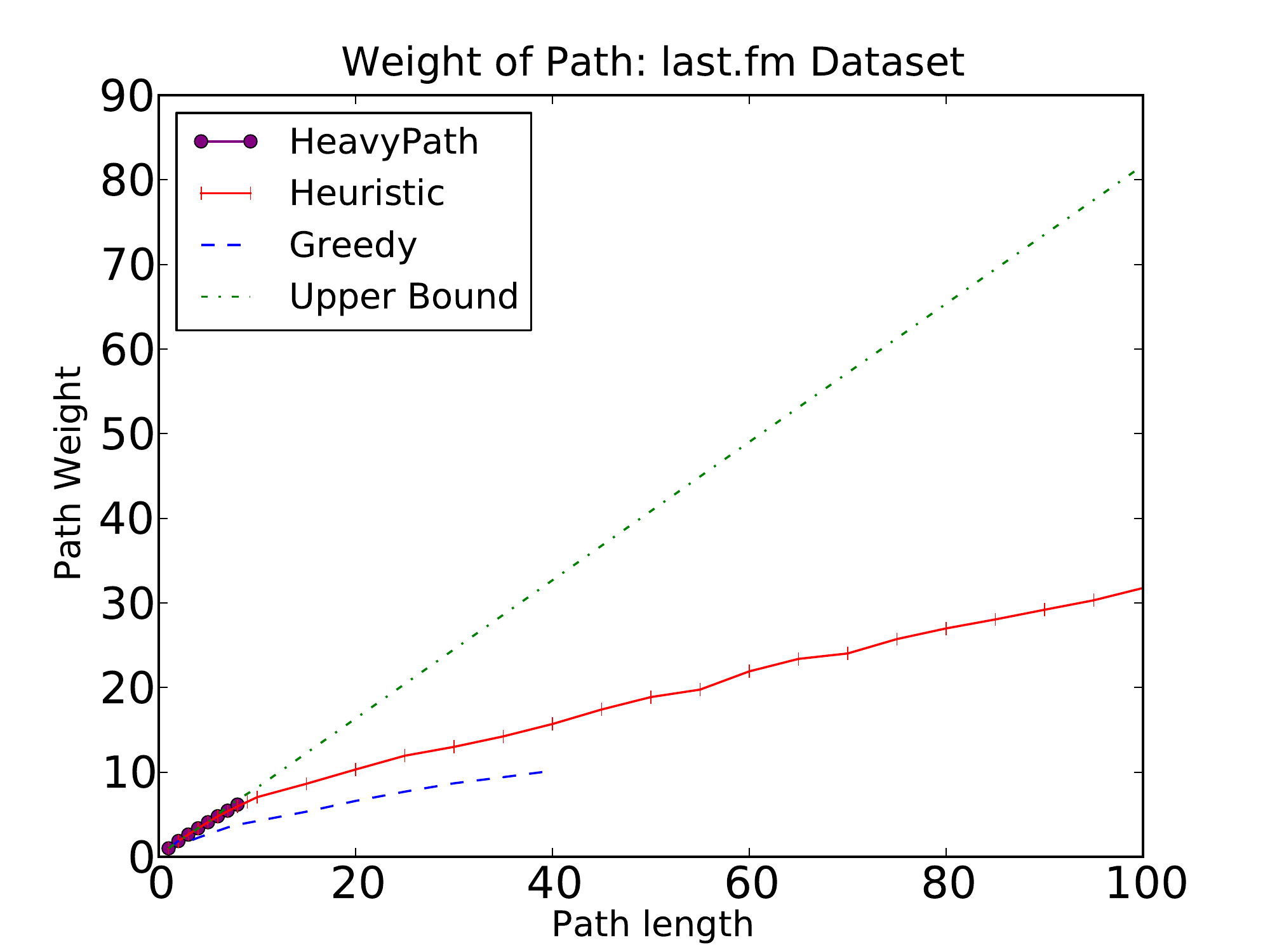}
\label{fig:Hlastfm}
}
\subfigure[Path quality using heuristics on Bay]{
\includegraphics[width=\figthree]{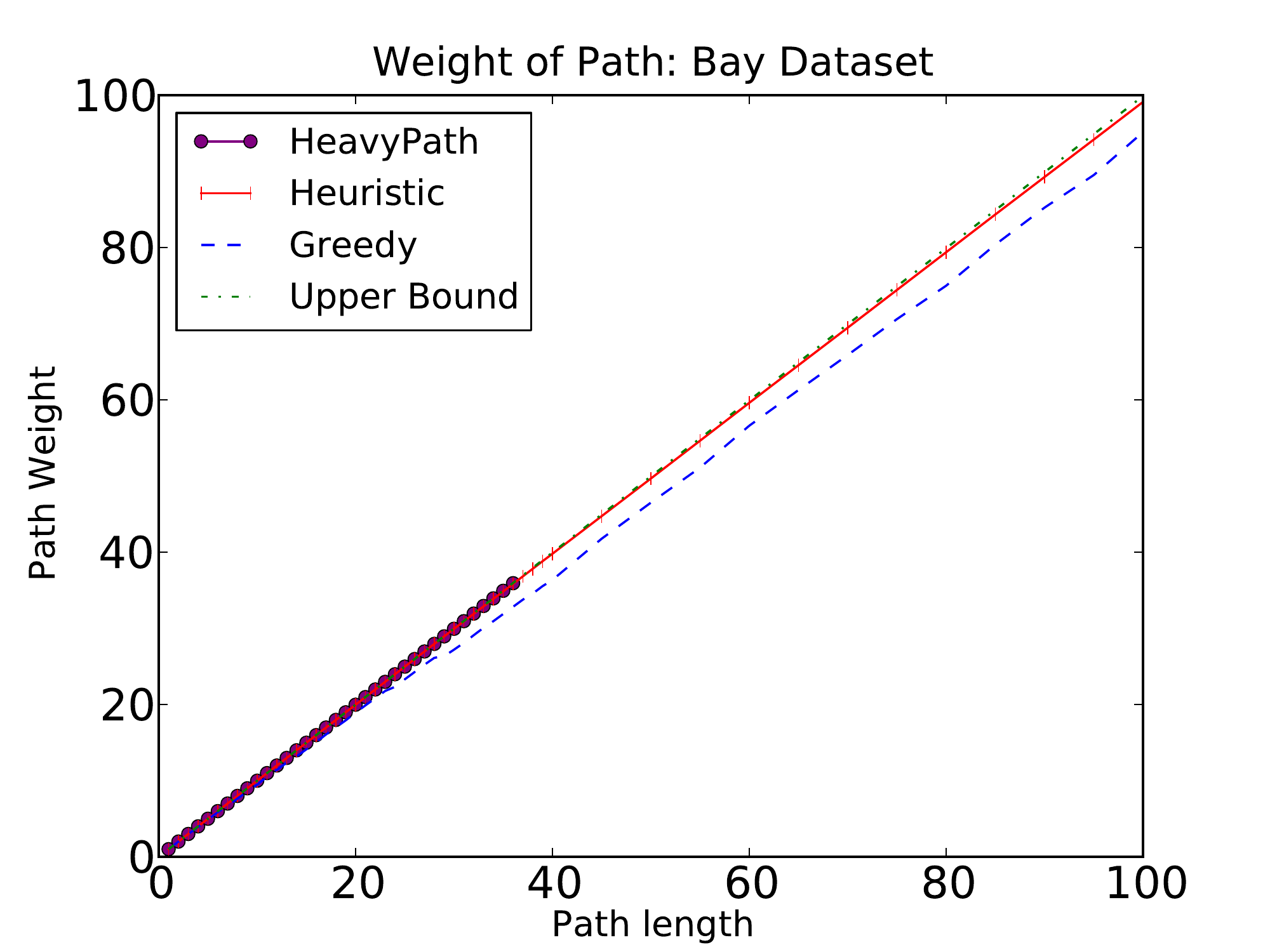}
\label{fig:Hbay}
}
\subfigure[Path length vs. $\rho$ on Cora]{
\includegraphics[width=\figthree]{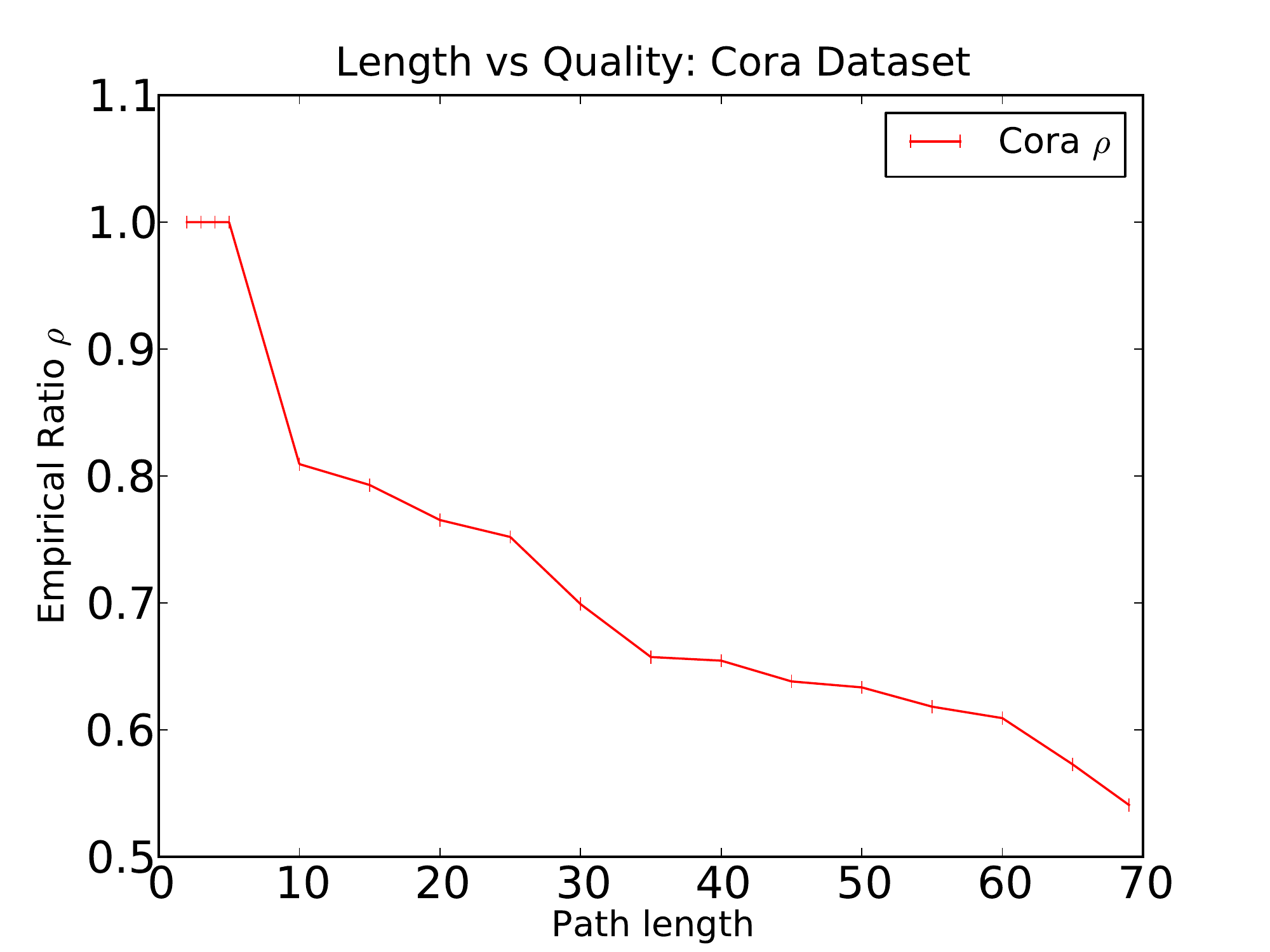}
\label{fig:HRcora}
} \subfigure[Memory vs. $\rho$ on Cora]{
\includegraphics[width=\figthree]{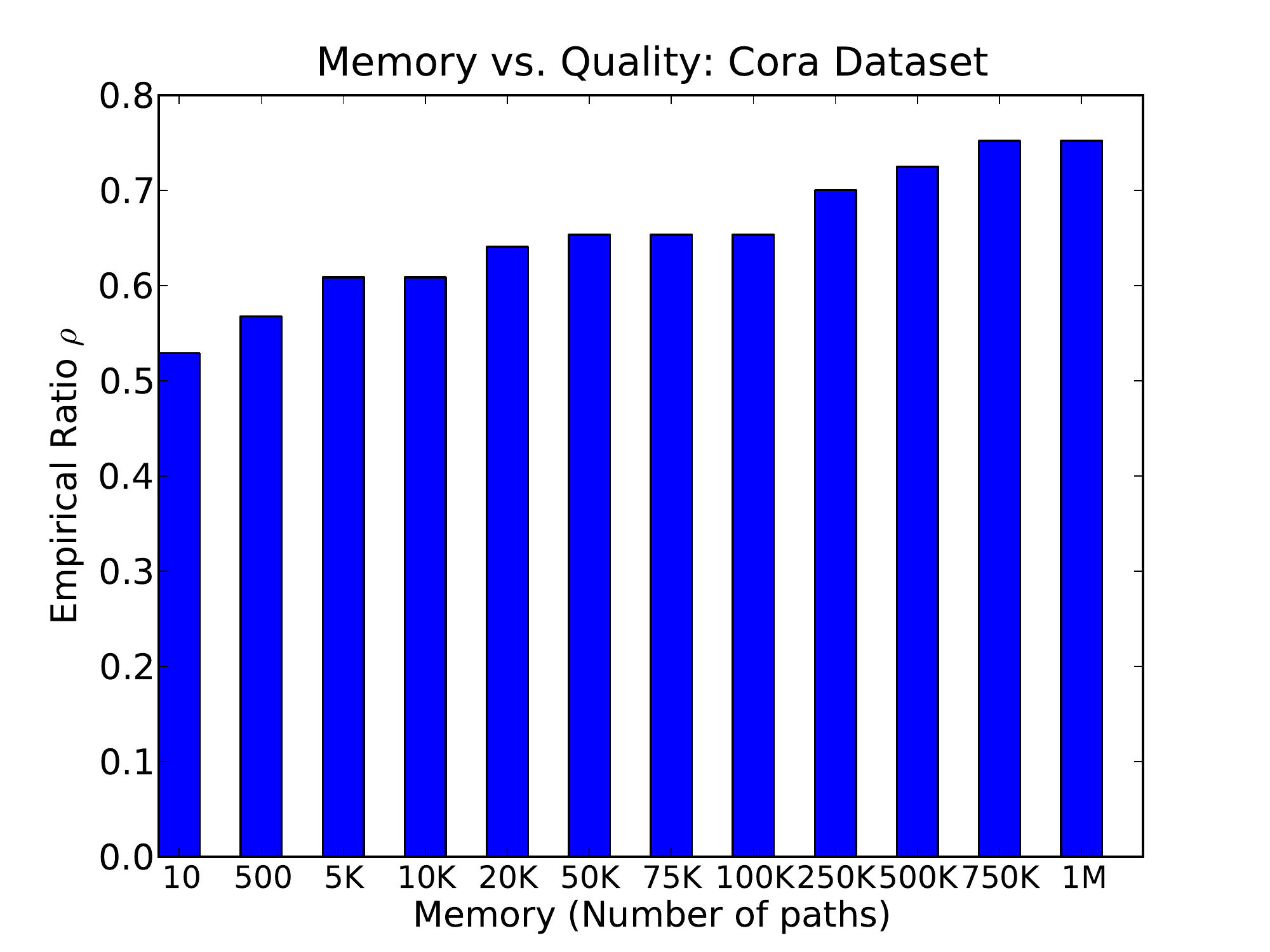}
\label{fig:MRcora}
} \subfigure[Running time of heuristics on last.fm]{
\includegraphics[width=\figthree]{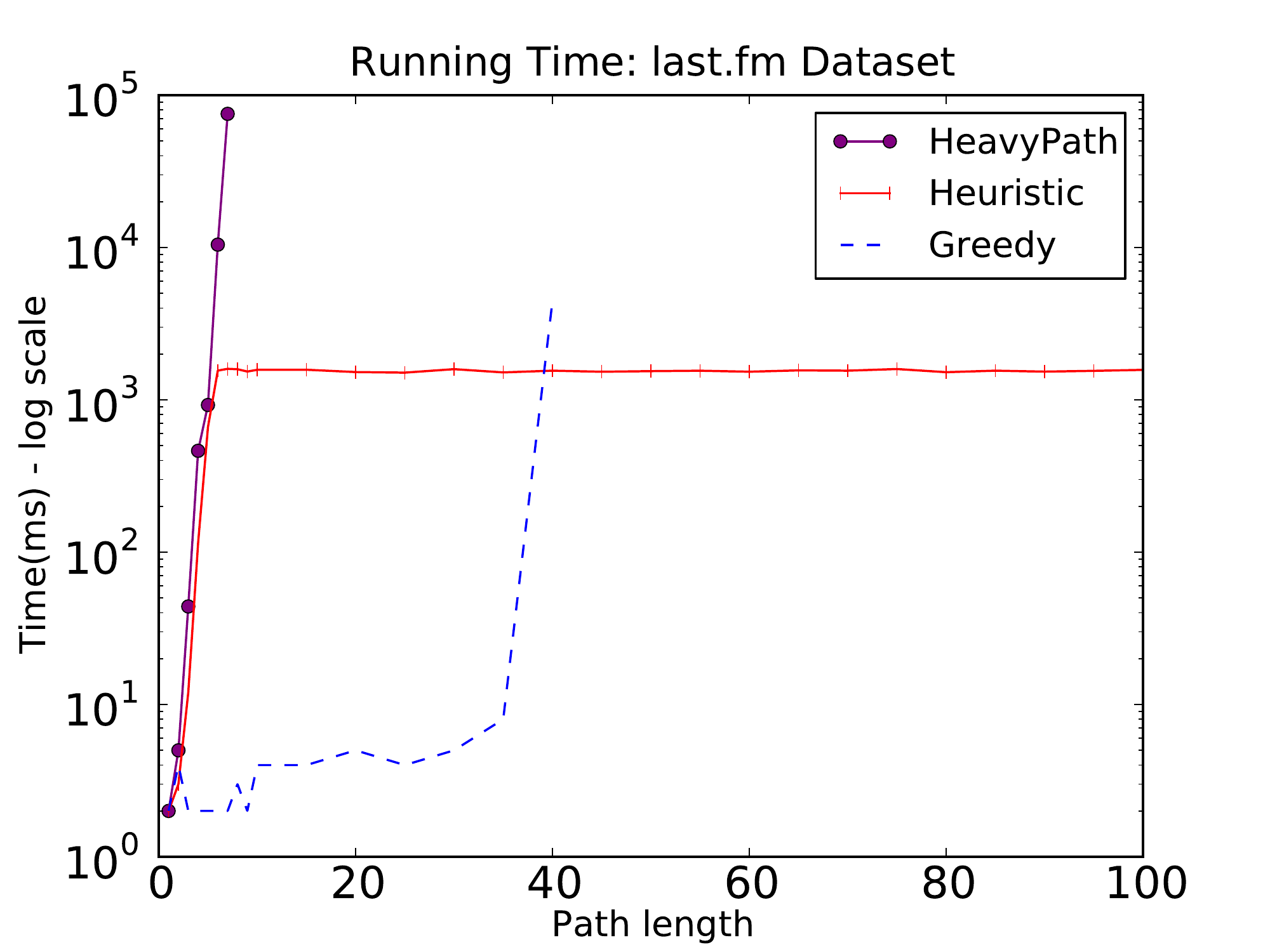}
\label{fig:HTlastfm}
}
\vspace{-2pt}
\caption{Performance using Heuristic approach}
\label{fig:HTime}
\label{fig:heuristics}
\vspace{-2pt}
\end{figure*}

\para{Running time for varying path lengths}
We compare the running times of the four algorithms proposed in Section~\ref{sec:algo} for
solving \hpp.  Figures~\ref{fig:Lcora},~\ref{fig:Llastfm},~\ref{fig:Lbay} show the running time for
finding the top-1 path of various lengths for Cora, last.fm and Bay datasets respectively.
The running time increases with the path length for all algorithms as expected.
For all three datasets, the \DP algorithm is orders of magnitude slower than the other algorithms
even for short paths of length 2. For Cora, \DP took over a day to compute the heaviest
path of length 6, albeit using only 4GB of the allocated 12GB of memory.
In all cases, except for $\len=3$ on Cora (see Figure~\ref{fig:Lcora}),
\Rank is slower than \Main, with the
difference in running times of the two methods increasing with path length.
We investigated the single instance where \Rank was faster than \Main and noticed
that for short paths that contain high degree nodes,
performing random accesses during \Main can result in extra costscompared with \Rank.
After $\len=4$ on Cora and last.fm, and $\len=11$ on Bay, \Rank runs out of
the allocated memory and quits.
In contrast, \Main is able to compute the heaviest paths of length $8$ on Cora, $7$ on last.fm, and
$36$ on Bay before running out of the allocated memory.
\red{\erase{The running times of \Main and \MainPlus are similar on Cora and Bay, while \MainPlus performs
slightly faster than \Main on last.fm graph.
It is also able to compute up to $\len=8$ on last.fm, that is one hop more than \Main. }}
All algorithms are faster and can compute longer paths on Bay as compared with
Cora and last.fm. Both the structure of the graph (especially the average degree), and the
distribution of the edge weights play a role in the running time and memory usage.
The Bay dataset is the largest of the three datasets, but has a small (i.e., 1.2) average
node degree, which makes it easier to traverse/construct paths.
The Cora dataset has only 70 nodes, but is a dense graph
that includes nodes which are connected to all other nodes.
In comparison, the last.fm graph has 40K nodes and average node degree $4.5$. 
\eat{
\from{laks}{i added a mention of memory usage, based on the mere fact that
algorithms run out of memory under some conditions. please check, esp. if we are not
going to have memory usage plots.}
}

On analyzing these results, we make the observation
that average node degree and edge weight distribution are the
main parameters that define the hardness of an instance. In fact,
our smallest dataset in terms of number of
nodes and edges, Cora,  is the most challenging of all. 
Polyzotis et al.~\cite{Schnaitter:10},
make similar observations in their experiments on the 
\Rank problem.

\para{Running time for varying number of \topk paths}
Figures~\ref{fig:kcora},~\ref{fig:klastfm},~\ref{fig:kbay} show the running times for finding top-$k$
paths of a fixed length, while varying $k$ from 1 to 100. We chose $\len=4$ for Cora and
last.fm and $\len=10$ for Bay as those were the longest path lengths for which
all algorithms produced an output within the allocated memory.
Since \DP is clearly very slow even for $k=1$, we focus on comparing the other algorithms in this
and subsequent experiments.
\Rank shows interesting behavior with increasing $k$. For smaller values of $k$,
e.g., $k<50$ for last.fm (see Figure~\ref{fig:klastfm}), the running time continues to increase
as $k$ increases. When $k$ is increased further, the running time does not change significantly.
When constructing paths, if the next heaviest path has already been constructed by \Rank, it can
output it immediately, and that seems to be the case for large values of $k$.
Recall that \Rank may build a subpath many times while constructing paths that subsume it.
The increase in running time for \Main \red{\erase{and \MainPlus}} as $k$ increases is insignificant.
Since the algorithm builds on shorter heavy paths,  
several paths of length less than
$\len$ may already be in the buffers and can be extended to compute the next heaviest path.
\red{\erase{Again, \MainPlus is slightly faster than \Main on Cora and last.fm datasets.}}

\para{\red{Impact of Random Access Strategy}}
\red{Figure~\ref{fig:LNRA} compares \Main without and with the random access strategy adopted 
and shows the additional benefit of employing the random access strategy. 
\eat{shows the benefit of adopting the random access strategy over an 
implementation of the adaptation of Rank Join with only sequential access, i.e., no random access.}%
Consistent across all datasets, we observe that using random access strategy 
speeds up the algorithm execution. The speedup is greater for longer paths.}

\para{Interpreting heavy paths}
We present a few representative example to show the application of \hpp to
generating playlists and finding flow of ideas, as discussed in Section~\ref{sec:intro}.

The heaviest path returned by \Main on last.fm with $\len=5$ has the following songs:
``Little Lion Man,
Sigh No More,
Timshel,
I Gave You All,
Winter Winds''.
We noticed that all these songs are from the ``folk rock'' genre and by the
band ``Mumford \& Sons''.
Interestingly, the heaviest path of $\len=7$ was comprised of a completely different set of songs:
``Put It On The Air, Beach House, Substitute For Murder,
Souvenir, Ease Your Mind, Novgorod, Sun God, Nemo'' from the
``art rock'' genre, by the band ``Gazpacho''.
On investigating further, we noticed a trend in how users of last.fm create playlists, which
is to add a collection of top/latest songs from a certain artist/band. Since the last.fm graph used
in our experiments was  abstracted from such playlists, we see paths corresponding to
songs by the same artist. It would be interesting to see the paths that emerge from
graphs abstracted from actual user listening history, which may potentially be more diverse.
We leave such exploration for future work.

We analyzed the ``topic paths'' that represent flow of ideas in the Cora graph of topics extracted
 from
research paper citations. The top path of $\len = 3$ had the topics:
``Cooperative Human Computer Interaction (HCI),
Distributed Operating Systems (OS),
Input, Output and Storage for Hardware and Architecture''.
It turns out that in this dataset, there are several papers about application like
group chat and distributed white boards that involve collaborative computing.
Such applications are concerned about both reliability of distributed servers
and real time end user experience. Hence, the topic path obtained from the
graph highlighted this\erase{not-so-} \blue{ less }obvious connection.
Longer heavy paths also exhibit interesting patterns, \erase{in the flow of
ideas across topics. }%
for instance, this heaviest topic path of length $7$:
``Object Oriented Programming,
Compiler Design,
Memory Management in OS\erase{ Operating Systems},
Distributed OS\erase{ Operating Systems},
Cooperative HCI\erase{Human Computer Interaction},
Multimedia and HCI\erase{Human Computer Interaction},
Networking Protocols,
Routing in Networks'' corresponds to a surprising,\erase{ interesting} and non-trivial flow
of ideas across a chain of topics.


\para{Quality of paths using the heuristic approach}
Figure~\ref{fig:heuristics} 
shows the path weight obtained using \hph
as $\len$ increases, \red{for $k=1$}.
The exact result as obtained by \Main is plotted for comparison, along with an ``upper bound''
that represents the best possible score of a path for a given $\len$ as detailed in
Section~\ref{sec:heuristic}.
As a baseline, we also plot the path resulting from a purely greedy approach that
builds the heaviest path of length $\len$ by starting from the heaviest edge and
repeatedly following heaviest edges out of the end nodes.
For the Cora dataset in Figure~\ref{fig:Hcora}, the
memory-bounded heuristic is no worse than $50\%$ of the theoretically
best possible path weight, and it
finds heavier paths of a given length compared with the greedy approach. Recall,
the theoretically best possible path weight is estimated on the upper
bound $U_{\len}$ as explained in Section~\red{\ref{sec:heuristic}}.
The approach can be used to find a path connecting all nodes in the graph ($\len=69$).
The results for the last.fm dataset in Figure~\ref{fig:Hlastfm} show that the heuristic approach is able to compute
paths of long lengths, with the accuracy decreasing with path length. The greedy approach stalled
after $\len=40$. The last.fm dataset has 6534 components, while other datasets have a
single connected component.
When the greedy starts with the heaviest edge, and the component to which it
belongs has fewer than 40 nodes, the greedy
algorithm restarts the process with the next heaviest edge. This results in a lot of exploration before
the algorithm can result in a heavy path of $\len>40$. The heuristic manages to find paths of
length up to $100$. Even at that length, weight of the path it finds is about $37.5\%$
of the theoretically best possible.
Interestingly, the heuristic approach finds paths whose weight is very close to that of the
best possible, for the
Bay dataset as seen in Figure~\ref{fig:Hbay}.

Figure~\ref{fig:HRcora} shows the empirical approximation ratio $\rho$ as the
path length increases, keeping the allocated memory fixed
(as with Figure~\ref{fig:Hcora}) at 12GB. As the path length increases,
$\rho$ decreases. This is expected, given the same bootstrap from the exact
algorithm, the estimate of the heuristic worsens with respect to the best possible weight.
It should be noted that {\sl $\rho$ is not a ratio\erase{n} w.r.t. the optimal solution, instead a conservative
estimate of the worst case ratio of the result of our algorithm on a given graph instance}. 
\eat{
\from{laks}{in the heuristic section, make sure to define $\rho$! it's undefined
currently. in fact, define the term empirical ratio and $\rho$ in a prominent visible way there.
secondly. be sure to say $\rho$ is not the ratio wrt the optimal solution.
rather it's wrt the best possible weight a path can possibly have. it's a conservative
approach that shows the worst case ratio for ou algorithm on a given instance.}
}

\para{Memory for the heuristic approach}
When the path length is fixed, and the memory
allocated to the heuristic approach is increased, $\rho$ increases.
Figure~\ref{fig:MRcora} illustrates this for the Cora dataset, with $\len = 25, k=1$
and memory allocation varied to store $10$ to $1$ million paths. It is worth noting that 
even with as little memory as needed to hold $5,000$ paths, $\rho$ is already at $0.6$ 
and with $250,000$ paths, it reaches $0.7$. Further improvement is limited.

\para{Running time for the heuristic approach}
Figure~\ref{fig:HTlastfm} shows the running time of the heuristic as well as of \Main and Greedy 
on the last.fm dataset.
\hph takes negligible amount of time for post-processing, for paths of
long length (shown up to $\len=100$). Greedy does not scale beyond $\len=40$. 
The other datasets show similar patterns for \Main and \hph, while Greedy 
takes only tens of milliseconds on other data sets. 

\subsection{Discussion}
We observed in Figures~\ref{fig:Lcora} that for one parameter setting of $\len=3$ on the
Cora dataset, \Rank outperformed \Main in terms of running time for finding the top-1 path.
We found that the random accesses performed for extending paths ending in
high degree nodes resulted in a slower termination. For instance, consider the sub-graph of the
graph in
Figure~\ref{fig:ex1} induced by the nodes $a', b', c', d_1', \ldots, d_n'$, but
with the following edge weights:
$w_{(a',b')}=1, w_{(b',c')}=1, w_{(c',d'_{1})}=1$, and all other edge weights to
nodes $d'_{2}\ldots d'_{n}$ the same at $0.01$. \Rank would scan the 3 top edges and terminate
(since the top path weight equals the threshold),
while \Main will scan $n-1$ additional edges by performing random accesses.
A potential enhancement to \Main is to perform the random accesses  in
a ``lazy'' fashion. In particular, perform random access on demand, and in a sorted order of
non-increasing edge weights. Such an approach would avoid wasteful random accesses to
edges that have very low weight.

While our exact algorithms \red{already} scale \red{much} better than the classical \Rank approach,
our heuristics take a only tens of milliseconds to compute results that are
no worse than 50\% of the optimal for path lengths up to 50 on all datasets we
tested.
It is worth noting that heuristics for the problem of TSP have been extensively
studied\footnote{\url{http://www2.research.att.com/~dsj/chtsp}} 
for decades. 
Although these are not directly applicable to \hpp,
it would be interesting to explore the ideas employed in those heuristics to
possibly get higher accuracy.

We conducted various additional experiments focusing on metrics such as 
number of edge reads performed, number of paths constructed, and the rates at 
which the termination thresholds used by the algorithms decay. The details 
can be found in Appendix~\ref{app:addexp}.

\section{Related work}
\label{sec:related}

The problem of finding heavy paths 
is well suited for enabling applications as described in 
Section~\ref{sec:intro}. 
Recently, there has been interest in the problem of 
itinerary planning~\cite{Choudhury:10, Xie:10}. 
Choudhury et al.~\cite{Choudhury:10} model it as a 
variation of the orienteering problem, and in their setting 
\blue{the end points of the tour are given,}
making the problem considerably simpler than ours.
Xie et al.~\cite{Xie:10} study it in the context of generating 
packages (sets of items) under user-defined constraints, e.g., an itinerary 
of points of interest with 2 museums and 2 parks in London.\erase{In \cite{Xie:10}, }
However, the recommendations that are returned to the user are sets of items 
which do not capture any order. 
In both these papers, the total cost of the POI tour is subject to a 
constraint and the objective is to maximize the ``value'' of the tour, 
where the value is determined by user ratings.\erase{Furthermore, 
in \cite{Choudhury:10}, the beginning and end points 
$s$ and $t$ of the tour are assumed to be given.} 
In contrast\erase{with these papers}, by modeling itinerary finding as a \hpp problem, we aim to 
minimize the cost of the tour \red{(through high value items)}, which is 
technically a different problem. 
Another related work is 
\cite{Ekstrand:10} on 
 generating a ranked list of papers, given query keywords. 

\eat{Both the machine learning and the music information retrieval 
communities have studied the problem of automatically generating playlists. 
One of the early works was a system called AutoDJ
by Platt et al.~\cite{Platt:01}
}
In \cite{Hansen:09}, Hansen and Golbeck make a case for 
recommending playlists and other collections of items.  
The AutoDJ system of Platt et al.~\cite{Platt:01} is one of the early
works on playlist generation 
that uses acoustic features 
of songs. 
\erase{Smart Radio~\cite{Hayes:02} is 
a popular system  
for music recommendation. }
There is a large number of similar services -- see~\cite{Fields:10}
for 
comparison of some of these services. 
To our knowledge, 
playlist generation as an optimization problem has not been studied before.

\erase{The problem of finding heavy paths in a graph }%
\blue{The \hpp} was studied
recently by~\cite{Bansal:07}  for the specific application of finding ``persistent
chatter'' in the blogosphere. 
Unlike our setting, the graph associated with their application is \len-partite and acyclic.
\erase{Their DFS and BFS based algorithms were suitable for this special structure of the graph. }%
As mentioned in the introduction, adaptations of their algorithms to general 
graphs lead to rather expensive solutions. 

As mentioned earlier, 
the \hpp problem can be mapped to a length restricted version of TSP 
known as $\len$-TSP, where $\len$ is the length of the tour. 
$\len$-TSP is NP-hard and inapproximable when triangular inequality 
doesn't hold~\cite{Arora:06}. For the special Euclidean case, there is a 
2-approximation algorithm due to Garg et al.~\cite{Garg:05}. 
We propose a practical solution based on the well established \Rank 
framework for efficiently finding the exact answer to this problem for 
reasonable path lengths. To the best our knowledge, this solution is novel. 

\Rank 
was first proposed 
by Ilyas et al.~\cite{Ilyas:03, Ilyas:04, Ilyas:05} to produce top-$k$ join tuples in
a relational database. They proposed the logical rank join
operator that enforces no restriction on the number of input relations. 
In addition, they describe HRJN, a binary 
operator for two input relations, and a pipelining solution 
for more than two relations \red{since} a multi-way join is
not supported well by traditional database engines. 
\erase{Polyzotis et al.~\cite{Schnaitter:10} argue  that  repeated application
of the binary operator has no optimality advantage
over the multi-way \Rank operator. }It is well known that the multi-way \Rank 
operator is instance optimal, 
however similar optimality results are not known for iterated applications of the 
binary operator~\cite{Schnaitter:10, Schnaitter:8}. \erase{We therefore compare}%
\red{Therefore our comparison of \Main with the multi-way 
\Rank approach, to establish both our theoretical and empirical claims, is well justified.} 
\eat{  
Although logical and physical rank join operators
have been proposed and studied in the past for the general 
case with multiple input relations, 
most research in this area has been focused on binary \Rank. 
An instance optimal rank join operator is proposed in~\cite{Schnaitter:8}
through maintaining a tight upper bound to improve HRJN~\cite{Ilyas:03}. 
However, the proposed algorithm is not efficient in CPU time
according to a later study~\cite{Schnaitter:10}.
A new algorithm is proposed which is both instance
optimal and efficient in~\cite{Schnaitter:10}. 
}

The seminal work on \Rank~\cite{Ilyas:03} already mentions 
the potential usefulness of random accesses, 
however it does not evaluate it theoretically or empirically. 
A recent study~\cite{RA:11} proposes a cost-based approach 
as a guideline for determining  when random access should be performed 
in the case of binary \Rank. \red{As such, their results are not directly 
applicable for \hpp.}  
%
In contrast to all the prior work on \Rank, 
we address the specific question of finding the \topk 
heaviest paths in a weighted graph and adapt the \Rank 
framework to\erase{and} develop more efficient algorithms.\erase{ based on 
that framework.} Our techniques leverage the fact that 
the join is a self-join, store 
intermediate 
results and \blue{use }random accesses to aggressively lower the 
thresholds and facilitate early termination. Besides, 
we develop techniques for carefully managing random accesses \erase{in order }%
to minimize duplicate derivations of paths. 
Furthermore, we establish theoretical results that are in favor of using random
access for repeated binary rank self-joins.

\section{Conclusions and Future Work}
\label{sec:concl}
Finding the \topk heaviest paths in a graph is \red{an} 
important  problem with many practical applications. This
problem is NP-hard. 
In this paper we focus on developing 
practical exact and heuristic algorithms for this problem.
We identify its connection with the well-known
\Rank paradigm and provide insights on how to
improve \Rank for this special setting.
To the best of our knowledge, we are
the first to identify this connection.
We present the \Main algorithm that significantly
improves a straightforward adaptation of \Rank
\erase{through controlled} \red{by employing and controlling} 
random accesses and via more aggressive
threshold updating. 
\red{\erase{We further improve \Main
and propose the \MainPlus algorithm that improves
\Main and experimentally found that 
\MainPlus works better than \Main on graphs 
with many small connected components.}}
We propose
a practical heuristic algorithm which is able to provide an empirical approximation
ratio and scales well both w.r.t. path length and number 
of paths. Our experimental
results suggest that our algorithms are both scalable and
reliable. 
Our future work
includes improving memory usage and scalability of our exact
algorithms, 
\red{exploration and adaptation of ideas in \cite{RA:11} for improving the 
performance further}, 
and
application of our algorithms in recommender systems
and social networks.
 It is also interesting to investigate \topk algorithms with probabilistic guarantees for the heavy path problem.

\eat{Our future work includes improving memory usage
and scalability of our exact algorithms through lazy random
access, other heuristic algorithms to improve the empirical
approximation ratio as well as applications of our algorithms
in recommender systems and social networks.
}

\bibliographystyle{abbrv}
\bibliography{recSys}

\begin{appendix}

\section{Dynamic Programming Details}
\label{app:dp}

Essentially, dynamic programming constructs all simple paths of length $\len$
in order to find the heaviest among them. But unlike DFS, it aggregates
path segments early, thus achieving some pruning.
\eat{For example, suppose there
are $p$ paths $P_1, ..., P_p$ of length $l-1$ that end at $y$, and all of them avoid the node $j$,
which is a neighbor of $y$.
Then for the purpose of finding the heaviest path of length $l$ ending at $j$, we just need
to remember the heaviest among $P_1, ..., P_p$. Another key observation exploited by the
dynamic program is that the start nodes of the paths we consider are immaterial and can be
dropped from further consideration.}
More precisely, here are the
equations of the dynamic program. 
Letters $i,j,y$
denote variables which
will be instantiated to actual graph nodes when the program is run. $S$ denotes the
avoidance set. The variable $l$ will be instantiated in the range $[2, \len]$.
\begin{align*}
P^l_{i,j,S} &= \textsc{MAX}\{P^{l-1}_{i,y,S\cup\{y\}}\circ e(y,j) \mid (y,j) \in E, \\
  & y\not\in S, j\in S\} \\
P^1_{i,j,S} &= \begin{cases}
		\textsc{create}(P, e(i,j)) & \mbox{if} (i,j) \in E \\
		\textsc{null} & \mbox{otherwise}
		\end{cases}
\end{align*}
In the above equations, we can think of $P^l_{i,j,S}$ as a ``path object''
with properties ${path}$ and ${weight}$. Here, $P^l_{i,j,S}.{path}$ denotes the
heaviest $S$-avoiding path from $i$ to $j$ of length $l$, and $P^l_{i,j,S}.{weight}$ denotes
its weight. The operator
\textsc{MAX} takes a collection of path objects and  finds the object
with maximum weight among them. The ``$\circ$'' operator takes
a $P$ object and an edge $e(u,v)$, concatenates the edge with the
$P.{path}$ and updates $P.{weight}$ by adding the weight of the
edge. Finally, $create(P, e(i,j))$ creates a
path object $P$ whose ${path}$ property is initialized to $(i,j)$ and whose
${weight}$ property is initialized to the edge weight of $(i, j)$.

We invoke the dynamic program above using $P^{\len}_{\$i, j, \{j\}}$
every node $j \in V$ , where we have left the start node as a variable $\$i$.
The heaviest path of length $\len$ in
the graph is the heaviest among the paths found above for various $j$.

\eat{
\begin{figure}
\vspace{-3mm}
  \centering
  \includegraphics[width=3in, height=1in]{example.eps}\\
\vspace{-2mm}
  \caption{Example Co-occurrence Graph. }\label{fig:example}
\vspace{-3mm}
\end{figure}

\begin{example}
\label{ex:ex1}
Figure~\ref{fig:example} shows an example that will be used as a running example
to illustrate various algorithms.
Consider $P^3_{I,5,\{5\}}$, i.e., asking for the heaviest simple path ending
at $5$. Note the start node is left as a variable and the avoidance set includes
only the end node $5$. A ``run'' of the dynamic program on this ``query'' is illustrated
in Figure~\ref{fig:dprun}. Notice the applications of \textsc{MAX} operator
(indicated via arcs under tree nodes) which allow multiple paths to be pruned
at multiple stages. Observe also the fact that start nodes are completely
``ignored'' in that among path segments starting at different start nodes, the
\textsc{MAX} operator picks only one, the heaviest one.
\end{example}
}

For finding \topk heaviest paths of a
given length for $k > 1$, all we need to do is
define
the \textsc{MAX} operator so that it works in an iterative mode and finds the next
heaviest path segment (ending at a certain node, of a given length, and avoiding
certain set of nodes). We can apply this idea recursively and easily extend the
dynamic program for finding \topk heaviest paths of a given length. 


\eat{
\begin{figure}
\begin{center}
\scalebox{0.5}{\input{dprun.pstex_t}}
\end{center}
\caption{\label{fig:dprun} A run of the dynamic program. Arcs under tree nodes
indicate application of \textsc{MAX} operation.}
\end{figure}
}

\section{Additional Experiments}
\label{app:addexp}
\eat{
- #edge reads.
- #paths (of any length) constructed and stored.
- rate of decay of stopping thresholds.
- separating compute time from GC time and showing
that compute(A) >> total(HP), for algorithms A
with which HP is compared.
}

\begin{figure*}[t]
\centering
\subfigure[Cora.]{
\includegraphics[width=\figthree]{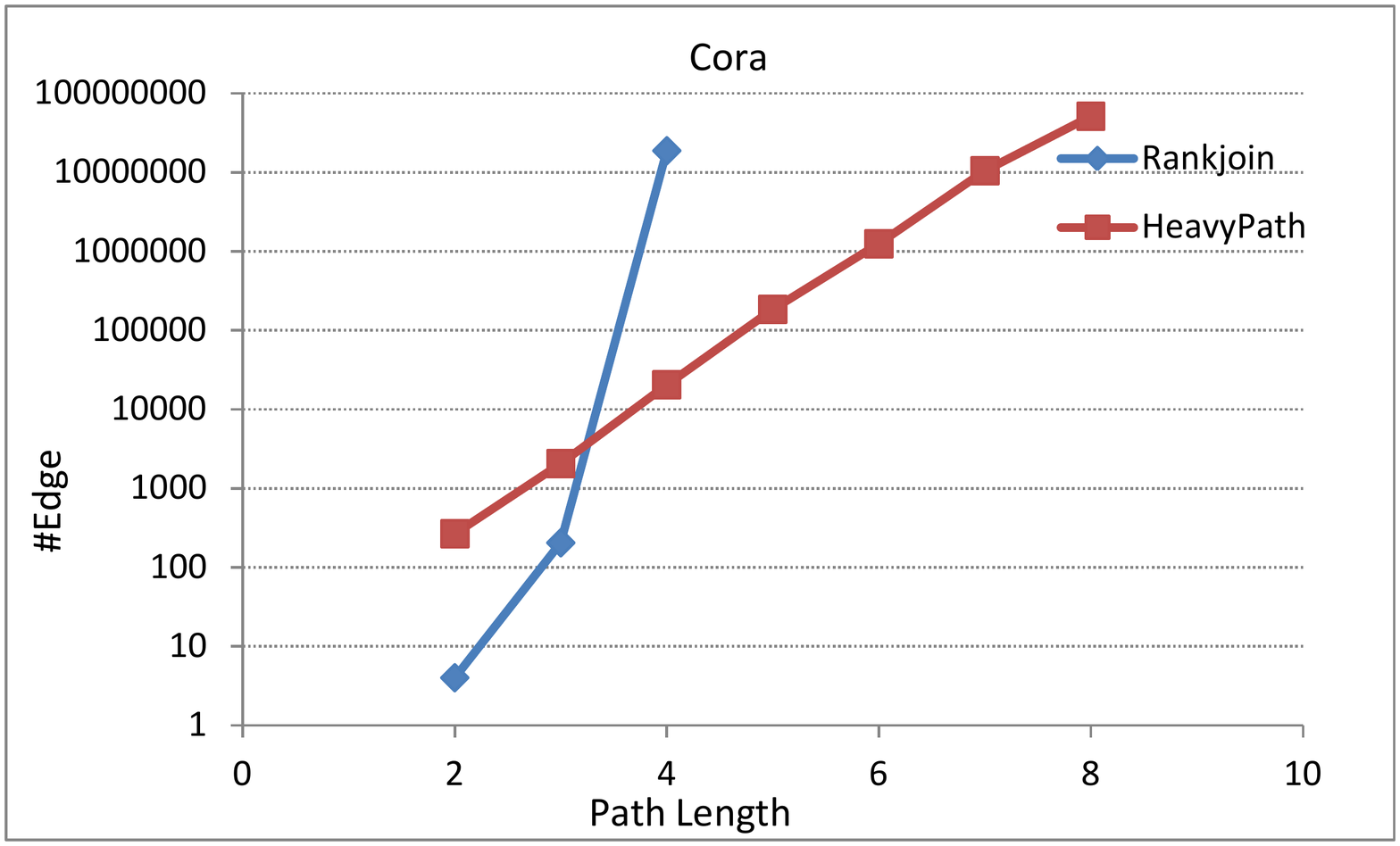}
\label{fig:Ecora}
}
\subfigure[last.fm.]{
\includegraphics[width=\figthree]{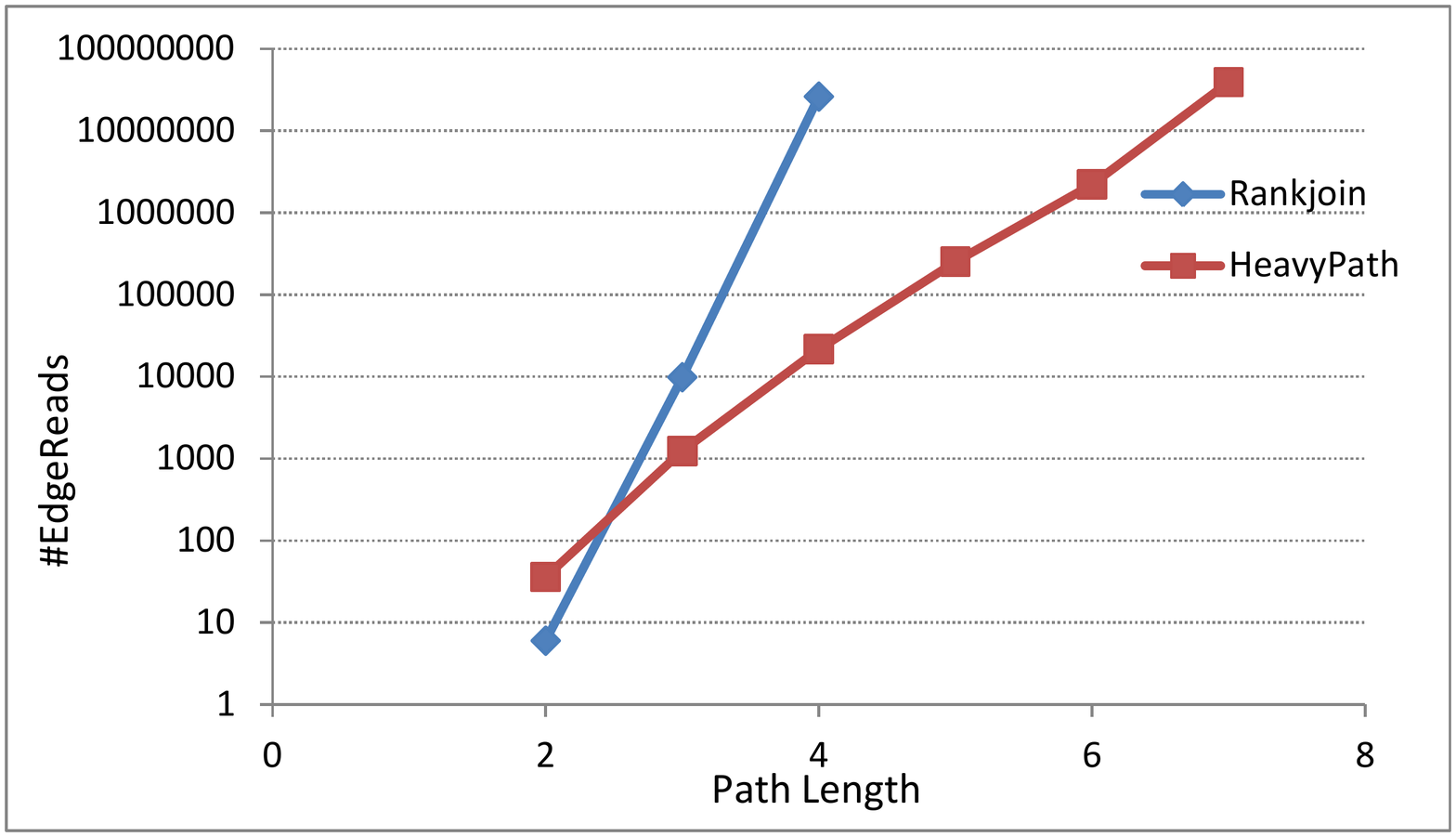}
\label{fig:Elastfm}
}
\subfigure[Bay.]{
\includegraphics[width=\figthree]{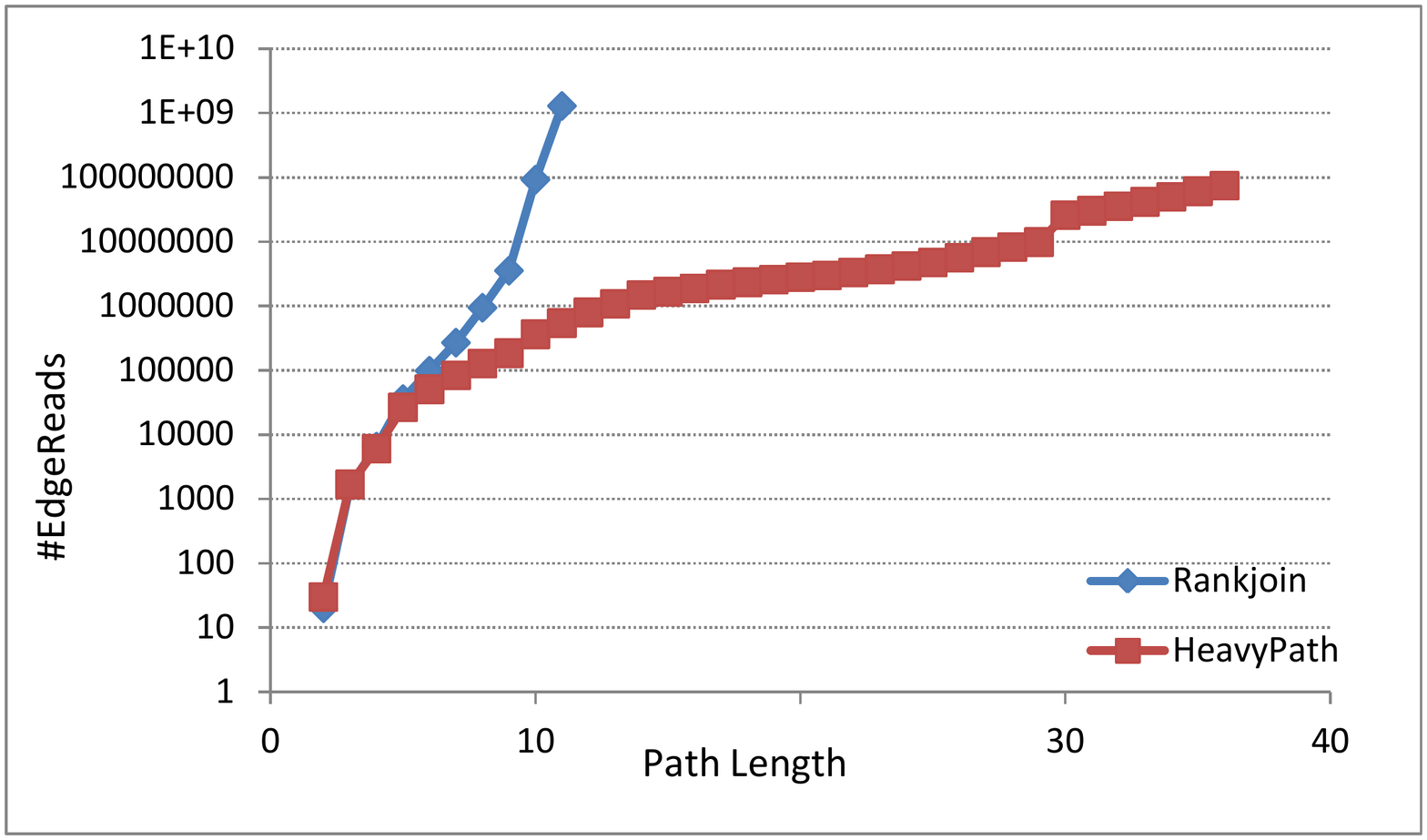}
\label{fig:Ebay}
}
\vspace{-6pt}
\caption{Comparing the number of Edge Reads.}
\label{fig:Ecomp}
\end{figure*}

\begin{figure*}[t!]
\centering
\subfigure[Cora.]{
\includegraphics[width=\figthree]{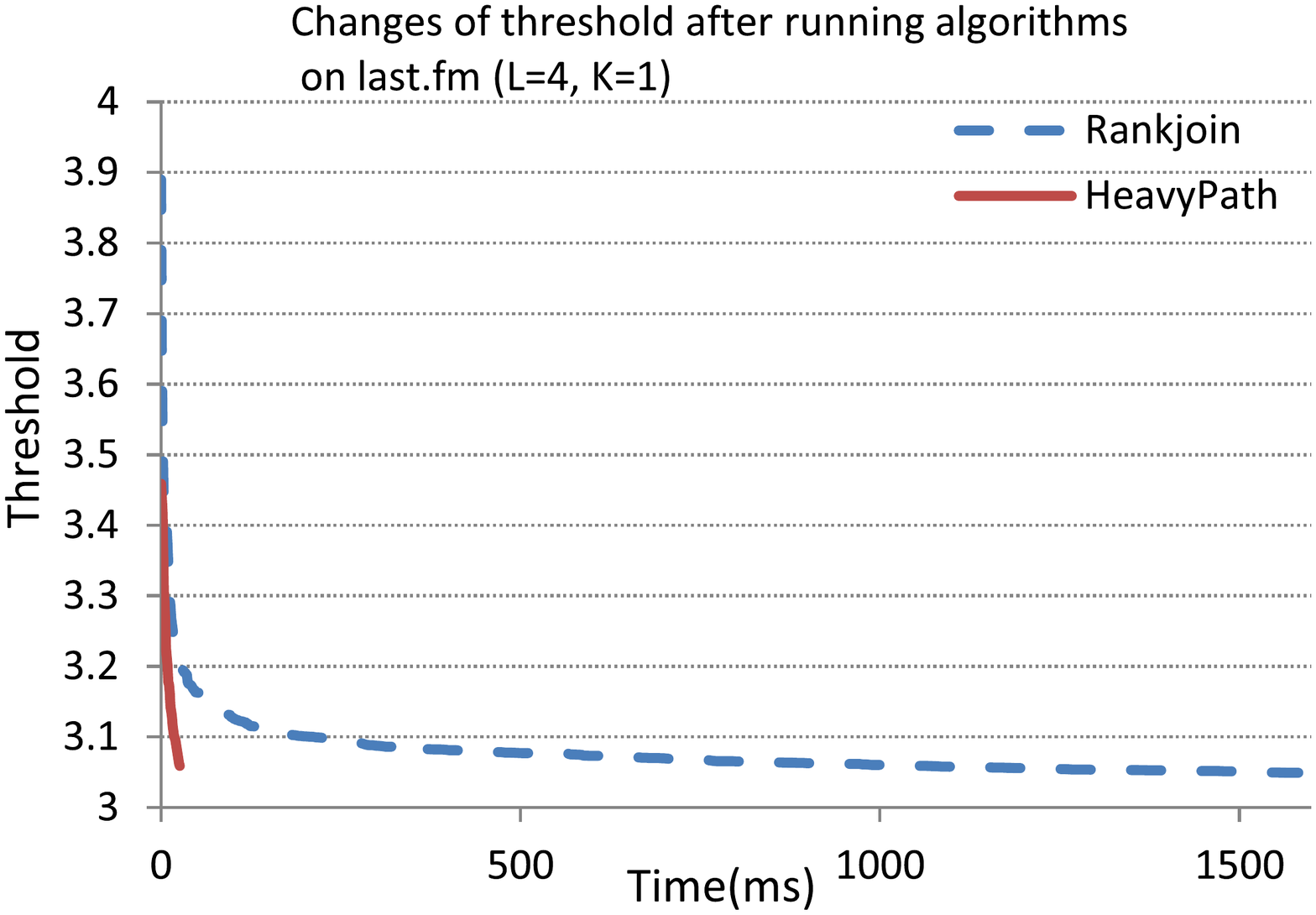}
\label{fig:Tcora}
}
\subfigure[last.fm.]{
\includegraphics[width=\figthree]{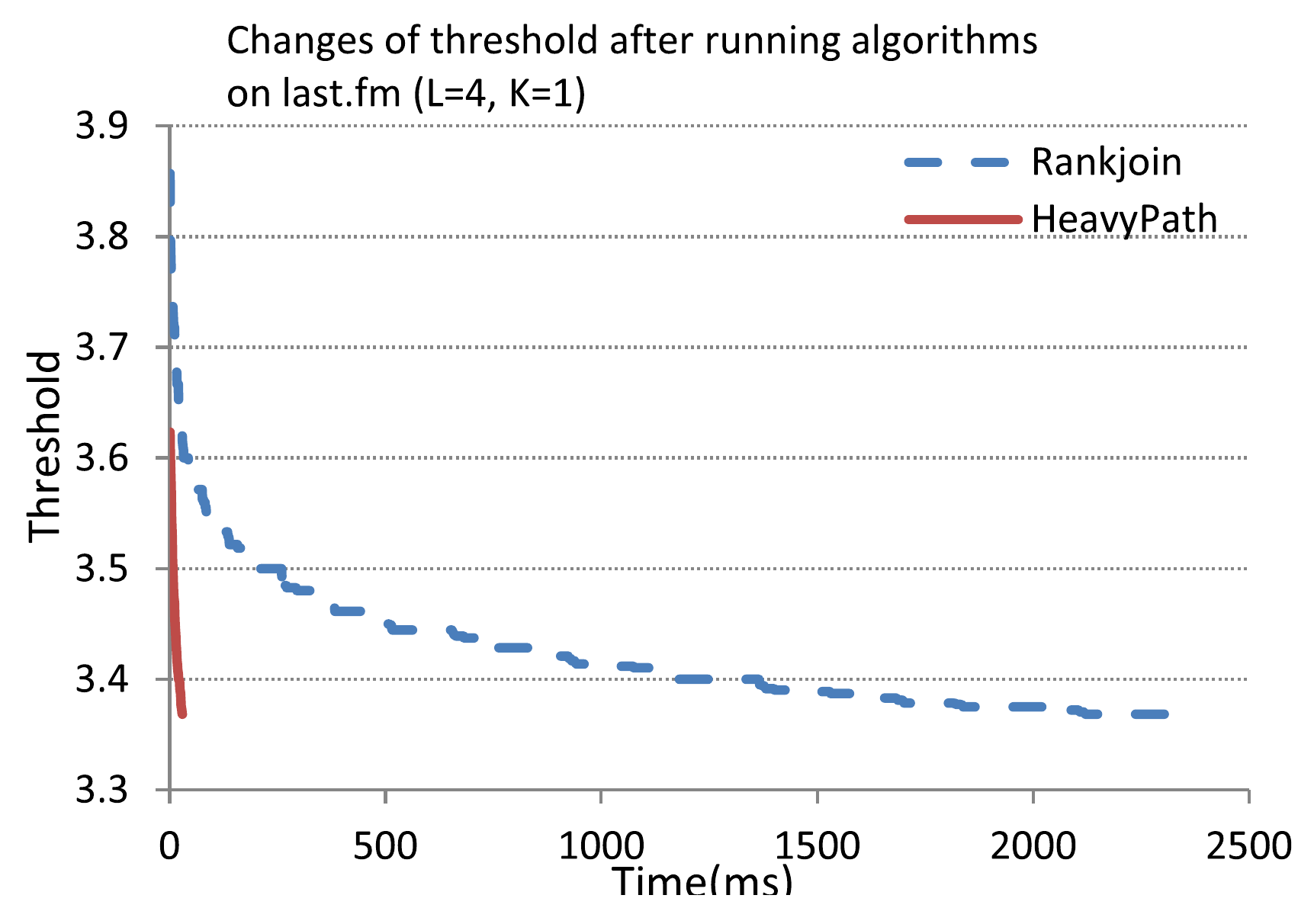}
\label{fig:Tlastfm}
}
\subfigure[Bay.]{
\includegraphics[width=\figthree]{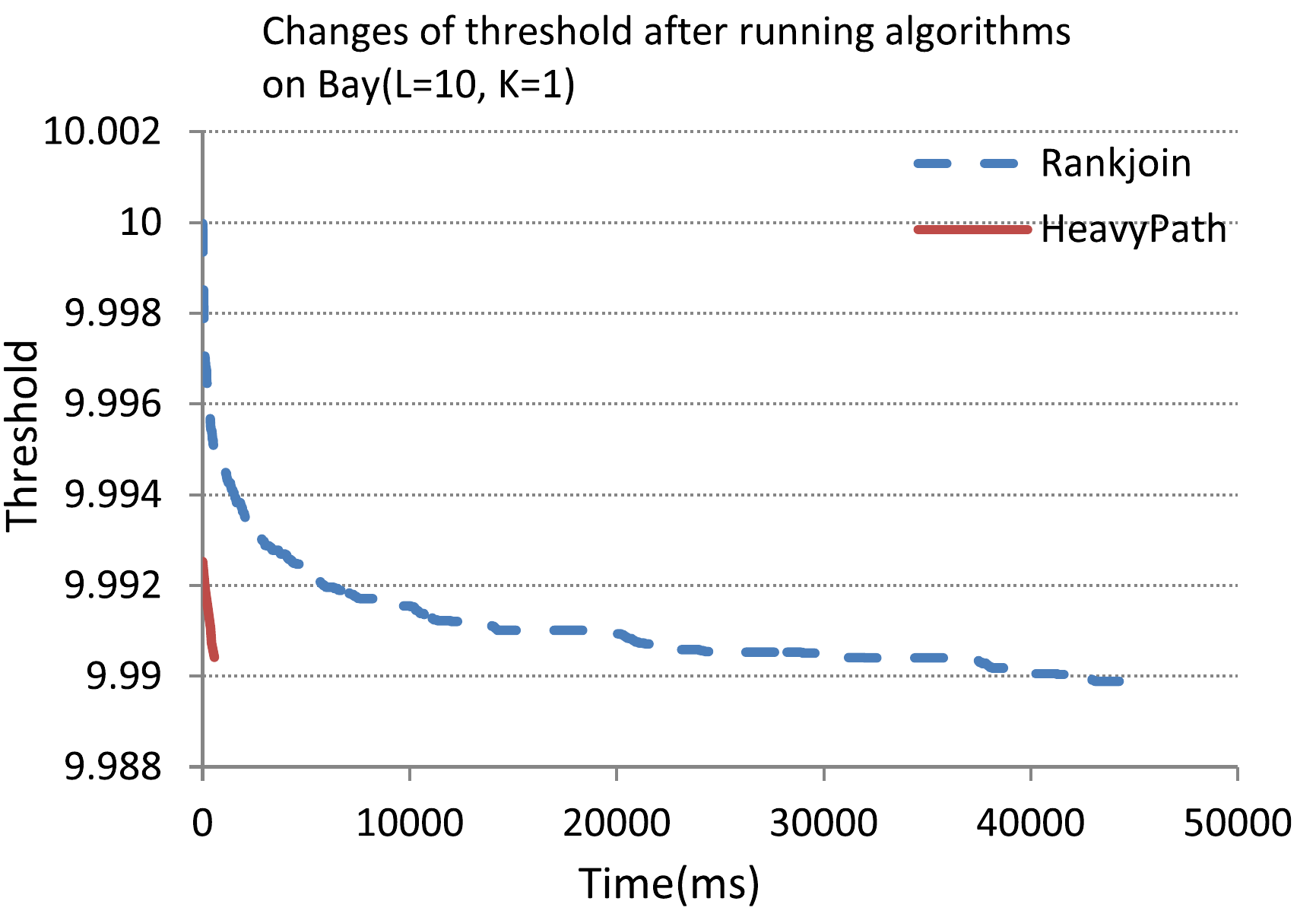}
\label{fig:Tbay}
}\vspace{-6pt}
\caption{Comparing Threshold Decays of Rank Join and HeavyPath}
\label{fig:Thres}
\end{figure*}
\begin{figure*}[t!]
\centering
\subfigure[Cora.]{
\includegraphics[width=\figthree]{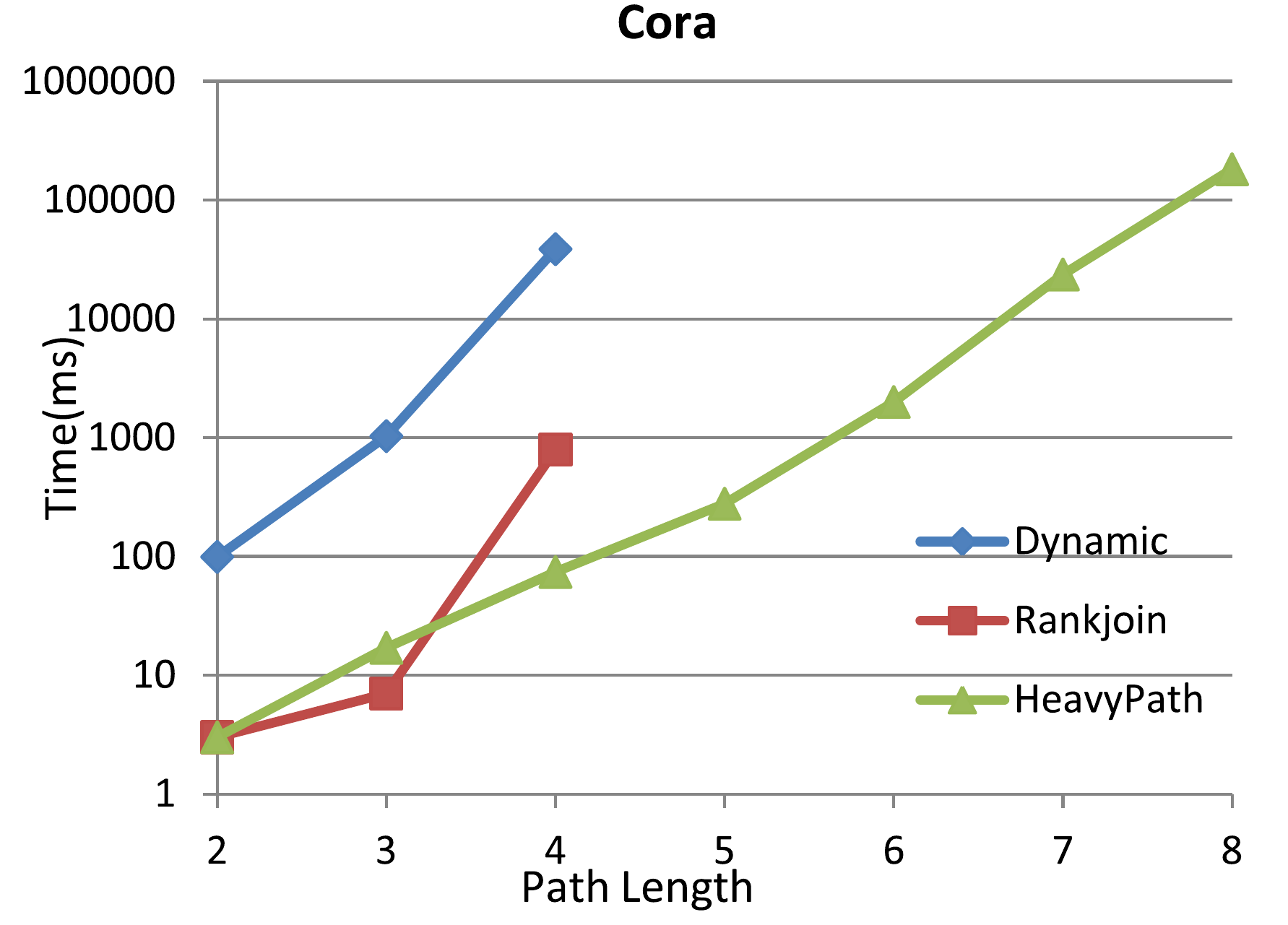}
\label{fig:Gcora}
}
\subfigure[last.fm.]{
\includegraphics[width=\figthree]{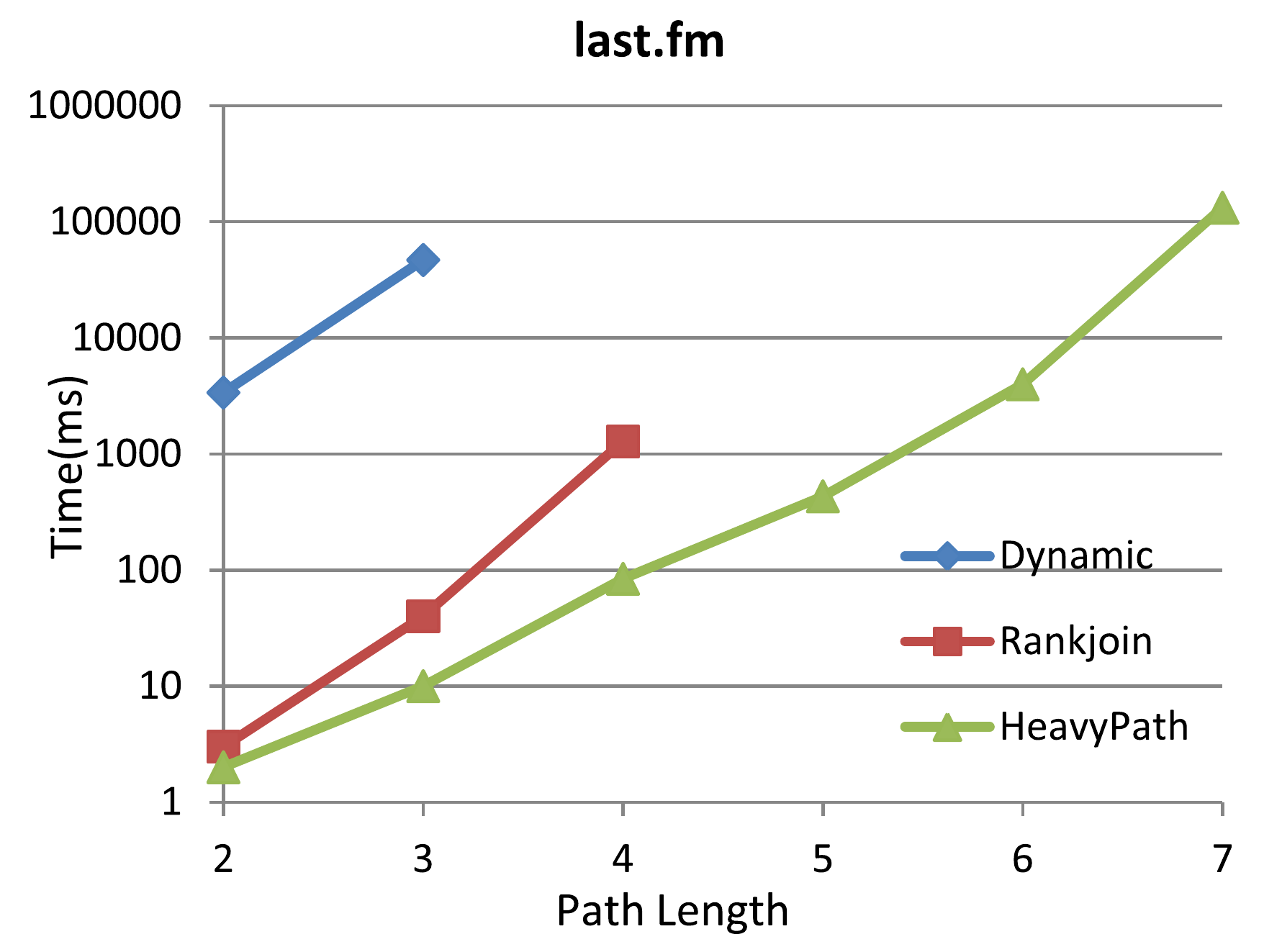}
\label{fig:Glastfm}
}
\subfigure[Bay.]{
\includegraphics[width=\figthree]{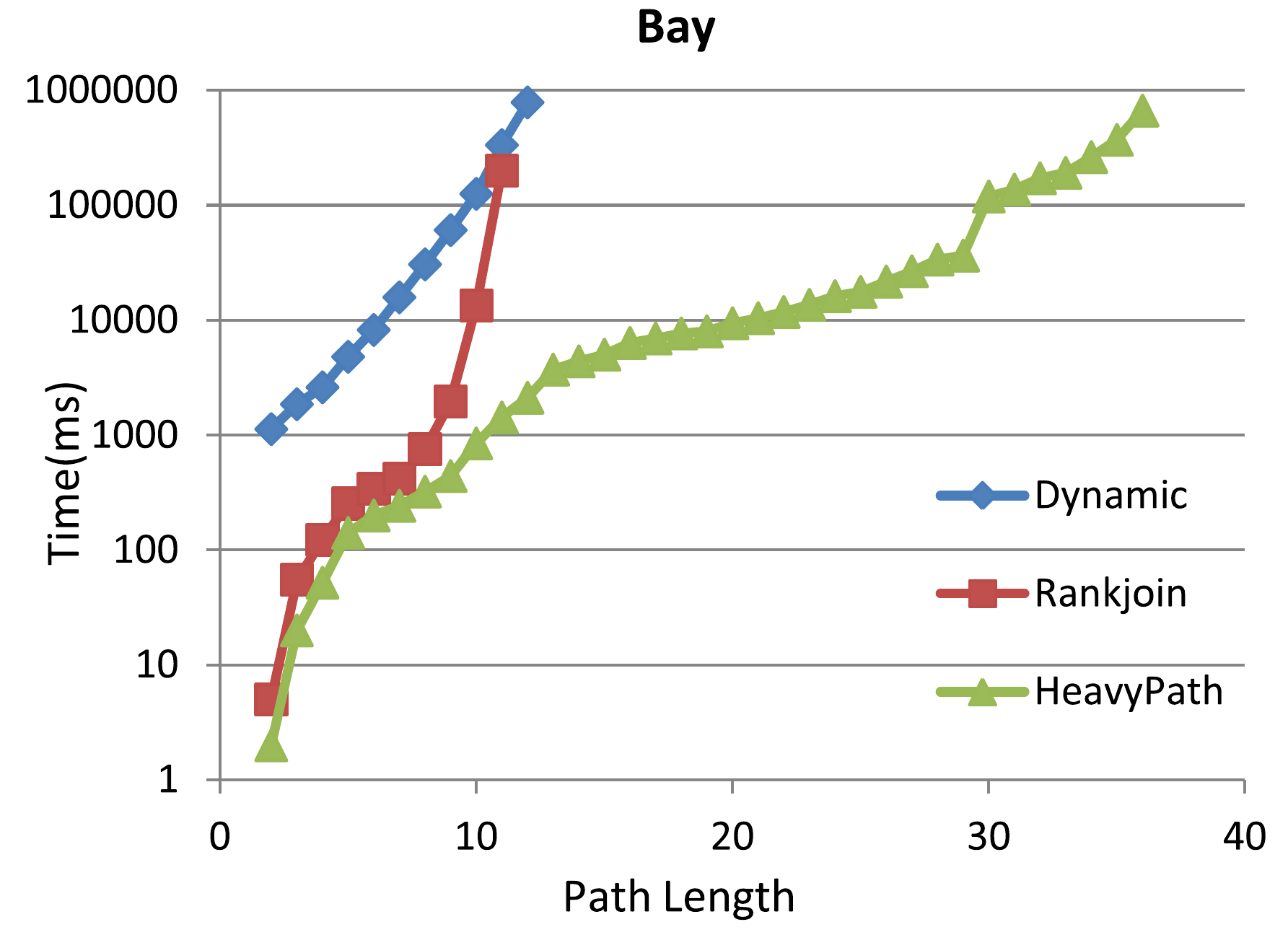}
\label{fig:Tbay}
}\vspace{-6pt}
\caption{Comparing total time of HeavyPath to compute time of other algorithms}
\label{fig:Garbage}
\end{figure*}

\begin{figure*}[t]
\centering
\subfigure[HeavyPath]{
\includegraphics[scale=1]{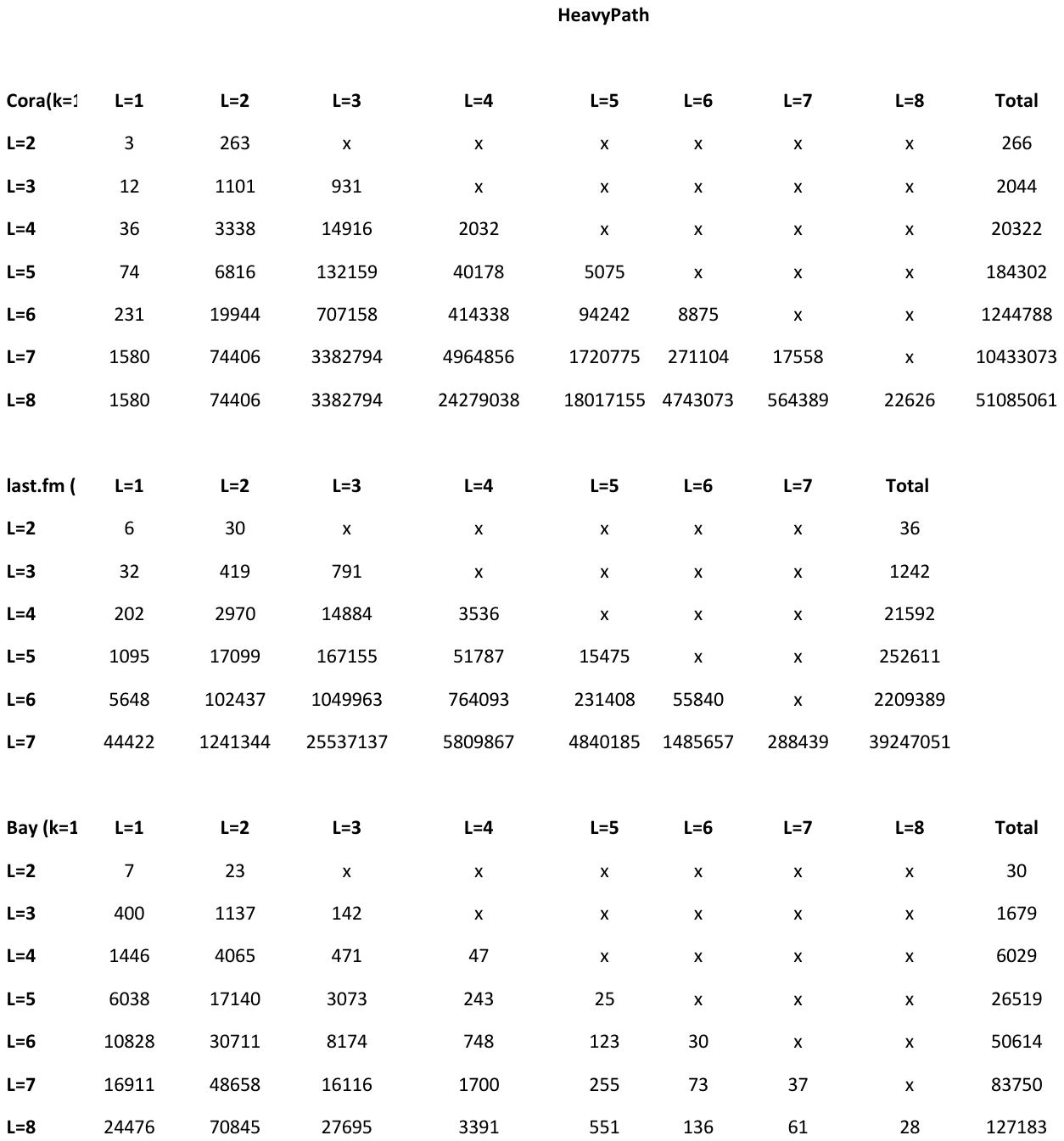}
}
\subfigure[Rank Join]{
\includegraphics[scale=0.7]{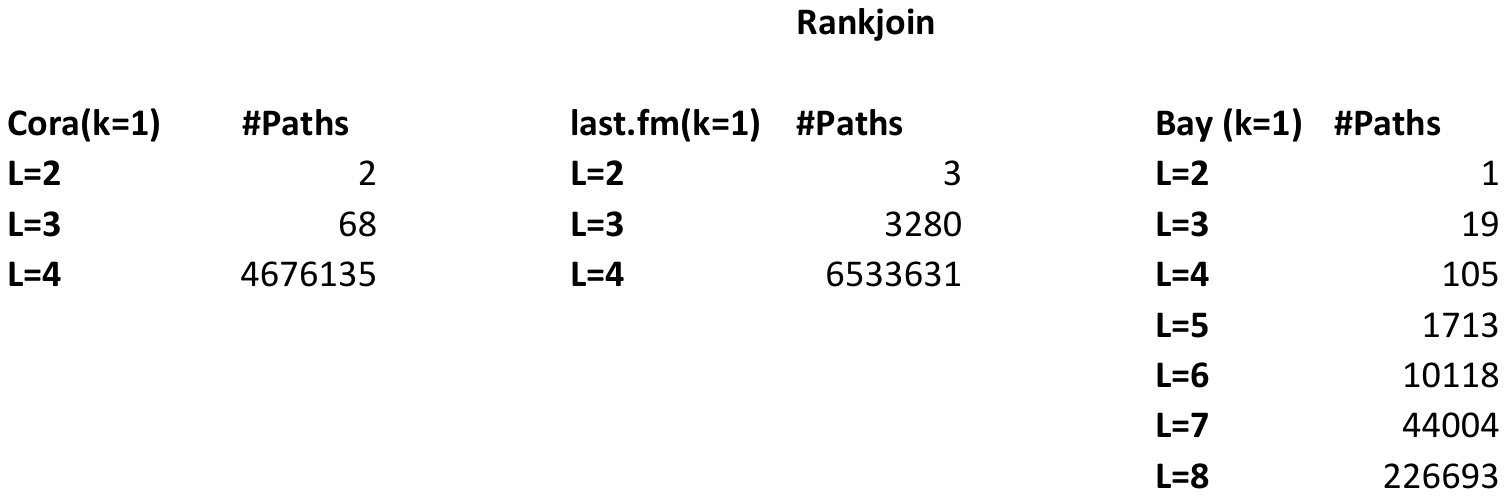}
}
\caption{Comparing the number of paths created by HeavyPath and Rank Join}
\label{fig:Pcomp}
\end{figure*}


Our main result is that \Main is algorithmically superior to all compared algorithms
including \Rank, in terms of the amount of work done. We establish this claim in
multiple levels, by performing additional experiments that include comparing
the number of edges read, the paths created, and the rate at which
the thresholds decay. In all these experiments, we set $k = 1$,
i.e., we focused on the top path.

\para{Edge Reads}
We measured the total number of edge reads performed by \Main and by
\RJ, where every time an edge is read, under sorted or random access or
while performing joins to construct paths, is counted as one. We show the
results in Figure~\ref{fig:Ecomp}. The \# edge reads for \RJ exceeds $10^{10}$
for path lengths beyond 4 on Cora and last.fm and beyond 11 on Bay, which
then runs out of memory. On the other hand, \Main is able to go up to lengths
8, 7, and 36 on the three respective data sets, incurring far fewer edge reads.
The reason for running out of memory has to do with the number of paths
computed and examined, investigated next.

\para{Paths Constructed}
We counted the number of paths. In case of \Rank, we only counted paths of
length $\len$ where \len is the required path length. In case of \Main, we counted
paths of all lengths $\le \len$ that are constructed and stored by it in its
buffers. The results are shown in Figure~\ref{fig:Pcomp}. The relative
performance of \Main and \Rank w.r.t. the number of paths computed by them is
consistent with the trend observed in the previous experiment on \# edge reads,
showing \Main ends up doing significantly less work than \Rank and hence is able
to scale to longer lengths.

\para{Threshold Decay}
Both algorithms rely on their stopping threshold, $\theta$ for \Rank and
$\theta_{\len}$ for \Main for early termination. We measured the rate at which the thresholds decay
since that gives an indication how quickly the algorithm will terminate. Figure~\ref{fig:Thres}
shows the results. For each algorithm, the threshold value is shown until the algorithm
terminates, successfully finding the heaviest path.
It is obvious that in all cases, the threshold $\theta_{\len}$ of \Main
decays extremely fast, whereas in comparison, the threshold $\theta$ used by \Rank decays
much more slowly. Since \Main employs random accesses whereas \Rank doesn't, it is
clear that random access is responsible for the rapid decay of the threshold,
resulting in significant gain in performance. Please notice, the three experiments
above are absolutely unaffected by systems issues, including garbage collection.

\para{Running time}
Finally, we separated the total running time into time spent on garbage collection and
the rest.\footnote{We did this using the \emph{Jstats} tool and would be happy to provide
additional details on this if required.}
Let's call the rest ``compute time'' for convenience. We compare
the \emph{total time} for \Main with the \emph{compute time} for the
other algorithms. It is worth noting here that for the majority of the
points reported in our results (which of course correspond to those
cases where the appropriate algorithm terminated) successfully, garbage collection time was
either $0$ or very small. The result is shown in Figure~\ref{fig:Garbage}. For simplicity,
we ``time out'', i.e., stop the experiment, whenever the time taken exceeds $10^6$ seconds.
It can be observed
that despite taking out the (small to negligible amount of) garbage collection from
the total times of other algorithms, the total time of \Main is still significantly less
than the compute times of other algorithms.

In each of these experiments, we can observe that \Main
significantly outperforms other algorithms. These experiments clearly establish
that \Main's superiority over the other algorithms in terms of the
amount of work done, i.e., in purely algorithmic terms, regardless of systems
issues, and that our findings  are not in any way compromised by garbage collection.

\end{appendix}
\end{document}